\documentclass[aps, oneprint, twoside, superscriptaddress, nofootinbib,
floatfix, notitlepage, onecolumn]{revtex4-1}

\usepackage{amsmath,amsthm}
\usepackage{amssymb,enumerate}
\usepackage{subcaption}
\usepackage{bm}
\usepackage[dvipsnames]{xcolor}
\usepackage{float}
\usepackage[T1]{fontenc}
\usepackage[pdftex]{graphicx}
\usepackage[pdftex,colorlinks=true,linkcolor=blue,citecolor=blue,urlcolor=blue]{hyperref}
\usepackage{comment}
\usepackage{multirow}
\usepackage{tabularx, booktabs}
\usepackage[normalem]{ulem}
\usepackage{tikz}
\usepackage{bm}
\usepackage{pifont}
\usepackage{array}
\usepackage{caption}
\usepackage{cancel}

\newcommand{\bra}[1]{\langle #1 |}
\newcommand{\ket}[1]{| #1 \rangle}
\newcommand{\braket}[1]{\langle #1 \rangle}

\newcolumntype{Y}{>{\centering\arraybackslash}X}

\makeatletter
\newcommand{\vast}{\bBigg@{3}}
\newcommand{\Vast}{\bBigg@{4}}
\newcommand{\VAST}{\bBigg@{6}}
\makeatother

\usepackage{booktabs}

\definecolor{darkgreen}{rgb}{0,0.6,0}
\newtheorem{assumption}{Assumption}
\newtheorem{corollary}{Corollary}
\newtheorem{definition}{Definition}
\newtheorem{lemma}{Lemma}
\newtheorem{notation}{Notation}
\newtheorem{proposition}{Proposition}
\newtheorem{remark}{Remark}
\newtheorem{theorem}{Theorem}

\begin{document}

\title{Applying the quantum approximate optimization algorithm to general constraint satisfaction problems}

\author{Sami Boulebnane\footnote{Now at J.\ P.\ Morgan}}
\affiliation{Phasecraft Ltd.}
\affiliation{University College London}
\author{Maria Ciudad-Alañón\footnote{Now at University of Waterloo}}
\affiliation{Phasecraft Ltd.}
\author{Lana Mineh}
\affiliation{Phasecraft Ltd.}
\author{Ashley Montanaro}
\affiliation{Phasecraft Ltd.}
\affiliation{University of Bristol}
\author{Niam Vaishnav}
\affiliation{Phasecraft Ltd.}

\date{\today}

\begin{abstract}
In this work we develop theoretical techniques for analysing the performance of the quantum approximate optimization algorithm (QAOA) when applied to random boolean constraint satisfaction problems (CSPs), and use these techniques to compare the complexity of a variety of CSPs, such as $k$-SAT, 1-in-$k$ SAT, and NAE-SAT. Our techniques allow us to compute the success probability of QAOA with one layer and given parameters, when applied to randomly generated instances of CSPs with $k$ binary variables per constraint, in time polynomial in $n$ and $k$. We apply this algorithm to all boolean CSPs with $k=3$ and a large number of CSPs with $k=4$, $k=5$, and compare the resulting complexity with the complexity of solving the corresponding CSP using the standard solver MapleSAT, determined experimentally. We find that random $k$-SAT seems to be the most promising of these CSPs for demonstrating a quantum-classical separation using QAOA.
\end{abstract}

\maketitle


\section{Introduction}

The quantum approximate optimization algorithm~\cite{1411.4028} (QAOA) is a prominent quantum algorithm for solving hard combinatorial optimization problems, and one which is particularly well-suited to execution on near-term quantum computers. A large body of work has developed investigating QAOA's performance on a variety of optimization problems \cite{Marwaha2022boundsapproximating,Farhi2022quantumapproximate,Basso2022,basso_et_al:LIPIcs.TQC.2022.7,2402.19456}, where despite some tantalizing numerical hints, there is no theoretical proof that QAOA can outperform the best classical methods. 
Here we instead consider applying QAOA to solve hard constraint satisfaction problems. This is the approach taken in~\cite{2208.06909}, which studied the performance of QAOA applied to random instances of the boolean satisfiability problem with $k$ variables per clause (random $k$-SAT) at the satisfiability phase transition. In that work, a combination of theoretical and numerical bounds gave evidence that QAOA could outperform the best classical algorithm tested, for sufficiently large problem instances.

Here, we generalise the approach of~\cite{2208.06909} by systematically determining the complexity of solving randomly generated instances of other boolean constraint satisfaction problems (CSPs) using QAOA. The framework we consider is as follows. An instance of a CSP on $n$ variables is made up of a number of constraints (also known as clauses), each of which depends on a small number of boolean variables and rules out certain potential solutions. All of the constraints must be satisfied for an instance to be a ``yes'' instance. Here we define a CSP in terms of the types of constraint that are allowed. For example, in $k$-SAT, each constraint is of the form $l_{i_1} \vee l_{i_2} \vee \dots \vee l_{i_k}$, where $i_1,\dots,i_k \in \{1,\dots,n\}$ are indices, and $l_i$ are literals: either variables $x_i$ directly, or negations of variables $\overline{x_i}$. Another example of a problem that fits into this framework is $k$-XOR-SAT, where each constraint is of the form $x_{i_1} \oplus \dots \oplus x_{i_k}$. Different CSPs can have very different levels of complexity; for example, $k$-SAT is NP-complete for $k \ge 3$, whereas $k$-XOR-SAT is in P.

We generate random instances of CSPs by picking the constraints at random, from the family of allowed constraints. We usually assume that the constraint itself is fixed, and what may vary is which literals it is applied to; and also that the subset of literals chosen is uniformly random. The number of constraints is usually chosen according to a Poisson distribution with a given mean.

The QAOA algorithm consists of $p$ layers, each of which is of the form $e^{i \beta B} e^{i \gamma C}$, where $B$ is a so-called \emph{mixer} Hamiltonian (often $B = \sum_j X_j$) and $C = \sum_{x \in \{0,1\}^n} C(x) \ket{x}\bra{x}$ is the diagonal \emph{cost} Hamiltonian, where $C(x)$ gives the cost of assignment $x \in \{0,1\}^n$. In the case of CSPs, usually $C(x)$ is defined as the number of clauses unsatisfied by $x$. Then $C(x) = 0$ if and only if $x$ is a satisfying assignment.

Here we will focus on the case of QAOA with only one layer ($p=1$). This is significantly easier to analyse than the general case, while still enabling us to compare the relative complexities of various CSPs. The main results we obtain are as follows:

\begin{enumerate}
    \item We develop a (classical) algorithm for evaluating the success probability of one-layer QAOA (averaged over random instances) which runs in time $\mathcal{O}(n^3)$. The algorithm takes as input the probabilities that a randomly chosen clause is simultaneously violated by up to three given bit-strings. These quantities can sometimes be written down analytically. In other cases, we give an algorithm for computing them in time $\mathcal{O}\left(k^7\right)$, where $k$ is the number of literals per clause.
    \item We compute the success probabilities for all CSPs based on constraints on 3 bits, and a selection of CSPs based on constraints on 4 and 5 bits, and use these to approximately compute the runtime scaling of QAOA for these problems when near the satisfiability threshold. We then compare the theoretical performance of QAOA with the runtime scaling of a leading classical solver MapleSAT, determined numerically. We find that, for all the problems considered, the scaling of MapleSAT appears to be more efficient than the scaling of QAOA. This is not unexpected, given that we only analyse $p=1$ QAOA, and in~\cite{2208.06909}, it was necessary to go to significantly larger $p$ to outperform classical approaches.
    Based on our numerical results, the most challenging problem for both classical and quantum approaches seems to be $k$-SAT.
\end{enumerate}

In addition, we apply our approach to the well-studied constraint satisfaction problems 1-in-$k$ SAT and NAE-SAT, where the required probabilities can be computed analytically. We compare the theoretical results with numerical predictions and observe an excellent level of agreement. These results are deferred to appendices.

If our results for $p=1$ are representative of the relative performance of QAOA and classical algorithms for larger $p$, they suggest that within the family of problems which can be expressed as randomly generated CSPs, the most promising problem for demonstrating a quantum-classical separation using QAOA is random $k$-SAT, as studied in~\cite{2208.06909}.

\section{Background}

\subsection{Definitions}

\subsubsection{Constraint satisfaction problems}
\label{sec:constraint_satisfaction_problems}

In this section, we introduce definitions and notations common to all constraint satisfaction problems considered in this paper. We start by giving a general description of the random constraint satisfaction problems considered in this work. Each of these problems is entirely characterized by a truth table over $k$ bits as defined below:

\begin{definition}[General truth table]
	\label{def:general_truth_table}
	A truth table on $k$ bits is a Boolean function: $T: \{0, 1\}^k \longrightarrow \{0, 1\}$ mapping each sequence of $k$ bits to $1$ (true) or $0$ (false).
\end{definition}

For notational convenience, we may interchangeably work with bits taking values in $\{0, 1\}$ or values in $\{1, -1\}$ according to the following convention:
\begin{notation}[Bits in $\{0, 1\}$ vs. bits in $\{1, -1\}$]
	The correspondence between bit values in the $\{0, 1\}$ convention and bit values in the $\{1, -1\}$ convention is given in the following table:
	$$
	\begin{array}{c|c|c}
		\mathrm{Value\,in\,} \{0, 1\} \mathrm{\,convention} & \mathrm{Value\,in\,} \{1, -1\} \mathrm{\,convention} & \mathrm{Boolean\,interpreation}\\
		\hline
		0 & 1 & \textsc{false}\\
		1 & -1 & \textsc{true}
	\end{array}
	$$
\end{notation}

Given a truth table, one can define a non-random instance of a constraint satisfaction problem based on this table:
\begin{definition}[Constraint satisfaction problem instance]
	\label{def:csp_instance}
	Let a truth table $T$ over $k$ bits be given as introduced in definition \ref{def:general_truth_table} and fix integers $n \geq 1$ (number of variables) and $m \geq 0$ (number of clauses). A constraint satisfaction problem instance with $n$ variables and $m$ clauses is specified by $m$ clauses $\sigma_0, \sigma_1, \ldots, \sigma_{m - 1}$. A clause $\sigma_j$ is an ordered $k$-tuple of pairs 
	\begin{align}
		\sigma_j & = \left(\left(l_{j,\,0},\,\nu_{j,\,0}\right),\,\left(l_{j,\,1},\,\nu_{j,\,1}\right),\,\ldots,\,\left(l_{j,\,k - 1},\,\nu_{j,\,k - 1}\right)\right),
	\end{align}
	where
	\begin{align}
		l_{j,\,q} & \in [n] && \forall j \in [m],\,\forall q \in [k],\\
		\nu_{j,\,q} & \in \{0, 1\} && \forall j \in [m],\,\forall q \in [k].
	\end{align}
	The $l_{j,\,q}$ represent variable indices, and the $\nu_{j,\,q}$ the negation associated to variable with index $l_{j,\,q}$. We may also occasionally denote
	\begin{align}
		\left(l, \nu\right) \in \sigma_j
	\end{align}
	for iterating $\left(l, \nu\right)$ over $\left(l_{j,\,0}, \nu_{j,\,0}\right)$, $\left(l_{j,\,1}, \nu_{j,\,1}\right)$, $\ldots$, $\left(l_{j,\,k - 1}, \nu_{j,\,k - 1}\right)$ when the order of the enumeration is irrelevant. The iteration should be understood as over a multiset since there may be repetitions in pairs $\left(l_{j,\,q}, \nu_{j,\,q}\right)$ as $q \in [k]$.
	
	Given an assignment
	\begin{align}
		\bm{x} = \left(x_0, x_1, \ldots, x_{n - 1}\right) \in \{0, 1\}^n
	\end{align}
	of variables, the truth value of clause $\sigma_j$ for this assignment is defined as:
	\begin{align}
		T\left(x_{l_{j,\,0}} \oplus \nu_{j,\,0},\,x_{l_{j,\,1}} \oplus \nu_{j,\,1},\,\ldots,\,x_{l_{j,\,k - 1}} \oplus \nu_{j,\,k - 1}\right) = T\left[\left(x_{l_{j,\,q}} \oplus \nu_{l_{j,\,q}}\right)_{q \in [k]}\right]
	\end{align}
	We use notation
	\begin{align}
		\bm{x} \vdash \sigma_j \iff T\left[\left(x_{l_{j,\,q}} \oplus \nu_{j,\,q}\right)_{q \in [k]}\right] = 1 \iff ``\bm{x} \mathrm{\,\,satisfies\,\,}\sigma_j",\\
		\bm{x} \not\vdash \sigma_j \iff T\left[\left(x_{l_{j,\,q}} \oplus \nu_{j,\,q}\right)_{q \in [k]}\right] = 0 \iff ``\bm{x} \mathrm{\,\,violates\,\,}\sigma_j".
	\end{align}
	to signify an assignment satisfies or violates a clause. We will frequently extend this notation to include several assignments, e.g.
    \begin{align}
        \bm{x},\,\bm{y} \vdash \sigma_j \iff \bm{x} \textrm{ satisfies } \sigma_j \textrm{ and } \bm{y} \textrm{ satisfies } \sigma_j\\
        \bm{x},\,\bm{y} \not\vdash \sigma_j \iff \bm{x} \textrm{ violates } \sigma_j \textrm{ and } \bm{y} \textrm{ violates } \sigma_j
    \end{align}
    for two assignments $\bm{x}, \bm{y} \in \{0, 1\}^n$.
    The truth of the problem instance defined by clauses $\sigma_0, \sigma_1, \ldots, \sigma_{m - 1}$ for some assignment is defined by simultaneous truth of all clauses for this assignment:
    \begin{align}
        \bm{x} \vdash \bm\sigma & \iff \bm{x} \vdash \sigma_j \quad \forall j \in [m], \qquad \textrm{ where } \bm\sigma = \left(\sigma_0, \ldots, \sigma_{m - 1}\right)
    \end{align}
\end{definition}

After defining a general constraint satisfaction problem instance, one may construct such an instance randomly:
\begin{definition}[Random problem instance]
	\label{def:csp_random_instance}
	Let be given a truth table $T$ on $k$ bits as introduced in definition \ref{def:general_truth_table} and integers $n \geq 1$ (number of variables) and $m \geq 0$ (number of clauses). A random instance of a constraint satisfaction problem with $n$ variables and $m$ clauses is constructed by sampling $m$ clauses $\sigma_j$, $0 \leq j < m$, identically independently. Each clause
	\begin{align}
		\sigma_j = \left(\left(l_{j,\,0},\,\nu_{j,\,0}\right),\,\left(l_{j,\,1},\,\nu_{j,\,1}\right),\,\ldots,\,\left(l_{j,\,k - 1},\,\nu_{j,\,k - 1}\right)\right)
	\end{align}
	is constructed by choosing each variable index $l_{j,\,q}$ ($0 \leq q < k$) 
	independently and uniformly in $\{0, \ldots, n - 1\}$, and each negation $\nu_{j,\,0}$ independently and with equal probability $0$ or $1$. We usually denote a single clause randomly sampled from this distribution by $\sigma$, and denote by
	\begin{align}
		\mathbf{E}_{\sigma}f\left(\sigma\right) \qquad \sigma: \mathrm{random\,clause}
	\end{align}
	the expectation of a function $f(\sigma)$ of this random clause.
    In fact, in most of this work, we will consider problem instances with a given \textit{expected} number of clauses rather than a fixed number of clauses. Given a integer $n \geq 1$ and a positive number $r > 0$, a random problem instance with $n$ variables and expected clauses-to-variables ratio $r$ is defined by sampling
    \begin{align}
        m \sim \mathrm{Poisson}\left(rn\right)
    \end{align}
    and sampling a random instance with $n$ variables and (exactly) $m$ clauses as specified above.
\end{definition}

\begin{remark}[Repetition of variables in clauses]
	Since the indices $l_{j,\,q}$ of variables occuring in the clause are chosen independently, Definition \ref{def:csp_random_instance} of a random constraint satisfaction problem instance allows for repeated variables inside a clause. Similarly, since choices of negations $\nu_{j,\,q}$ are also independent, a repeated variable may occur with two different negations (true and false) in the clause. If this situation occurs it can lead to the clause being trivially true or false for some choices of truth tables $T$. It would be possible to enforce choice of variable indices without repetition to avoid these edge cases, but at least empirically, both choices give comparable results, and both seem to be accepted conventions in the optimization literature.
\end{remark}

\subsubsection{The Quantum Approximate Optimization Algorithm}

The Quantum Approximate Optimization Algorithm (QAOA), introduced by Farhi et al.~\cite{1411.4028} (although an essentially identical algorithm was previously proposed by Hogg~\cite{hogg00}), is a variational quantum algorithm for addressing combinatorial optimization or constraint satisfaction problems. This variational circuit consists of a certain number of layers, commonly denoted by $p$ and also occasionally called the \textit{QAOA depth} (which may be distinct from the quantum circuit depth). Each layer depends on two variational parameters, also known as \textit{QAOA angles}, to be trained. In this work, we specialize to $p = 1$ QAOA, meaning the variational circuit depends on only two parameters
\begin{align}
\label{eq:qaoa_angles}
    \gamma, \beta \in \mathbf{R}
\end{align}
In its simplest form (sufficient for this work), QAOA aims at finding a bitstring $\bm{x} \in \{0, 1\}^n$ minimizing a classical cost function
\begin{align}
    C\left(\bm{x}\right), \qquad \bm{x} \in \{0, 1\}^n
\end{align}
It does so by preparing a variational quantum state and measuring the latter in the computational basis, providing candidate solution bitstrings:

\begin{definition}[$p = 1$ QAOA variational state]
\label{def:qaoa_variational_state}
    Given a classical cost function $C$ of an $n$-bit bitstring, $p = 1$ QAOA with angles $\gamma, \beta \in \mathbf{R}$ prepares the following variational state:
    \begin{align}
        \ket{\Psi(\gamma, \beta)} & = \exp\left(-\frac{i\beta}{2}H_B\right)\exp\left(-\frac{i\gamma}{2}H_C\right)\ket{+}^{\otimes n},
    \end{align}
    where
    \begin{align}
        H_B & = \sum_{j \in [n]}X_j
    \end{align}
    is a transverse field Hamiltonian and
    \begin{align}
        H_C & = \sum_{\bm{x} \in \{0, 1\}^n}C(\bm{x})\ket{\bm{x}}\bra{\bm{x}}
    \end{align}
    is the Hamiltonian diagonal in the computational basis corresponding to cost function $C$.
\end{definition}

Note that due to its variational nature, QAOA is not an entirely specified algorithm: one should find $\gamma, \beta$ producing satisfying solutions. One may consider optimizing $\gamma, \beta$ for each problem instance, using a loss function that can be efficiently estimated from bitstrings sampled from the variational state. In this work, we pursue a different approach~\cite{fixed_parameters_maxcut_qaoa} of sampling problem instances from a random ensemble, and using common parameters for all of these. Parameters are trained by analytically computing the instance-averaged success probability.

More precisely, in this paper we apply QAOA to a constraint satisfaction problem as defined in section \ref{sec:constraint_satisfaction_problems}, defined by a set of clauses
\begin{align}
    \bm\sigma = \left(\sigma_0, \ldots, \sigma_{m - 1}\right)
\end{align}
The cost function $C_{\bm\sigma}\left(\bm{x}\right)$ to optimize over bitstrings $\bm{x} \in \{0, 1\}^n$ counts the number of constraints violated by $\bm{x}$:
\begin{align}
    C_{\bm\sigma}\left(\bm{x}\right) & = \sum_{j \in [m]}\mathbf{1}\left[\bm{x} \not\vdash \sigma\right]
\end{align}
The corresponding QAOA variational trial state is denoted
\begin{align}
    \ket{\Psi\left(\gamma, \beta, \bm\sigma\right)} & = \exp\left(-\frac{i\beta}{2}H_B\right)\exp\left(-\frac{i\gamma}{2}C_{\bm\sigma}\right)\ket{+}^{\otimes n},
\end{align}
where we have committed a slight abuse of notation identifying $C_{\bm\sigma}$ and its corresponding diagonal Hamiltonian. Then, assuming the instance is satisfiable, the sought optimum of $C_{\bm\sigma}$ is $0$ (no violated constraint), and the success probability for a specific instance can be written as:
\begin{align}
\label{eq:qaoa_success_probability}
    \sum_{\substack{\bm{x} \in \{0, 1\}^n\\C_{\bm\sigma}}(\bm{x}) = 0}\left|\braket{\bm{x}|\Psi(\gamma, \beta, \bm\sigma)}\right|^2
\end{align}

\subsubsection{Analysis of QAOA}

In this paragraph, we now introduce notations and definitions specific to the analysis of QAOA. We start by recalling the definition of configuration basis numbers, first considered in \cite{Farhi2022quantumapproximate} to obtain closed-form expressions for the performance of QAOA in the average-instance case. Note that compared to the previous work, we specialize the definition straightaway to $p = 1$ QAOA which is the focus of the current paper:

\begin{definition}[Configuration basis numbers]
\label{def:configuration_basis_numbers}
Let $q \geq 1$ be an integer. A set of configuration basis numbers of weight $q$ is a family of non-negative integers
\begin{align}
    \bm{q} = \left(q_s\right)_{s \in \{0, 1\}^3}
\end{align}
indexed by $3$-bit bitstrings $s \in \{0, 1\}$ and summing to $q$:
\begin{align}
    \sum_{s \in \{0, 1\}^3}q_s & = q.
\end{align}
The set of configuration basis numbers of weight is denoted $\mathcal{P}(q)$.
\end{definition}

\begin{remark}
\label{rem:configuration_basis_number_notation}
Note definition \ref{def:configuration_basis_numbers} does not require the vector of configuration basis numbers:
\begin{align}
    \bm{q} & = \left(q_{000}, q_{001}, q_{010}, q_{011}, q_{100}, q_{101}, q_{110}, q_{111}\right)
\end{align}
to share the same letter as partitioned integer $q$. However, in practice, we will always make this convenient notational choice in the present work. We will consider configuration basis numbers partitioning either the number of Boolean variables in the problem ($q = n$) or the number of variables per clause ($q = k$).
\end{remark}

Note there are $\binom{q + 6}{7} = \mathcal{O}\left(q^7\right)$ sets of configuration basis numbers of weight $q$. To each triplet of bitstrings of length $q$, one may associate a set of configuration basis numbers according to the following definition:

\begin{definition}[Configuration basis numbers associated to bitstring triplet]
\label{def:bitstring_triplet_configuration_basis_numbers}
Let be given an integer $q \geq 1$ and $3$ length-$q$ bitstrings $\bm{w}^{[1]}, \bm{w}^{[0]}, \bm{w}^{[-1]} \in \{0, 1\}^q$. We will often refer to these as a bitstring triplet $\left(\bm{w}^{[1]},\,\bm{w}^{[0]},\,\bm{w}^{[-1]}\right)$. Then, the weight-$q$ configuration basis numbers $\left(q_s\right)_{s \in \{0, 1\}^3}$ associated to this triplet are:
\begin{align}
    q_{s^{[1]}s^{[0]}s^{[-1]}} & := \left|\left\{j \in [q]\,:\,\left(w^{[1]}_j, w^{[0]}_j, w^{[-1]}_j\right) = \left(s^{[1]}, s^{[0]}, s^{[-1]}\right)\right\}\right| \qquad \left(s^{[1]}, s^{[0]}, s^{[-1]}\right) \in \{0, 1\}^3.
\end{align}
Conversely, given a set of configuration basis numbers $\left(q_s\right)_{s \in \{0, 1\}^3}$, the bitstring triplets $\left(\bm{w}^{[1]}, \bm{w}^{[0]}, \bm{w}^{[-1]}\right)$ having these configuration basis numbers associated are said to \textnormal{satisfy} configuration basis numbers $\left(q_s\right)_{s \in \{0, 1\}^3}$.
\end{definition}

\begin{remark}[Interpretation of configuration basis numbers]
Configuration basis numbers can intuitively be regarded as a higher-dimensional generalization of the Hamming weight. Indeed, given weight-$q$ configuration basis numbers $\left(q_s\right)_{s \in \{0, 1\}^3}$, one may construct the linear combination of computational basis states defined by satisfying these numbers:
\begin{align}
    \sum_{\substack{\bm{w}^{[1]},\,\bm{w}^{[0]},\,\bm{w}^{[-1]} \in \{0, 1\}^q\\\left(\bm{w}^{[1]}, \bm{w}^{[0]}, \bm{w}^{[-1]}\right)\mathrm{\,satisfies\,}\left(q_s\right)_{s \in \{0, 1\}^3}}}\hspace*{-50px}\ket{\bm{w}^{[1]}} \otimes \ket{\bm{w}^{[0]}} \otimes \ket{\bm{w}^{[-1]}} \in \left(\left(\mathbf{C}^2\right)^{\otimes q}\right)^{\otimes 3},
\end{align}
which one regards as a state over $3$ systems of $q$ qubits. These states are easily checked to be pairwise orthogonal as indexed by $\left(q_s\right)_{s \in \{0, 1\}^3}$, and invariant under tensor product permutations
\begin{align}
    P_{\tau}^{\otimes 3},
\end{align}
where $\tau$ is an arbitrary permutation of $[q]$ and $P_{\tau}$ is the permutation of $q$ qubits induced by $\tau$. Conversely, it is easily checked that any state invariant under all such tensor product permutations is in the span of the states constructed above. These states can then be seen as generalizations of Dicke states, with configuration basis numbers generalizing the Hamming weight under this analogy. Note that since these basis states are orthogonal and in one-to-one correspondence with the $\mathcal{O}\left(q^7\right)$ configuration basis numbers of weight $q$, they span a space of the same dimension ---polynomial rather than exponential in $q$. As shown more explicitly in e.g. \cite{2110.10685}, this dimension reduction naturally occurs as a result of the averaging over random problem instances. This fact provides some insight on why computing QAOA expectations averaged over instances is much easier than computing the same expectations on special instances. On the other hand, it does not lead to a practical method of evaluating the expectation at QAOA depth $p > 1$ since the degree of the polynomial grows too large \cite{Farhi2022quantumapproximate,Basso2022,2208.06909}.
\end{remark}

We now give a few combinatorial lemmas on configuration basis numbers. In this context, we start by recalling the multinomial theorem, generalizing the standard binomial theorem. This requires the notion of multinomial coefficients:
\begin{definition}[Multinomial coefficient]
\label{def:multinomial_coefficient}
Let be given integers $q \geq 0, r \geq 1$ and a partition of $q$ into $r$ non-negative integers:
\begin{align}
    q & = \sum_{j \in [r]}q_j,\\
    q_j & \geq 0 \qquad \forall j \in [r].
\end{align}
We then define the multinomial coefficient
\begin{align}
\label{eq:multinomial_coefficient}
    \binom{q}{\left(q_j\right)_{j \in [r]}} & := \binom{q}{q_0, q_1, \ldots, q_{r - 1}} := \frac{q!}{\prod\limits_{j \in [r]}q_j!}.
\end{align}
When integers $\bm{q}$ are collected into a vector
\begin{align}
    \bm{q} = \left(q_0, q_1, \ldots, q_{r - 1}\right),
\end{align}
we may also use notation
\begin{align}
    \binom{q}{\bm{q}} & := \binom{q}{\left(q_j\right)_{j \in [r]}}.
\end{align}
Similar to binomial coefficients, multinomial coefficients have an intuitive combinatorial interpretation. Namely, given $q$ objects and $r$ labels $0, 1, \ldots, r - 1$, they count the number of ways of labelling $q_0$ objects $0$, $q_1$ objects $1$, $\ldots$, $q_{r - 1}$ objects $r - 1$. Similar to the convention with binomial coefficients, the multinomial coefficient is defined to vanish if any of the integers $q_0, q_1, \ldots, q_{r - 1}$ is negative.
\end{definition}
Multinomial coefficients allow to generalize Newton's binomial theorem to the \textit{multinomial theorem}:

\begin{theorem}[Multinomial theorem]
\label{th:multinomial_theorem}
Let $q \geq 0, r \geq 1$, be integers and $z_0, z_1, \ldots, z_{r - 1}$ arbitrary complex numbers. Then the following generalization of Newton's binomial identity holds:
\begin{align}
    \left(\sum_{j \in [r]}z_j\right)^q & = \sum_{\substack{q_0, q_1, \ldots, q_{r - 1} \geq 0\\q_0 + q_1 + \ldots + q_{r - 1} = q}}\binom{q}{\left(q_j\right)_{j \in [r]}}\prod_{j \in [r]}z_j^{q_j}.
\end{align}
\end{theorem}

After introducing multinomial coefficients and the related multinomial theorem, we can now state the following combinatorial lemma counting the bitstring triplets satisfying given configuration basis numbers. This is the key results to convert summation over bitstrings over more tractable summations over configuration numbers:

\begin{lemma}[Counting bitstring triplets with prescribed configuration basis numbers]
\label{lemma:counting_bitstring_triplets}
Let $q \geq 1$ an integer and $\bm{q} = \left(q_s\right)_{s \in \{0, 1\}^3}$ a set of weight-$q$ configuration basis numbers. There are
\begin{align}
    \binom{q}{\bm{q}} := \binom{q}{\left(q_s\right)_{s \in \{0, 1\}^3}},
\end{align}
triplets of $q$-bitstrings $\left(\bm{w}^{[1]}, \bm{w}^{[0]}, \bm{w}^{[-1]}\right)$ satisfying configuration basis numbers $\left(q_s\right)_{s \in \{0, 1\}^3}$.
\begin{proof}
Choosing a bitstring triplet $\left(\bm{w}^{[1]}, \bm{w}^{[0]}, \bm{w}^{[-1]}\right)$ satisfying some configuration basis numbers $\left(q_s\right)_{s \in \{0, 1\}^3}$ is equivalent to choosing, for each $s$, the set of indices $J_s \subset [q]$ such that $\left(z^{[1]}_j, z^{[0]}_j, z^{[-1]}_j\right) = s$ for all $j \in J_s$. This is equivalent to labelling each index $j \in [q]$ with some label $s \in \{0, 1\}^3$, where there must be $q_s$ indices labelled $s$. It follows from the combinatorial interpretation of multinomial coefficients in definition \ref{def:multinomial_coefficient} that there are
\begin{align}
    \binom{q}{\bm{q}}
\end{align}
possible choices.
\end{proof}
\end{lemma}

We next introduce the notion of sub-bitstring triplet:

\begin{definition}[Sub-bitstring triplet]
\label{def:subbitstring_triplet}
    Let $q \geq 1$ and $0 \leq q' \leq q$ integers. Let $\left(\bm{w}^{[1]}, \bm{w}^{[0]}, \bm{w}^{[-1]}\right)$ in $q$-bitstring triplet. A $q'$-sub-bitstring triplet of $\left(\bm{w}^{[1]}, \bm{w}^{[0]}, \bm{w}^{[-1]}\right)$ is a $q'$-bitstring triplet $\left(\bm{x}^{[1]}, \bm{x}^{[0]}, \bm{x}^{[-1]}\right)$, where bitstrings $\bm{x}^{[1]}, \bm{x}^{[0]}, \bm{x}^{[-1]}$ result from evaluating $\bm{w}^{[1]}, \bm{w}^{[0]}, \bm{w}^{[-1]}$ at a common multi-set of indices. More precisely, a $q'$-bitstring triplet is defined by a $q'$-tuple of indices
    \begin{align}
        J & = \left(j_0, j_1, \ldots, j_{q' - 1}\right) \qquad j_0, j_1, \ldots, j_{q' - 1} \in [q],
    \end{align}
    giving
    \begin{align}
        \bm{x}^{[1]} := \left(z^{[1]}_{j_0}, z^{[1]}_{j_1}, \ldots, z^{[1]}_{j_{q' - 1}}\right), && \bm{x}^{[0]} & := \left(z^{[0]}_{j_0}, z^{[0]}_{j_1}, \ldots, z^{[0]}_{j_{q' - 1}}\right), && \bm{x}^{[-1]} := \left(z^{[-1]}_{j_0}, z^{[-1]}_{j_1}, \ldots, z^{[-1]}_{j_{q' - 1}}\right).
    \end{align}
    The $q'$-sub-bitstring triplet defined above will also be called restriction of $q$-bitstring triplet to $J$, and denoted
    \begin{align}
        \left(\bm{w}^{[1]}_J,\,\bm{w}^{[0]}_J,\,\bm{w}_J^{[-1]}\right).
    \end{align}
\end{definition}
Note definition \ref{def:subbitstring_triplet} allows for repetition in tuple $J$ defining sub-bitstrings $\bm{x}^{[1]}, \bm{x}^{[0]}, \bm{x}^{[-1]}$. Similar to lemma \ref{lemma:counting_bitstring_triplets} counting bitstring triplets satisfying a given configuration basis number, we now state a lemma to count sub-bitstring triplets:

\begin{lemma}[Counting choices of sub-bitstring triplets]
\label{lemma:counting_subbitstring_triplets}
Let $n \geq 0, k \geq 0$ integers. Let $\left(\bm{z}^{[1]}, \bm{z}^{[0]}, \bm{z}^{[-1]}\right)$ an $n$-bitstring triplet with configuration basis numbers $\left(n_s\right)_{s \in \{0, 1\}^3}$ (see definition \ref{def:bitstring_triplet_configuration_basis_numbers}). Let be given configuration basis numbers $\left(k_s\right)_{s \in \{0, 1\}^3}$ of weight $k$.  Then, there are
\begin{align}
    \binom{k}{\bm{k}}\prod_{s \in \{0, 1\}^3}n_s^{k_s}
\end{align}
$k$-sub-bitstring triplets of $\left(\bm{z}^{[1]}, \bm{z}^{[0]}, \bm{z}^{[-1]}\right)$ satisfying configuration $\left(k_s\right)_{s \in \{0, 1\}^3}$.
\begin{proof}
Let us partition bitstring indices $j \in [n]$ according to their configuration $s \in \{0, 1\}^3$:
\begin{align}
    J_s & := \left\{j \in [n]\,:\,\left(z^{[1]}_j, z^{[0]}_j, z^{[-1]}_j\right) = s\right\}. 
\end{align}
By definition of configuration basis numbers,
\begin{align}
    \left|J_s\right| & = n_s.
\end{align}
The choice of a $k$-sub-bitstring triplet of $\left(\bm{z}^{[1]}, \bm{z}^{[0]}, \bm{z}^{[-1]}\right)$ can be described as the choice of a $k$-tuple of indices
\begin{align}
    J' & = \left(j_0, j_1, \ldots, j_{k - 1}\right) \in [n]^k.
\end{align}
The additional constraint that the resulting $k$-sub-bitstring triplet $\left(\bm{w}^{[1]}, \bm{w}^{[0]}, \bm{w}^{[-1]}\right)$, where
\begin{align}
    \bm{w}^{[t]} & = \left(w^{[t]}_{j_0}, w^{[t]}_{j_1}, \ldots, w^{[t]}_{j_{k - 1}}\right) \qquad \forall t \in \{1, 0, -1\},
\end{align}
satisfies configuration basis numbers $\left(k_s\right)_{s \in \{0, 1\}^3}$ means that exactly $k_s$ indices $j_0, j_1, \ldots, j_{k - 1}$ must belong to $J_s$ for all $s \in \{0, 1\}^3$. The choice of a tuple $J'$ satisfying these constraints can then be described as making the following two choices, with each combination of choices giving a different tuple:
\begin{itemize}
    \item Define positions from tuple $J'$ that will have indices $j \in J_s$. This is equivalent to labelling positions from the tuple with $s \in \{0, 1\}^3$ such that there are exacly $k_s$ positions labelled $s$ for all $s$.
    \item For each position labelled $s$, choose an index $j \in J_s$.
\end{itemize}
According to the combinatorial interpretation of multinomial coefficients (definition \ref{def:multinomial_coefficient}), there are
\begin{align}
    \binom{k}{\bm{k}}
\end{align}
possibilities for the first choice. Besides, there are
\begin{align}
    \prod_{s \in \{0, 1\}^3}n_s^{k_s}
\end{align}
possibilities for the second choice. The result follows.
\end{proof}
\end{lemma}

In particular, combining lemma \ref{lemma:counting_subbitstring_triplets} just proven as the multinomial theorem \ref{th:multinomial_theorem}, one can recover the number of $k$-sub-bitstring triplets (i.e. sub-bitstring triplets of any configuration) of an $n$-bitstring triplet:
\begin{align*}
    \sum_{\bm{k} \in \mathcal{P})(k)}\binom{k}{\bm{k}}\prod_{s \in \{0, 1\}^3}n_s^{k_s} & = \left(\sum_{s \in \{0, 1\}^3}n_s\right)^k \qquad \textrm{(multinomial theorem)}\\
    & = n^k.
\end{align*}

As we will prove later, computing the instance-averaged success probability of $p = 1$ QAOA will require to sum over all configuration basis numbers of weight $n$, where $n$ is the problem instance size (number of Boolean variables in the constraint satisfaction problem). Since there are $\mathcal{O}\left(n^7\right)$ such sets numbers, this is in practice intractable. Fortunately, the functions of configuration basis numbers we deal with will only depend on the configuration basis numbers in a special way, leading to define \textit{reduced configuration basis numbers}:
\begin{definition}[Reduced configuration basis numbers]
\label{def:reduced_configuration_basis_numbers}
Let be given an integer $q \geq 1$. A set of reduced configuration basis numbers of weight $q$ is a set of non-negative integers 
\begin{align}
    \bm{q'} = \left(q'_s\right)_{s \in \{0, 1\}^2} = \left(q'_{00}, q'_{01}, q'_{10}, q'_{11}\right)
\end{align}
indexed by $2$-bit bitstrings $s \in \{0, 1\}^2$ and summing to $q$:
\begin{align}
    \sum_{s \in \{0, 1\}^2}q'_s = q'_{00} + q'_{01} + q'_{10} + q'_{11} = q.
\end{align}
Similar to notation $\mathcal{P}(q)$ introduced in definition \ref{def:configuration_basis_numbers} for configuration basis numbers, we denote by
\begin{align}
    \mathcal{P}'(q)
\end{align}
the set of reduced configuration basis numbers of weight $q$.

Besides, given (non-reduced) configuration basis numbers $\left(q_s\right)_{s \in \{0, 1\}^3}$ of weight $q$ as introduced in definition \ref{def:configuration_basis_numbers}, the reduced configuration basis numbers associated to these are defined as:
\begin{align}
    q'_{s^{[0]}\,s^{[-1]}} & := q_{0\,s^{[0]}\,s^{[-1]}} + q_{1\,\overline{s^{[0]}}\,\overline{s^{[-1]}}},
\end{align}
i.e.
\begin{align}
	q'_{00} := q_{000} + q_{111}, && q'_{01} := q_{001} + q_{110}, && q'_{10} = q_{010} + q_{101}, && q'_{11} := q_{011} + q_{100}.
\end{align}
\end{definition}
Given a weight $q \geq 1$, there are $\binom{q + 3}{3} = \mathcal{O}\left(q^3\right)$ reduced configuration basis numbers. The following simple combinatorial lemma specifies how a sum over configuration basis number can be simplified to one over reduced configuration basis numbers if the summed function only depends on the latter:

\begin{lemma}[Converting to summation over reduced configuration basis numbers]
\label{lemma:summation_original_to_reduced_configuration_basis_numbers}
Let $q \geq 1$ an integer $f\left(\cdot, \cdot, \cdot, \cdot\right)$ a function defined (at least) on $4$-tuple of non-negative integers summing to $q$, i.e. on weight-$q$ reduced configuration basis numbers as introduced in \ref{def:reduced_configuration_basis_numbers}. Then the following identity holds:
\begin{align}
    \sum_{\bm{q} \in \mathcal{P}(q)}\binom{q}{\bm{q}}f\left(q_{000} + q_{111}, q_{001} + q_{110}, q_{010} + q_{101}, q_{011} + q_{100}\right) & = 2^q\sum_{\substack{\bm{q'} \in \mathcal{P}'\left(q\right)}}\binom{q}{\bm{q'}}f\left(q'_{00}, q'_{01}, q'_{10}, q'_{11}\right).
\end{align}
\begin{proof}
This results from pairing together original configuration basis numbers adding up to a reeduced configuration basis numbers, i.e. pairing up
\begin{align}
	q_{0\,s^{[0]}\,s^{[-1]}}, \quad q_{1\,\overline{s^{[0]}}\,\overline{s^{[-1]}}},
\end{align}
where
\begin{align}
	q'_{s^{[0]}\,s^{[-1]}} & := q_{0\,s^{[0]}\,s^{[-1]}} + q_{1\,\overline{s^{[0]}}\,\overline{s^{[-1]}}}.
\end{align}
Namely, expanding the multinomial coefficient, the summand reads:
\begin{align}
	& \binom{q}{\bm{q}}f\left(q_{000} + q_{111},\,q_{001} + q_{110},\,q_{010} + q_{101},\,q_{011} + q_{100}\right)\nonumber\\
	& = \frac{q!}{\prod\limits_{s \in \{0, 1\}^3}q_s!}f\left(q_{000} + q_{111},\,q_{001} + q_{110},\,q_{010} + q_{101},\,q_{011} + q_{100}\right)\\
	& = \frac{q!}{\prod\limits_{s \in \{0, 1\}^2}q'_s!}\left(\prod_{s^{[0]},\,s^{[-1]} \in \{0, 1\}}\frac{q'_{s^{[0]}\,s^{[-1]}}!}{q_{0\,s^{[0]}\,s^{[-1]}}!q_{1\,s^{[0]}\,s^{[-1]!}}}\right)f\left(q'_{00},\,q'_{01},\,q'_{10},\,q'_{11}\right)
\end{align}
One may next for each $s^{[0]}, s^{[-1]} \in \{0, 1\}$ and value of $q'_{s^{[0]}\,s^{[-1]}}$, sum over $q_{0\,s^{[0]}\,s^{[-1]}}$ and $q_{1\,s^{[0]}\,s^{[-1]}}$ such that their sum $q'_{s^{[0]}\,s^{[-1]}}$ is fixed; this is done using the standard binomial theorem:
\begin{align}
	\sum_{\substack{q_{0\,s^{[0]}\,s^{[-1]}},\, q_{1\,s^{[0]}\,s^{[-1]}}\\q_{0\,s^{[0]}\,s^{[-1]}} + q_{1\,s^{[0]}\,s^{[-1]}} = q'_{s^{[0]}\,s^{[-1]}}}}\frac{q'_{s^{[0]}\,s^{[-1]}}!}{q_{0\,s^{[0]}\,s^{[-1]}}!q_{1\,s^{[0]}\,s^{[-1]}}!} & = 2^{q'_{s^{[0]}\,s^{[-1]}}}
\end{align}
The result follows from also using
\begin{align}
	\prod_{s^{[0]},\,s^{[-1]} \in \{0, 1\}}2^{q'_{s^{[0]},\,s^{[-1]}}} & = 2^{\sum\limits_{s^{[0]},\,s^{[-1]} \in \{0, 1\}}q'_{s^{[0]}\,s^{[-1]}}}\\
	& = 2^{q}
\end{align}
\end{proof}
\end{lemma}

We finally introduce a specific notation for swapping between bitstrings and their negation in configuration basis numbers:

\begin{notation}[Negation of configuration basis numbers]
	\label{notation:configuration_basis_numbers_negation}
	Let $q \geq 1$ an integer and $\bm{q} \in \mathcal{P}(q)$ a weight-$q$ configuration basis number. We denote by
	\begin{align}
		\bm{\overline{q}} & := \left(\overline{q}_s\right)_{s \in \{0, 1\}^3},\\
		\overline{q}_s & := q_{\overline{s}} \qquad \forall s \in \{0, 1\}^3
	\end{align}
	the new weight-$q$ configuration basis numbers obtained from swapping the original configuration basis numbers of $s$ and $\overline{s}$ for all $s \in \{0, 1\}^3$.
\end{notation}

\section{Calculating QAOA success probabilities}
\label{sec:calculating_qaoa_success_probabilities}

In this section, we develop a formula to calculate the expected success probability for QAOA with $p=1$ when applied to any random constraint satisfaction problem (as specified in definition \ref{def:csp_random_instance}). The calculation in this section, using configuration basis numbers introduced in definition \ref{def:configuration_basis_numbers}, is adapted from the derivation of the QAOA energy on the Sherrington-Kirkpatrick model \cite{Farhi2022quantumapproximate} (see also \cite{Claes2021instance}). Recall that in definition \ref{def:csp_random_instance}, a random CSP instance is constructed as a sequence of i.i.d random clauses. To obtain a tractable formula, we will need a technical assumption about averages over a single clause, which essentially captures the permutation invariance of the random clause ensemble.

\begin{assumption}[Probabilities of simultaneous clause violations are expressible from configuration basis numbers]
	\label{assp:single_clause_expectation_configuration_basis_numbers}
	Consider a (fixed, non-random) triplet of $n$-bit strings
	\begin{align}
		\left(\bm{z}^{[1]}, \bm{z}^{[0]}, \bm{z}^{[-1]}\right)
	\end{align}
	and let be given the weight-$n$ configuration basis numbers (definition \ref{def:bitstring_triplet_configuration_basis_numbers})
	\begin{align}
		\bm{n} & = \left(n_s\right)_{s \in \{0, 1\}^3}
	\end{align}
	of this triplet. Then, denoting by $\mathbf{E}_{\sigma}\left[\,\cdot\,\right]$ the expectation over a single random clause $\sigma$ as constructed in definition \ref{def:csp_random_instance}, the following probabilities of simultaneous clause violations by bitstrings $\bm{z}^{[1]}, \bm{z}^{[0]}, \bm{z}^{[-1]}$:
	\begin{align}
	\label{eq:expectation1}	\mathbf{E}_{\sigma}\mathbf{1}\left[\bm{z}^{[1]} \not\vdash \sigma\right], && \mathbf{E}_{\sigma}\mathbf{1}\left[\bm{z}^{[0]} \not\vdash \sigma\right], && \mathbf{E}_{\sigma}\mathbf{1}\left[\bm{z}^{[-1]} \not\vdash \sigma\right],\\
		\label{eq:expectation2}\mathbf{E}_{\sigma}\mathbf{1}\left[\bm{z}^{[1]},\,\bm{z}^{[0]} \not\vdash \sigma\right], && \mathbf{E}_{\sigma}\mathbf{1}\left[\bm{z}^{[1]},\,\bm{z}^{[-1]} \not\vdash \sigma\right], && \mathbf{E}_{\sigma}\mathbf{1}\left[\bm{z}^{[0]},\,\bm{z}^{[-1]} \not\vdash \sigma\right],\\
		\mathbf{E}_{\sigma}\mathbf{1}\left[\bm{z}^{[1]}, \bm{z}^{[0]}, \bm{z}^{[-1]}\not\vdash \sigma\right]
	\end{align}
	only depend on the configuration basis numbers $\bm{n}$ of the bitstring triplet. More specifically, this function is a multivariate polynomial in the reduced configuration basis number:
	\begin{align}
		n_s + n_{\overline{s}} \qquad s \in \{0, 1\}^3.
	\end{align}
	As a consequence (through Boolean polynomial expansion), averaging over $\sigma$ any function of $\left(\bm{z}^{[1]},\bm{z}^{[0]},\bm{z}^{[-1]}\right)$ depending on the bitstrings only through the above indicator functions gives a polynomial in configuration basis numbers.
\end{assumption}

Note that in Assumption \ref{assp:single_clause_expectation_configuration_basis_numbers} it would have been sufficient to only include one each of the expectations in (\ref{eq:expectation1}), (\ref{eq:expectation2}), but we include them all for clarity.
Assumption \ref{assp:single_clause_expectation_configuration_basis_numbers}, which is a statement on the probability (over random choice of clause), that a set of bitstrings violates a clause, has a simple consequence for expectations of more general functions of bitstrings:

\begin{corollary}[Single-clause average of functions of bitstrings depending only on indicator function]
	\label{cor:single_clause_expectation_configuration_basis_numbers}
Let
\begin{align}
	f: \{0, 1\}^3 \longrightarrow \mathbf{C}
\end{align}
an arbitrary function of $3$ Boolean variables, and consider the random function (over random choice of clause $\sigma$) of a triplet of $n$-bit strings $\left(\bm{z}^{[1]}, \bm{z}^{[0]}, \bm{z}^{[-1]}\right)$
\begin{align}
	f\left(\mathbf{1}\left[\bm{z}^{[1]} \not\vdash \sigma\right], \mathbf{1}\left[\bm{z}^{[0]} \not\vdash \sigma\right], \mathbf{1}\left[\bm{z}^{[-1]} \not\vdash \sigma\right]\right).
\end{align}
Then, the average over random clause $\sigma$ of this function:
\begin{align}
	\mathbf{E}_{\sigma}f\left(\mathbf{1}\left[\bm{z}^{[1]} \not\vdash \sigma\right], \mathbf{1}\left[\bm{z}^{[0]} \not\vdash \sigma\right], \mathbf{1}\left[\bm{z}^{[-1]} \not\vdash \sigma\right]\right)
\end{align}
is a polynomial in the reduced configuration basis numbers
\begin{align}
	n_s + n_{\overline{s}}.
\end{align}
\begin{proof}
This results from writing $f$ ---a function of $3$ Boolean variables--- as a Boolean polynomial:
\begin{align*}
	f\left(x^{[1]}, x^{[0]}, x^{[-1]}\right) & = \sum_{T \subset \{0, 1\}^3}\hat{f}_T\prod_{t \in T}x^{[t]} \qquad \forall \left(x^{[1]}, x^{[0]}, x^{[-1]}\right) \in \{0, 1\}^3,
\end{align*}
where the $\hat{f}_T$, $T \subset \{0, 1\}^3$ are complex numbers. Applying this Boolean expansion to the random function of single clause $\sigma$ gives
\begin{align*}
	f\left(\mathbf{1}\left[\bm{z}^{[1]} \not\vdash \sigma\right], \mathbf{1}\left[\bm{z}^{[0]} \not\vdash \sigma\right], \mathbf{1}\left[\bm{z}^{[-1]} \not\vdash \sigma\right]\right) & =  \sum_{T \subset \{0, 1\}^3}\hat{f}_T\prod_{t \in T}\mathbf{1}\left[\bm{z}^{[t]} \not\vdash \sigma\right]\\
	& =  \sum_{T \subset \{0, 1\}^3}\hat{f}_T\mathbf{1}\left[\bm{z}^{[t]} \not\vdash \sigma\,\forall t \in T\right].
\end{align*}
Now, by assumption \ref{assp:single_clause_expectation_configuration_basis_numbers}, the single-clause average of each
\begin{align}
	\mathbf{1}\left[\bm{z}^{[t]} \not\vdash \sigma\,\forall t \in T\right]
\end{align}
is a polynomial in the reduced configuration basis numbers
\begin{align}
	n_s + n_{\overline{s}} \qquad s \in \{0, 1\}^3.
\end{align}
The average of the random function is therefore a sum of such polynomials, hence also a polynomial of this kind.
\end{proof}
\end{corollary}

In section \ref{sec:general_formulae_general_truth_table}, we will see assumption \ref{assp:single_clause_expectation_configuration_basis_numbers} holds for any random constraint satisfaction problem obeying definition \ref{def:csp_instance}, \ref{def:csp_random_instance}. In the meantime, we explicitly verify this assumption on a case-by-case basis for more specific problems.

\begin{proposition}[Instance-averaged success probability for general CSP]
\label{prop:success_probability_general_formula}
Consider a family of random constraint satisfaction problem as specified in definition \ref{def:csp_random_instance} and suppose assumption \ref{assp:single_clause_expectation_configuration_basis_numbers} on single-clause averages holds. Then, if $m$ is fixed, the instance-averaged success probability of $p = 1$ QAOA on this random CSP is given by
\begin{align}
    & \mathbf{E}_{\bm\sigma = \left(\sigma_0, \ldots, \sigma_{m - 1}\right)}\braket{\Psi\left(\gamma, \beta, \bm\sigma\right)|\mathbf{1}\left[H\left[\bm\sigma\right] = 0\right]|\Psi\left(\gamma, \beta, \bm\sigma\right)}\nonumber\\
    & = \sum_{\bm{n'} \in \mathcal{P}'(n)}\binom{n}{\bm{n'}}\left(\cos^2\frac{\beta}{2}\right)^{n'_{000}}\left(\sin^2\frac{\beta}{2}\right)^{n'_{010}}\left(\frac{i\sin\beta}{2}\right)^{n'_{011}}\left(-\frac{i\sin\beta}{2}\right)^{n'_{001}}P_{\mathrm{single}}\left(\bm{n'}\right)^m.\label{eq:success_probability_general_formula_no_poisson}
    \end{align}
If $m$ is chosen from a Poisson distribution, we have
\begin{align}
    & \mathbf{E}_{m \sim \mathrm{Poisson}(rn)}\mathbf{E}_{\bm\sigma = \left(\sigma_0, \ldots, \sigma_{m - 1}\right)}\braket{\Psi\left(\gamma, \beta, \bm\sigma\right)|\mathbf{1}\left[H\left[\bm\sigma\right] = 0\right]|\Psi\left(\gamma, \beta, \bm\sigma\right)}\nonumber\\
    & = \sum_{\bm{n'} \in \mathcal{P}'(n)}\binom{n}{\bm{n'}}\left(\cos^2\frac{\beta}{2}\right)^{n'_{000}}\left(\sin^2\frac{\beta}{2}\right)^{n'_{010}}\left(\frac{i\sin\beta}{2}\right)^{n'_{011}}\left(-\frac{i\sin\beta}{2}\right)^{n'_{001}}\exp\left(rn\left(P_{\mathrm{single}}\left(\bm{n'}\right) - 1\right)\right),\label{eq:success_probability_general_formula}
\end{align}
where the sum is over reduced configuration basis numbers
\begin{align}
    \bm{n'} & = \left(n'_{000}, n'_{001}, n'_{010}, n'_{011}\right)
\end{align}
of weight $n$, and $P_{\mathrm{single}}$ is the function given by corollary \ref{cor:single_clause_expectation_configuration_basis_numbers}, applied to function
\begin{align}
    f\left(x^{[1]}, x^{[0]}, x^{[-1]}\right) & = \exp\left(-\frac{i\gamma}{2}\left(x^{[1]} - x^{[-1]}\right)\right)\left(1 - x^{[0]}\right).
\end{align}
More explicitly, $P_{\mathrm{single}}$ is the unique 4-variate polynomial such that
\begin{align}
    & \mathbf{E}_{\sigma}\left\{\exp\left(-\frac{i\gamma}{2}\left(\mathbf{1}\left[\bm{z}^{[1]} \not\vdash \sigma\right] - \mathbf{1}\left[\bm{z}^{[-1]} \not\vdash \sigma\right]\right)\right)\left(1 - \mathbf{1}\left[\bm{z}^{[0]} \not\vdash \sigma\right]\right)\right\}\nonumber\\
    & = \mathbf{E}_{\sigma}\left\{\exp\left(\frac{i\gamma}{2}\left(\mathbf{1}\left[\bm{z}^{[1]} \vdash \sigma\right] - \mathbf{1}\left[\bm{z}^{[-1]} \vdash \sigma\right]\right)\right)\mathbf{1}\left[\bm{z}^{[0]} \vdash \sigma\right]\right\}\nonumber\\
    & = P_{\mathrm{single}}\left(n_{000} + n_{111}, n_{001} + n_{110}, n_{010} + n_{101}, n_{011} + n_{100}\right),
\end{align}
whenever $n$-bitstring triplet $\left(\bm{z}^{[1]}, \bm{z}^{[0]}, \bm{z}^{[-1]}\right)$ satisfies configuration basis numbers
\begin{align}
    \bm{n} = \left(n_{000}, n_{001}, n_{010}, n_{011}, n_{100}, n_{101}, n_{110}, n_{111}\right).
\end{align}
The instance-averaged success probability can be computed using $O(n^3)$ evaluations of $P_{\mathrm{single}}$ together with $O(n^3)$ other operations.
\begin{proof}
    The method is an adaptation of \cite{Farhi2022quantumapproximate,Basso2022}. We start by expanding the QAOA success probability (first, without averaging over CSP instances) as a path integral in the computational basis:
    \begin{align}
        & \braket{\Psi\left(\gamma, \beta, \bm\sigma\right)|\mathbf{1}\left[H[\bm\sigma] = 0\right]|\Psi\left(\gamma, \beta, \bm\sigma\right)}\nonumber\\
        & = \bra{+}^{\otimes n}e^{\frac{i\gamma}{2}H[\bm\sigma]}e^{\frac{i\beta}{2}H_B}\mathbf{1}\left[H[\bm\sigma] = 0\right]e^{-\frac{i\beta}{2}H_B}e^{-\frac{i\gamma}{2}H[\bm\sigma]}\ket{+}^{\otimes n}\nonumber\\
        & = \bra{+}^{\otimes n}e^{\frac{i\gamma}{2}H[\bm\sigma]}\left(\sum_{\bm{z}^{[1]} \in \{1, -1\}^n}\hspace*{-10px}\ket{\bm{z}^{[1]}}\bra{\bm{z}^{[1]}}\right)e^{\frac{i\beta}{2}H_B}\left(\sum_{\bm{z}^{[0]} \in \{1, -1\}^n}\mathbf{1}\left[H[\bm\sigma]\left(\bm{z}^{[0]}\right) = 0\right]\ket{\bm{z}^{[0]}}\bra{\bm{z}^{[0]}}\right)\nonumber\\
        & \hspace*{40px} e^{-\frac{i\beta}{2}H_B}\left(\sum_{\bm{z}^{[-1]} \in \{1, -1\}^n}\hspace*{-15px}\ket{\bm{z}^{[-1]}}\bra{\bm{z}^{[-1]}}\right)e^{-\frac{i\gamma}{2}H[\bm\sigma]}\ket{+}^{\otimes n}\nonumber\\
        & = \sum_{\substack{\bm{z}^{[1]} \in \{1, -1\}^n\\\bm{z}^{[0]} \in \{1, -1\}^n\\\bm{z}^{[-1]} \in \{1, -1\}^n}}\hspace*{-10px}e^{\frac{i\gamma}{2}H[\bm\sigma]\left(\bm{z}^{[1]}\right)}\mathbf{1}\left[H[\bm\sigma]\left(\bm{z}^{[0]}\right) = 0\right]e^{-\frac{i\gamma}{2}H[\bm\sigma]\left(\bm{z}^{[-1]}\right)}\nonumber\\
        & \hspace*{80px} \times \braket{+|^{\otimes n}\bm{z}^{[1]}}\braket{\bm{z}^{[1]}|e^{\frac{i\beta}{2}H_B}|\bm{z}^{[0]}}\braket{\bm{z}^{[0]}|e^{-\frac{i\beta}{2}H_B}|\bm{z}^{[-1]}}\braket{\bm{z}^{[-1]}|+}^{\otimes n}\label{eq:success_probability_instance_step_1}
    \end{align}
    We now evaluate the matrix elements (in the computational basis) thanks to the tensor product structure of the mixer unitary $e^{-\frac{i\beta}{2}H_B}$ and initial state $\ket{+}^{\otimes n}$. Indeed, the matrix elements involving the initial state are trivial:
    \begin{align}
        \braket{+|^{\otimes n}\bm{z}^{[1]}} = 2^{-n/2}, && \braket{\bm{z}^{[1]}|+}^{\otimes n} = 2^{-n/2}.
    \end{align}
    Plugging these matrix elements into equation \ref{eq:success_probability_instance_step_1} for the success probability on instance $\bm\sigma$, we obtain
    \begin{align}
        & \braket{\Psi\left(\gamma, \beta, \bm\sigma\right)|\mathbf{1}\left[H\left[\bm\sigma\right] = 0\right]|\Psi\left(\gamma, \beta, \bm\sigma\right)}\nonumber\\
        & = 2^{-n}\sum_{\substack{\bm{z}^{[1]} \in \{1, -1\}^n\\\bm{z}^{[0]} \in \{1, -1\}^n\\\bm{z}^{[-1]} \in \{1, -1\}^n}}\hspace*{-10px}e^{\frac{i\gamma}{2}H[\bm\sigma]\left(\bm{z}^{[1]}\right)}\mathbf{1}\left[H[\bm\sigma]\left(\bm{z}^{[0]}\right) = 0\right]e^{-\frac{i\gamma}{2}H[\bm\sigma]\left(\bm{z}^{[-1]}\right)}\prod_{j \in [n]}\braket{z^{[1]}_j|e^{\frac{i\beta}{2}X}|z^{[0]}_j}\braket{z^{[0]}_j|e^{-\frac{i\beta}{2}X}|z^{[-1]}_j}\label{eq:success_probability_instance_step_2}
    \end{align}
    We now consider the part involving the cost function $H[\bm\sigma]$; this contribution factorizes over clauses $\bm\sigma = \left(\sigma_0, \ldots, \sigma_{m - 1}\right)$:
    \begin{align}
        & e^{\frac{i\gamma}{2}H\left[\bm\sigma\right]\left(\bm{z}^{[1]}\right)}\mathbf{1}\left[H\left[\bm\sigma\right]\left(\bm{z}^{[0]}\right) = 0\right]e^{-\frac{i\gamma}{2}H\left[\bm\sigma\right]\left(\bm{z}^{[-1]}\right)}\nonumber\\
        & = e^{\frac{i\gamma}{2}\sum_{j \in [m]}\mathbf{1}\left[\bm{z}^{[1]} \not \vdash \sigma_j\right]}\left(\prod_{j \in [m]}\mathbf{1}\left[\bm{z}^{[0]} \vdash \sigma_j\right]\right)e^{-\frac{i\gamma}{2}\sum_{j \in [m]}\mathbf{1}\left[\bm{z}^{[-1]} \not\vdash \sigma_j\right]}\nonumber\\
        & = e^{\frac{i\gamma}{2}\sum_{j \in [m]}\mathbf{1}\left[\bm{z}^{[1]} \not \vdash \sigma_j\right]}\left(\prod_{j \in [m]}\left(1 - \mathbf{1}\left[\bm{z}^{[0]} \not\vdash \sigma_j\right]\right)\right)e^{-\frac{i\gamma}{2}\sum_{j \in [m]}\mathbf{1}\left[\bm{z}^{[-1]} \not\vdash \sigma_j\right]}\nonumber\\
        & = \prod_{j \in [m]}e^{\frac{i\gamma}{2}\mathbf{1}\left[\bm{z}^{[1]} \not\vdash \sigma_j\right] - \frac{i\gamma}{2}\mathbf{1}\left[\bm{z}^{[-1]} \not\vdash \sigma_j\right]}\left(1 - \mathbf{1}\left[\bm{z}^{[0]} \not\vdash \sigma_j\right]\right).
    \end{align}
    Plugging this into equation \ref{eq:success_probability_instance_step_2}, the instance-wise success probability becomes:
    \begin{align}
        & \braket{\Psi\left(\gamma, \beta, \bm\sigma\right)|\mathbf{1}\left[H\left[\bm\sigma\right] = 0\right]|\Psi\left(\gamma, \beta, \bm\sigma\right)}\nonumber\\
        & = 2^{-n}\hspace*{-10px}\sum_{\substack{\bm{z}^{[1]} \in \{1, -1\}^n\\\bm{z}^{[0]} \in \{1, -1\}^n\\\bm{z}^{[-1]} \in \{1, -1\}^n}}\left(\prod_{j \in [m]}e^{\frac{i\gamma}{2}\mathbf{1}\left[\bm{z}^{[1]} \not\vdash \sigma_j\right] - \frac{i\gamma}{2}\mathbf{1}\left[\bm{z}^{[-1]} \not\vdash \sigma_j\right]}\left(1 - \mathbf{1}\left[\bm{z}^{[0]} \not\vdash \sigma_j\right]\right)\right)\prod_{j \in [n]}\braket{z^{[1]}_j|e^{\frac{i\beta}{2}X}|z^{[0]}_j}\braket{z^{[0]}_j|e^{-\frac{i\beta}{2}X}|z^{[-1]}_j}\label{eq:success_probability_instance_step_3}
    \end{align}
    We now average the above success probability over random instances with exactly $m$ clauses:
    \begin{align}
        & \mathbf{E}_{\bm\sigma = \left(\sigma_0, \ldots, \sigma_{m - 1}\right)}\braket{\Psi\left(\gamma, \beta, \bm\sigma\right)|\mathbf{1}\left[H\left[\bm\sigma\right] = 0\right]|\Psi\left(\gamma, \beta, \bm\sigma\right)}\nonumber\\
        & = 2^{-n}\hspace*{-10px}\sum_{\substack{\bm{z}^{[1]} \in \{1, -1\}^n\\\bm{z}^{[0]} \in \{1, -1\}^n\\\bm{z}^{[-1]} \in \{1, -1\}^n}}\left(\mathbf{E}_{\bm\sigma = \left(\sigma_0, \ldots, \sigma_{m - 1}\right)}\prod_{j \in [m]}e^{\frac{i\gamma}{2}\mathbf{1}\left[\bm{z}^{[1]} \not\vdash \sigma_j\right] - \frac{i\gamma}{2}\mathbf{1}\left[\bm{z}^{[-1]} \not\vdash \sigma_j\right]}\left(1 - \mathbf{1}\left[\bm{z}^{[0]} \not\vdash \sigma_j\right]\right)\right)\nonumber\\
        & \hspace*{90px} \times \prod_{j \in [n]}\braket{z^{[1]}_j|e^{\frac{i\beta}{2}X}|z^{[0]}_j}\braket{z^{[0]}_j|e^{-\frac{i\beta}{2}X}|z^{[-1]}_j}\nonumber\\
        & = 2^{-n}\hspace*{-10px}\sum_{\substack{\bm{z}^{[1]} \in \{1, -1\}^n\\\bm{z}^{[0]} \in \{1, -1\}^n\\\bm{z}^{[-1]} \in \{1, -1\}^n}}\left(\prod_{j \in [m]}\mathbf{E}_{\sigma_j}\left[e^{\frac{i\gamma}{2}\mathbf{1}\left[\bm{z}^{[1]} \not\vdash \sigma_j\right] - \frac{i\gamma}{2}\mathbf{1}\left[\bm{z}^{[-1]} \not\vdash \sigma_j\right]}\left(1 - \mathbf{1}\left[\bm{z}^{[0]} \not\vdash \sigma_j\right]\right]\right)\right)\prod_{j \in [n]}\braket{z^{[1]}_j|e^{\frac{i\beta}{2}X}|z^{[0]}_j}\braket{z^{[0]}_j|e^{-\frac{i\beta}{2}X}|z^{[-1]}_j}\nonumber\\
        & = 2^{-n}\hspace*{-10px}\sum_{\substack{\bm{z}^{[1]} \in \{1, -1\}^n\\\bm{z}^{[0]} \in \{1, -1\}^n\\\bm{z}^{[-1]} \in \{1, -1\}^n}}\left(\mathbf{E}_{\sigma}\left[e^{\frac{i\gamma}{2}\mathbf{1}\left[\bm{z}^{[1]} \not\vdash \sigma\right] - \frac{i\gamma}{2}\mathbf{1}\left[\bm{z}^{[-1]} \not\vdash \sigma\right]}\left(1 - \mathbf{1}\left[\bm{z}^{[0]} \not\vdash \sigma\right]\right)\right]\right)^m\prod_{j \in [n]}\braket{z^{[1]}_j|e^{\frac{i\beta}{2}X}|z^{[0]}_j}\braket{z^{[0]}_j|e^{-\frac{i\beta}{2}X}|z^{[-1]}_j},\label{eq:success_probability_instance_averaged_step_1}
    \end{align}
    where to go from the second to the third line, we used that choice of random choices $\sigma_j$ were i.i.d. We now wish to turn this sum over bitstring triplets, including $8^n$ terms, to one over weight-$n$ configuration basis numbers, involving a number of terms polynomial in $n$ only. Reasoning over a fixed $n$-bitstring triplet $\left(\bm{z}^{[1]}, \bm{z}^{[0]}, \bm{z}^{[-1]}\right)$, let us then consider the configuration basis numbers $\bm{n} = \left(n_s\right)_{s \in \{0, 1\}^3}$ of this triplet. The product over $j \in [n]$, coming from the QAOA mixer unitary, can be expressed in terms of configuration basis numbers according to lemma \ref{lemma:success_probability_mixer_unitary_contribution} proven after this proposition:
    \begin{align}
        \prod_{j \in [n]}\braket{z^{[1]}_j|e^{\frac{i\beta}{2}X}|z^{[0]}_j}\braket{z^{[0]}_j|e^{-\frac{i\beta}{2}X}|z^{[-1]}_j} & = \left(\cos^2\frac{\beta}{2}\right)^{n_{000} + n_{111}}\left(\sin^2\frac{\beta}{2}\right)^{n_{010} + n_{101}}\left(\frac{i\sin\beta}{2}\right)^{n_{011} + n_{100}}\left(-\frac{i\sin\beta}{2}\right)^{n_{001} + n_{110}}.
    \end{align}
    Let us now focus on the factor involving the single-clause average $\mathbf{E}_{\sigma}\left[\,\cdot\,\right]$. We note the quantity averaged over clause $\sigma$:
    \begin{align}
        e^{\frac{i\gamma}{2}\mathbf{1}\left[\bm{z}^{[1]} \not\vdash \sigma\right] - \frac{i\gamma}{2}\mathbf{1}\left[\bm{z}^{[-1]} \not\vdash \sigma\right]}\left(1 - \mathbf{1}\left[\bm{z}^{[0]} \not\vdash \sigma\right]\right)
    \end{align}
    only depends on indicator functions checking whether one of bitstrings $\bm{z}^{[1]}, \bm{z}^{[0]}, \bm{z}^{[-1]}$ violates $\sigma$. Hence, one may invoke corollary \ref{cor:single_clause_expectation_configuration_basis_numbers} of assumption \ref{assp:single_clause_expectation_configuration_basis_numbers} to express the single-clause expectation as a function of the configuration basis numbers only:
    \begin{align}
        \mathbf{E}_{\sigma}\left[e^{\frac{i\gamma}{2}\mathbf{1}\left[\bm{z}^{[1]} \not\vdash \sigma\right] - \frac{i\gamma}{2}\mathbf{1}\left[\bm{z}^{[-1]} \not\vdash \sigma\right]}\left(1 - \mathbf{1}\left[\bm{z}^{[0]} \not\vdash \sigma\right]\right)\right] & = P_{\mathrm{single}}\left(n_{000} + n_{111}, n_{001} + n_{110}, n_{010} + n_{101}, n_{011} + n_{100}\right),
    \end{align}
    where $P_{\mathrm{single}}$ is a polynomial function whose existence was guaranteed by assumption \ref{assp:single_clause_expectation_configuration_basis_numbers}. Alternatively, one can explicitly write the Boolean polynomial expansion of the single-clause averaged function:
    \begin{align}
        & e^{\frac{i\gamma}{2}\mathbf{1}\left[\bm{z}^{[1]} \not\vdash \sigma\right] - \frac{i\gamma}{2}\mathbf{1}\left[\bm{z}^{[-1]} \not\vdash \sigma\right]}\left(1 - \mathbf{1}\left[\bm{z}^{[0]} \not\vdash \sigma\right]\right)\nonumber\\
        & = \left(1 + \left(e^{i\gamma/2} - 1\right)\mathbf{1}\left[\bm{z}^{[1]} \not\vdash \sigma\right]\right)\left(1 + \left(e^{-i\gamma/2} - 1\right)\mathbf{1}\left[\bm{z}^{[-1]} \not\vdash \sigma\right]\right)\left(1 - \mathbf{1}\left[\bm{z}^{[0]} \not\vdash \sigma\right]\right)\nonumber\\
        & = 1 + \left(e^{i\gamma/2} - 1\right)\mathbf{1}\left[\bm{z}^{[1]} \not\vdash \sigma\right] + \left(e^{-i\gamma/2} - 1\right)\mathbf{1}\left[\bm{z}^{[-1]} \not\vdash \sigma\right] + 4\sin^2\frac{\gamma}{4}\mathbf{1}\left[\bm{z}^{[1]} \not\vdash \sigma\right]\mathbf{1}\left[\bm{z}^{[-1]} \not\vdash \sigma\right]\nonumber\\
        & \hspace*{15px} - \mathbf{1}\left[\bm{z}^{[0]} \not\vdash \sigma\right] - \left(e^{i\gamma/2} - 1\right)\mathbf{1}\left[\bm{z}^{[0]} \not\vdash \sigma\right]\mathbf{1}\left[\bm{z}^{[1]} \not\vdash \sigma\right] - \left(e^{-i\gamma/2} - 1\right)\mathbf{1}\left[\bm{z}^{[0]} \not\vdash \sigma\right]\mathbf{1}\left[\bm{z}^{[-1]} \not\vdash \sigma\right]\nonumber\\
        & \hspace*{15px} - 4\sin^2\frac{\gamma}{4}\mathbf{1}\left[\bm{z}^{[1]} \not\vdash \sigma\right]\mathbf{1}\left[\bm{z}^{[0]} \not\vdash \sigma\right]\mathbf{1}\left[\bm{z}^{[-1]} \not\vdash \sigma\right]\nonumber\\& = 1 + \left(e^{i\gamma/2} - 1\right)\mathbf{1}\left[\bm{z}^{[1]} \not\vdash \sigma\right] + \left(e^{-i\gamma/2} - 1\right)\mathbf{1}\left[\bm{z}^{[-1]} \not\vdash \sigma\right] + 4\sin^2\frac{\gamma}{4}\mathbf{1}\left[\bm{z}^{[1]},\,\bm{z}^{[-1]} \not\vdash \sigma\right]\nonumber\\
        & \hspace*{15px} - \mathbf{1}\left[\bm{z}^{[0]} \not\vdash \sigma\right] - \left(e^{i\gamma/2} - 1\right)\mathbf{1}\left[\bm{z}^{[0]},\,\bm{z}^{[1]} \not\vdash \sigma\right] - \left(e^{-i\gamma/2} - 1\right)\mathbf{1}\left[\bm{z}^{[0]},\,\bm{z}^{[-1]} \not\vdash \sigma\right]\nonumber\\
        & \hspace*{15px} - 4\sin^2\frac{\gamma}{4}\mathbf{1}\left[\bm{z}^{[1]},\,\bm{z}^{[0]} ,\,\bm{z}^{[-1]} \not\vdash \sigma\right].
    \end{align}
    Given this expansion as a linear combination of indicator function, one can likewise invoke assumption \ref{assp:single_clause_expectation_configuration_basis_numbers} to postulate the existence of $P_{\mathrm{single}}$. Having expressed all contributions from equation \ref{eq:success_probability_instance_averaged_step_1} in terms of configuration basis numbers $\bm{n}$, it remains to account for the number of bitstring triplets satisfying a given set of configuration basis numbers; according to lemma \ref{lemma:counting_bitstring_triplets}, there are exactly
    \begin{align}
        \binom{n}{\bm{n}} = \binom{n}{n_{000}, n_{001}, n_{010}, n_{011}, n_{100}, n_{101}, n_{110}, n_{111}}
    \end{align}
    triplets satisfying $\bm{n}$. All in all, the instance-averaged success probability (for a random instance with a fixed number of clauses $m$) has been recast to
    \begin{align}
        & \mathbf{E}_{\bm\sigma = \left(\sigma_0, \ldots, \sigma_{m - 1}\right)}\braket{\Psi\left(\gamma, \beta, \bm\sigma\right)|\mathbf{1}\left[H\left[\bm\sigma\right] = 0\right]|\Psi\left(\gamma, \beta, \bm\sigma\right)}\nonumber\\
        & = 2^{-n}\sum_{\bm{n} \in \mathcal{P}(n)}\binom{n}{\bm{n}}P_{\mathrm{single}}\left(n_{000} + n_{111}, n_{001} + n_{110}, n_{010} + n_{101}, n_{011} + n_{100}\right)^m\nonumber\\
        & \hspace*{75px} \times \left(\cos^2\frac{\beta}{2}\right)^{n_{000} + n_{111}}\left(\sin^2\frac{\beta}{2}\right)^{n_{010} + n_{101}}\left(\frac{i\sin\beta}{2}\right)^{n_{011} + n_{100}}\left(-\frac{i\sin\beta}{2}\right)^{n_{001} + n_{110}}\label{eq:success_probability_instance_averaged_step_1.5}
    \end{align}
    as claimed in the theorem.

    Observe that the summand of equation \ref{eq:success_probability_instance_averaged_step_1.5}, except for the multinomial coefficient, only depends on the reduced configuration basis numbers
    \begin{align}
        n'_{000} := n_{000} + n_{111}, && n'_{001} := n_{001} + n_{110}, && n'_{010} := n_{010} + n_{101}, && n'_{011} := n_{011} + n_{100}.
    \end{align}
    We will denote the collection of reduced configuration basis numbers by
    \begin{align}
        \bm{n'} & := \left(n'_{000}, n'_{001}, n'_{010}, n'_{011}\right)
    \end{align}
    for brevity.
    Summing over these variables first, equation \ref{eq:success_probability_instance_averaged_step_1.5} becomes:
    \begin{align}
        & \mathbf{E}_{\bm\sigma = \left(\sigma_0, \ldots, \sigma_{m - 1}\right)}\braket{\Psi\left(\gamma, \beta, \bm\sigma\right)|\mathbf{1}\left[H\left[\bm\sigma\right] = 0\right]|\Psi\left(\gamma, \beta, \bm\sigma\right)}\nonumber\\
        & = 2^{-n}\sum_{\bm{n'} \in \mathcal{P}(n)}P_{\mathrm{single}}\left(\bm{n'}\right)^m\left(\cos^2\frac{\beta}{2}\right)^{n'_{000}}\left(\sin^2\frac{\beta}{2}\right)^{n'_{010}}\left(\frac{i\sin\beta}{2}\right)^{n'_{011}}\left(-\frac{i\sin\beta}{2}\right)^{n'_{001}}\nonumber\\
        & \hspace*{20px} \times \sum_{\substack{n_{000},\,n_{111}\\n_{000} + n_{111} = n'_{000}}}\sum_{\substack{n_{001},\,n_{110}\\n_{001} + n_{100} = n'_{001}}}\sum_{\substack{n_{010},\,n_{101}\\n_{010} + n_{101} = n'_{010}}}\sum_{\substack{n_{011},\,n_{100}\\n_{011} + n_{100} = n'_{011}}}\hspace*{-10px}\binom{n}{n_{000},\,n_{001}, n_{010}, n_{011}, n_{100}, n_{101}, n_{110}, n_{111}}\label{eq:success_probability_instance_averaged_step_3}
    \end{align}
    Decomposing the multinomial coefficient as follows:
    \begin{align}
        \binom{n}{n_{000},\,n_{001}, n_{010}, n_{011}, n_{100}, n_{101}, n_{110}, n_{111}} & = \frac{n!}{n'_{000}!n'_{001}!n'_{010}!n'_{011}!} \times \frac{n'_{000}!}{n_{000}!n_{111}!} \times \frac{n'_{001}!}{n_{001}!n_{110}!} \times \frac{n'_{010}!}{n_{010}!n_{101}!}\nonumber\\
        & \hspace*{20px} \times \frac{n'_{011}!}{n_{011}!n_{100}!}\nonumber\\
        & = \binom{n}{\bm{n'}}\binom{n'_{000}}{n_{000}, n_{111}}\binom{n'_{001}}{n_{001}, n_{110}}\binom{n'_{010}}{n_{010}, n_{101}}\binom{n'_{011}}{n_{011}, n_{100}},
    \end{align}
    and using the standard binomial theorem $4$ times, for instance
    \begin{align}
        \sum_{\substack{n_{000},\,n_{111}\\n_{000} + n_{111} = n'_{000}}}\binom{n'_{000}}{n_{000}, n_{111}} & = 2^{n'_{000}},
    \end{align}
    equation \ref{eq:success_probability_instance_averaged_step_3} becomes
    \begin{align}
        & \mathbf{E}_{\bm\sigma = \left(\sigma_0, \ldots, \sigma_{m - 1}\right)}\braket{\Psi\left(\gamma, \beta, \bm\sigma\right)|\mathbf{1}\left[H\left[\bm\sigma\right] = 0\right]|\Psi\left(\gamma, \beta, \bm\sigma\right)}\nonumber\\
        & = 2^{-n}\sum_{\bm{n'} \in \mathcal{P}(n)}P_{\mathrm{single}}\left(\bm{n'}\right)^m\left(\cos^2\frac{\beta}{2}\right)^{n'_{000}}\left(\sin^2\frac{\beta}{2}\right)^{n'_{010}}\left(\frac{i\sin\beta}{2}\right)^{n'_{011}}\left(-\frac{i\sin\beta}{2}\right)^{n'_{001}}\binom{n}{\bm{n'}}2^{n'_{000} + n'_{001} + n'_{010} + n'_{011}}\nonumber\\
        & = \sum_{\bm{n'} \in \mathcal{P}(n)}\binom{n}{\bm{n'}}\left(\cos^2\frac{\beta}{2}\right)^{n'_{000}}\left(\sin^2\frac{\beta}{2}\right)^{n'_{010}}\left(\frac{i\sin\beta}{2}\right)^{n'_{011}}\left(-\frac{i\sin\beta}{2}\right)^{n'_{001}}P_{\mathrm{single}}\left(\bm{n'}\right)^m,
    \end{align}
    which is the claimed result for the fixed-number of clauses $m$ case. Finally, averaging over the number of clauses $m$ according to $m \sim \mathrm{Poisson}(rn)$, one obtains:
    \begin{align}
        & \mathbf{E}_{m \sim \mathrm{Poisson}(rn)}\mathbf{E}_{\bm\sigma = \left(\sigma_0, \ldots, \sigma_{m - 1}\right)}\braket{\Psi\left(\gamma, \beta, \bm\sigma\right)|\mathbf{1}\left[H\left[\bm\sigma\right] = 0\right]|\Psi\left(\gamma, \beta, \bm\sigma\right)}\nonumber\\
        & = \sum_{\bm{n'} \in \mathcal{P}(n)}\binom{n}{\bm{n'}}\left(\cos^2\frac{\beta}{2}\right)^{n'_{000}}\left(\sin^2\frac{\beta}{2}\right)^{n'_{010}}\left(\frac{i\sin\beta}{2}\right)^{n'_{011}}\left(-\frac{i\sin\beta}{2}\right)^{n'_{001}}\exp\left(rn\left(P_{\mathrm{single}}\left(\bm{n'}\right) - 1\right)\right),
    \end{align}
    which is the claimed result.
\end{proof}
\end{proposition}

To complete the proof of proposition \ref{prop:success_probability_general_formula}, we state below a lemma expressing contributions from the mixer unitaries in equation \ref{eq:success_probability_instance_step_3} in terms of configuration basis numbers:

\begin{lemma}[Mixer unitary contributions from configuration basis numbers]
\label{lemma:success_probability_mixer_unitary_contribution}
Let be given a $n$-bitstring triplet $\left(\bm{z}^{[1]},\,\bm{z}^{[0]},\,\bm{z}^{[-1]}\right)$ with configuration basis numbers (definition \ref{def:configuration_basis_numbers})
\begin{align}
    \bm{n} = \left(n_s\right)_{s \in \{0, 1\}^3} = \left(n_{000}, n_{001}, n_{010}, n_{011}, n_{100}, n_{101}, n_{110}, n_{111}\right).
\end{align}
Then, the contribution from the mixer unitaries in equation \ref{eq:success_probability_instance_step_3} can be expressed in terms of these numbers as follows:
\begin{align}
    \prod_{j \in [n]}\braket{z^{[1]}_j|e^{\frac{i\beta}{2}X}|z^{[0]}_j}\braket{z^{[0]}_j|e^{-\frac{i\beta}{2}X}|z^{[-1]}_j} & = \left(\cos^2\frac{\beta}{2}\right)^{n_{000} + n_{111}}\left(\sin^2\frac{\beta}{2}\right)^{n_{010} + n_{101}}\left(\frac{i\sin\beta}{2}\right)^{n_{011} + n_{100}}\left(-\frac{i\sin\beta}{2}\right)^{n_{001} + n_{110}}.
\end{align}
\begin{proof}
This follows simply from partitioning qubit indices $j \in [n]$ according to the joint configuration of bits $\left(z^{[1]}_j, z^{[0]}_j, z^{[-1]}_j\right)$. Let then
\begin{align}
    J_s & := \left\{j \in [n]\,:\,\left(z^{[1]}_j, z^{[0]}_j, z^{[-1]}_j\right) = s\right\} \qquad \forall s = \left(s^{[1]}, s^{[0]}, s^{[-1]}\right) \in \{0, 1\}^3.
\end{align}
Then,
\begin{align*}
    & \prod_{j \in [n]}\braket{z^{[1]}_j|e^{\frac{i\beta}{2}X}|z^{[0]}_j}\braket{z^{[0]}_j|e^{-\frac{i\beta}{2}X}|z^{[-1]}_j}\\
    & = \prod_{s \in \{0, 1\}^3}\prod_{j \in J_s}\braket{z^{[1]}_j|e^{\frac{i\beta}{2}X}|z^{[0]}_j}\braket{z^{[0]}_j|e^{-\frac{i\beta}{2}X}|z^{[-1]}_j}\\
    & = \prod_{s \in \{0, 1\}^3}\prod_{j \in J_s}\braket{s^{[1]}|e^{\frac{i\beta}{2}X}|s^{[0]}}\braket{s^{[0]}|e^{-\frac{i\beta}{2}X}|s^{[-1]}}\\
    & = \prod_{s \in \{0, 1\}^3}\left(\braket{s^{[1]}|e^{\frac{i\beta}{2}X}|s^{[0]}}\braket{s^{[0]}|e^{-\frac{i\beta}{2}X}|s^{[-1]}}\right)^{|J_s|}\\
    & = \prod_{s \in \{0, 1\}^3}\left(\braket{s^{[1]}|e^{\frac{i\beta}{2}X}|s^{[0]}}\braket{s^{[0]}|e^{-\frac{i\beta}{2}X}|s^{[-1]}}\right)^{n_s},
\end{align*}
where in the final line we used that $n_s = |J_s|$ by definition of configuration basis numbers. The claimed formula then follows from evaluating explicitly the product of matrix elements
\begin{align}
    \braket{s^{[1]}|e^{\frac{i\beta}{2}X}|s^{[0]}}\braket{s^{[0]}|e^{-\frac{i\beta}{2}X}|s^{[-1]}}
\end{align}
for all $8$ values of $s \in \{0, 1\}^3$.
\end{proof}
\end{lemma}

Formula \ref{eq:success_probability_general_formula} from proposition \ref{prop:success_probability_general_formula} for the instance-averaged success probability of QAOA depends on a polynomial $P_{\mathrm{single}}$, which we call \textit{single-clause polynomial} as it is defined by an average over a single random clause. Given this polynomial, the sum evaluating the success probability has $\mathcal{O}(n^3)$ terms. To apply this formula to a specific constraint satisfaction problem, one needs to compute the single-clause polynomial, assuming it is well-defined. We collect the definition of this polynomial hereafter for convenience:

\begin{definition}[Single-clause polynomial]
\label{def:single_clause_polynomial}
    Let be given a truth table $T$ (defining a constraint satisfaction problem) as introduced in definition \ref{def:general_truth_table} and $n \geq 1$ an integer. The single-clause expectation polynomial associated to the constraint satisfaction problem is the polynomial in configuration basis numbers $\bm{n} = \left(n_s\right)_{s \in \{0, 1\}^3}$ defined as follows:
	\begin{align}	
		& P_{\mathrm{single}}\left(n_{000} + n_{111}, n_{001} + n_{110}, n_{010} + n_{101}, n_{011} + n_{100}\right)\nonumber\\
        & := \mathbf{E}_{\sigma}\left[\exp\left(-\frac{i\gamma}{2}\left(\mathbf{1}\left[z^{[1]} \vdash \sigma\right] - \mathbf{1}\left[z^{[-1]} \vdash \sigma\right]\right)\right)\mathbf{1}\left[z^{[0]} \vdash \sigma\right]\right]\\
        & = \mathbf{E}_{\sigma}\left[\exp\left(-\frac{i\gamma}{2}\left(\mathbf{1}\left[z^{[-1]} \not\vdash \sigma\right] - \mathbf{1}\left[z^{[1]} \not\vdash \sigma\right]\right)\right)\left(1 - \mathbf{1}\left[z^{[0]} \not\vdash \sigma\right]\right]\right)\label{eq:single_clause_expectation_polynomial}
	\end{align}
	where $\left(\bm{z}^{[1]}, \bm{z}^{[0]}, \bm{z}^{[-1]}\right)$ is a triplet of bitstrings satisfying configuration basis numbers $\bm{n}$. The definition makes sense only if the expectation depends on bitstrings $z^{[1]}, z^{[0]}, z^{[-1]}$ through their sole reduced configuration basis numbers
	\begin{align}
		n'_{00} := n_{000} + n_{111}, && n'_{01} := n_{001} + n_{110}, && n'_{10} := n_{010} + n_{101}, && n'_{11} := n_{011} + n_{100},	
	\end{align}
	and the resulting function of configuration basis numbers is a polynomial. As we will see in sections \ref{sec:general_formulae_hamming_weight_truth_table} and \ref{sec:general_formulae_general_truth_table}, this is indeed the case for the most general form of truth table when the random instance is constructed from the table according to definition \ref{def:csp_random_instance}. Finally, as a mild abuse of notation, we may occasionally write polynomial $P_{\mathrm{single}}$ as a function of the $8$ configuration basis numbers rather than the $4$ reduced ones, even though it only depends explicitly on the latters.
\end{definition}

Note the definition of the single-clause polynomial can be expressed in terms of probabilities of clause violations, or probability of clause satisfaction. It may be more convenient to work with one point of view rather than the other depending on the constraint satisfaction problem under consideration.

\section{General formulae for QAOA success probabilities}
\label{sec:generalformulae}

In the previous section, we described a formula for the instance-averaged success probability of QAOA. Recall that evaluating this formula requires one to compute the single-clause polynomial, after proving it is well-defined, i.e.\ depending on a bitstring triplet only via its (reduced) configuration basis numbers. In the present section, we show this for a large family of random constraint satisfaction problems. Before dealing in section \ref{sec:general_formulae_general_truth_table} with the most general problem satisfying definition \ref{def:csp_random_instance}, we start in section \ref{sec:general_formulae_hamming_weight_truth_table} with a more specialized family of problems, defined by what we call a \textit{Hamming weight truth table}. Hamming weight truth tables have the advantage of being easier to enumerate compared to fully generic ones, while encompassing many common satisfaction problems such as $k$-SAT, NAE-SAT, and $1$-in-$k$-SAT.

\subsection{Hamming weight truth table}
\label{sec:general_formulae_hamming_weight_truth_table}

We now show that the single-clause polynomial introduced in definition \ref{def:single_clause_polynomial} is indeed a polynomial in the configuration basis variables for a special family of truth tables. Namely, we assume a truth table where the truth value only depends on the Hamming weight of the clause literals:

\begin{definition}[``Hamming weight'' truth table]
\label{def:hamming_weight_truth_table}
A truth table $T: \{0, 1\}^k \longrightarrow \{0, 1\}$ is said to be of Hamming weight type if the truth value depends only on the Hamming weight of the bitstring:
\begin{align}
    T\left(x_0, \ldots, x_{n - 1}\right) & = \widetilde{T}\left(x_0 + \ldots, x_{n - 1}\right)
\end{align}
for some function
\begin{align}
    \widetilde{T}: [k + 1] & \longrightarrow \{0, 1\}. 
\end{align}
\end{definition}

This family of CSPs encompasses, for example, standard $k$-SAT, NAE-SAT, and 1-in-$k$-SAT.

By abuse of notation, we may still use $T$ for the function of Hamming weight $\widetilde{T}$ introduced in definition \ref{def:hamming_weight_truth_table}. The goal of this section is to compute the single-clause polynomial (definition \ref{def:single_clause_polynomial}) for such a truth table, which is done in proposition \ref{prop:single_clause_polynomial_hamming_weight_truth_table_explicit_expression}.

To arrive there, we first plug the definition of the Hamming weight truth table considered here into the defining expression of the single-clause polynomial:
\begin{align}
    P_{\mathrm{single}}\left(\bm{n}\right) & = \mathbf{E}_{\sigma}\left[\exp\left(-\frac{i\gamma}{2}\left(\mathbf{1}\left[\bm{z}^{[1]} \vdash \sigma\right] - \mathbf{1}\left[\bm{z}^{[-1]} \vdash \sigma\right]\right)\right)\mathbf{1}\left[\bm{z}^{[0]} \vdash \sigma\right]\right]\\
    & = \mathbf{E}_{\sigma}\left[\exp\left(-\frac{i\gamma}{2}\left(T\left(\sum_{\left(l, \nu\right) \in \sigma}z^{[1]}_l \oplus \nu\right) - T\left(\sum_{\left(l, \nu\right) \in \sigma}z^{[-1]}_l \oplus \nu\right)\right)\right)T\left(\sum_{\left(l, \nu\right) \in \sigma}z^{[0]}_l \oplus \nu\right)\right],\label{eq:single_clause_polynomial_hamming_weight_truth_table}
\end{align}
where $\left(z^{[1]}, z^{[0]}, z^{[-1]}\right)$ is a triplet of $n$-bitstrings satisfying configuration basis numbers $\bm{n}$. We view the above as a special case of the expectation
\begin{align}
    \mathbf{E}_{\sigma}f\left(\sum_{\left(l, \nu\right)}z^{[1]}_l \oplus \nu, \sum_{\left(l, \nu\right) \in \sigma}z^{[0]}_l \oplus \nu, \sum_{\left(l, \nu\right) \in \sigma}z^{[-1]}_l \oplus \nu\right),
\end{align}
where $f: [k + 1]^3 \longrightarrow \mathbf{C}$ is an arbitrary function. The function is evaluated at the Hamming weights of bitstrings $\bm{z}^{[1]}, \bm{z}^{[0]}, \bm{z}^{[-1]}$ restricted to the clause's scope and weighted by common negations. The following lemma computes the expectation of such a function:

\begin{lemma}[Averaging function of Hamming weights over random clauses]
\label{lemma:averaging_hamming_weights_function_hamming_weight_truth_table}
Let
\begin{align}
    f: [k + 1]^3 \longrightarrow \mathbf{C}
\end{align}
an arbitrary function. Then, for any triplet of $n$-bitstrings $\left(\bm{z}^{[1]}, \bm{z}^{[0]}, \bm{z}^{[-1]}\right)$, the average over a single random clause $\sigma$ can be expressed:
\begin{align}
    & \mathbf{E}_{\sigma}f\left(\sum_{\left(l, \nu\right) \in \sigma}z^{[1]}_l \oplus \nu, \sum_{\left(l, \nu\right) \in \sigma}z^{[0]}_l \oplus \nu, \sum_{\left(l, \nu\right) \in \sigma}z^{[-1]}_l \oplus \nu\right)\\
    & = 2^{-k}\sum_{\bm{k} \in \mathcal{P}(k)}\binom{k}{\bm{k}}\left(\prod_{s \in \{0, 1\}^3}\left(\frac{n_s + n_{\overline{s}}}{n}\right)^{k_s}\right)f\left(\left(\sum_{\substack{z^{[1]},\,z^{[0]},\,z^{[-1]} \in \{0, 1\}\\z^{[t]} = 1}}k_{z^{[1]}z^{[0]}z^{[-1]}}\right)_{t \in \{1, 0, -1\}}\right).
\end{align}
where $\bm{n}$ are the configuration basis numbers of $\left(\bm{z}^{[1]}, \bm{z}^{[0]}, \bm{z}^{[-1]}\right)$. In particular, the average is a polynomial in $\bm{n}$.
\begin{proof}
Recalling definition \ref{def:csp_instance} a clause is described by $k$ pairs
\begin{align}
    \left(l_0, \nu_0\right), \left(l_1, \nu_1\right), \ldots, \left(l_{k - 1}, \nu_{k - 1}\right).
\end{align}
For a random clause (definition \ref{def:csp_random_instance}), these pairs are generated identically independently, which each pair $\left(l, \nu\right)$ obtained drawing $l \in [n]$ (variable index) and $\nu \in \{0, 1\}$ (negation) independently and uniformly. We may then reason by considering first a random choice of negations
\begin{align}
    \bm\nu = \left(\nu_0, \nu_1, \ldots, \nu_{k - 1}\right)
\end{align}
and then compute the conditional expectation over choices of variable indices:
\begin{align}
    \bm{l} & = \left(l_0, l_1, \ldots, l_{k - 1}\right)
\end{align}
conditioned on the choice of negation. That is, we decompose the expectation:
\begin{align}
    \mathbf{E}_{\sigma}\left[\,\cdot\,\right] & = \mathbf{E}_{\bm\nu}\left[\mathbf{E}_{\bm{l}}\left[\,\cdot\,|\,\bm\nu\right]\right].
\end{align}
Let us now fix a specific choice of negations $\bm{\nu}$ and evaluate the inner conditional expectation. Defining the set of indices $\subset [k]$ where the negations are $0$ (no negation applied) or $1$ (negation applied):
\begin{align}
    J' & := \left\{j \in [k]\,:\,\nu_j = 0\right\},\\
    J'' & := \left\{j \in [k]\,:\,\nu_j = 1\right\},
\end{align}
the conditional expectation reads:
\begin{align}
    & \mathbf{E}_{\bm{l}}\left[f\left(\sum_{\left(l, \nu\right) \in \sigma}z^{[1]}_l \oplus \nu, \sum_{\left(l, \nu\right) \in \sigma}z^{[0]}_l \oplus \nu, \sum_{\left(l, \nu\right) \in \sigma}z^{[-1]} \oplus \nu\right)\,\bigg|\,\bm\nu\right]\nonumber\\
    & = \mathbf{E}_{\bm{l}}\left[f\left(\sum_{j \in J'}z^{[1]}_{l_j} + \sum_{j \in J''}\neg z^{[1]}_{l_j}, \sum_{j \in J'}z^{[1]}_{l_j} + \sum_{j \in J''}\neg z^{[0]}_{l_j}, \sum_{j \in J'}z^{[1]}_{l_j} + \sum_{j \in J''}\neg z^{[-1]}_{l_j}\right)\,\bigg|\,\bm\nu\right],\label{eq:hamming_weight_truth_table_single_clause_expectation_f_step_1}
\end{align}
 We now introduce the configuration basis numbers of bitstrings triplet $\left(\bm{z}^{[1]}, \bm{z}^{[0]}, \bm{z}^{[-1]}\right)$ restricted to variables of indices $J'$ and $J''$:
\begin{align}
    \bm{k'} & := \left(k'_s\right)_{s \in \{0, 1\}^3},\\
    k'_s & := \left|\left\{j \in J'\,:\,\left(z^{[1]}_{l_j}, z^{[0]}_{l_j}, z^{[-1]}_{l_j}\right) = s\right\}\right| \qquad \forall s \in \{0, 1\}^3,\label{eq:hamming_weight_truth_table_single_clause_expectation_kprime}\\
    \bm{k''} & := \left(k''_s\right)_{s \in \{0, 1\}^3},\\
    k''_s & := \left|\left\{j \in J''\,:\,\left(z^{[1]}_{l_j}, z^{[0]}_{l_j}, z^{[-1]}_{l_j}\right) = s\right\}\right| \qquad \forall s \in \{0, 1\}^3.\label{eq:hamming_weight_truth_table_single_clause_expectation_kdoubleprime}
\end{align}
We also denote
\begin{align}
    k' := |J'|, && k'' := |J''|,
\end{align}
so that $\bm{k'} \in \mathcal{P}(k'), \bm{k''} \in \mathcal{P}(k'')$. Then, the arguments of function $f$ can be expressed as follows from the configuration basis numbers:
\begin{align*}
    \sum_{j \in J'}z^{[1]}_{l_j} + \sum_{j \in J''}\neg z^{[1]}_{l_j} & = \sum_{z^{[0]},\,z^{[-1]} \in \{0, 1\}}k'_{1z^{[0]}z^{[-1]}} + \sum_{z^{[0]},\,z^{[-1]} \in \{0, 1\}}k''_{0z^{[0]}z^{[-1]}},\\
    \sum_{j \in J'}z^{[0]}_{l_j} + \sum_{j \in J''}\neg z^{[0]}_{l_j} & = \sum_{z^{[1]},\,z^{[-1]} \in \{0, 1\}}k'_{z^{[1]}1z^{[-1]}} + \sum_{z^{[1]},\,z^{[-1]} \in \{0, 1\}}k''_{z^{[1]}0z^{[-1]}},\\
    \sum_{j \in J'}z^{[-1]}_{l_j} + \sum_{j \in J''}\neg z^{[-1]}_{l_j} & = \sum_{z^{[1]},\,z^{[0]} \in \{0, 1\}}k'_{z^{[1]}z^{[0]}1} + \sum_{z^{[1]},\,z^{[0]} \in \{0, 1\}}k''_{z^{[1]}z^{[0]}0}.
\end{align*}
These expressions can be unified as:
\begin{align}
    \sum_{j \in J'}z^{[t]}_{l_j} + \sum_{j \in J''}\neg z^{[t]}_j & = \sum_{\substack{z^{[1]},\,z^{[0]},\,z^{[-1]} \in \{0, 1\}\\z^{[t]} = 1}}k'_{z^{[1]}z^{[0]}z^{[-1]}} + \sum_{\substack{z^{[1]},\,z^{[0]},\,z^{[-1]} \in \{0, 1\}\\z^{[t]} = 0}}k''_{z^{[1]}z^{[0]}z^{[-1]}} \qquad \forall t \in \{1, 0, -1\}.
\end{align}
Plugging this into equation \ref{eq:hamming_weight_truth_table_single_clause_expectation_f_step_1}, we have shown
\begin{align}
    & \mathbf{E}_{\bm{l}}\left[f\left(\sum_{\left(l, \nu\right) \in \sigma}z^{[1]}_l \oplus \nu, \sum_{\left(l, \nu\right) \in \sigma}z^{[0]}_l \oplus \nu, \sum_{\left(l, \nu\right) \in \sigma}z^{[-1]} \oplus \nu\right)\,\bigg|\,\bm\nu\right]\nonumber\\
    & = \mathbf{E}_{\bm{l}}\left[f\left(\left(\sum_{\substack{z^{[1]},\,z^{[0]},\,z^{[-1]} \in \{0, 1\}\\z^{[t]} = 1}}k'_{z^{[1]}z^{[0]}z^{[-1]}} + \sum_{\substack{z^{[1]},\,z^{[0]},\,z^{[-1]} \in \{0, 1\}\\z^{[t]} = 0}}k''_{z^{[1]}z^{[0]}z^{[-1]}}\right)_{t \in \{1, 0, -1\}}\right)\,\bigg|\,\bm{\nu}\right]\label{eq:hamming_weight_truth_table_single_clause_expectation_f_step_2}
\end{align}
This reduces the problem to computing the distribution of configuration basis numbers $\bm{k'}, \bm{k''}$ for a random choice of variable indices, given a choice of negations. For that, we observe that by invariance of configuration basis numbers under simultaneous permutations of bitstrings, $\bm{k'}$ defined in equation \ref{eq:hamming_weight_truth_table_single_clause_expectation_kprime} coincides with the configuration basis numbers of $k'$-bitstring triplets:
\begin{align}
    \left(\left(z^{[1]}_{l_j}\right)_{j \in J'}, \left(z^{[0]}_{l_j}\right)_{j \in J'}, \left(z^{[-1]}_{l_j}\right)_{j \in J'}\right),
\end{align}
where in the last equation we view set $J'$ as the ordered tuple of its element (order matters between the bitstrings of a triplet). Likewise, $\bm{k''}$ defined in equation \ref{eq:hamming_weight_truth_table_single_clause_expectation_kdoubleprime} coincides with the configuration basis numbers of bitstring:
\begin{align}
    \left(\left(z^{[1]}_{l_j}\right)_{j \in J''}, \left(z^{[0]}_{l_j}\right)_{j \in J''}, \left(z^{[-1]}_{l_j}\right)_{j \in J''}\right).
\end{align}
Now, by independence and uniformity of the choices of $l_j$, the first sub-bitstring triplet is distributed as a uniformly random $k'$-sub-bitstring triplet of $\left(\bm{z}^{[1]}, \bm{z}^{[0]}, \bm{z}^{[-1]}\right)$. According to lemma \ref{lemma:counting_subbitstring_triplets}, the probability of the subtriplet satisfying configuration $\bm{k'}$ is then:
\begin{align}
    \binom{k'}{\bm{k'}}\prod_{s \in \{0, 1\}^3}\left(\frac{n_s}{n}\right)^{k'_s}.
\end{align}
Similarly, the second sub-bitstring triplet is distributed as a uniformly random $k''$-sub-bitstring triplet of the original triplet, and is besides independent of the first ---recall that $J'$ and $J''$ are disjoint, implying independence of $\left(l_j\right)_{j \in J'}$ and $\left(l_j\right)_{j \in J''}$. It follows $\bm{k''}$ is independent of $\bm{k'}$ with distribution:
\begin{align}
    \binom{k''}{\bm{k''}}\prod_{s \in \{0, 1\}^3}\left(\frac{n_s}{n}\right)^{k''_s}.
\end{align}
The above means $\bm{k''} = \left(k''_s\right)_{s \in \{0, 1\}^3}$ is distributed according to a multinomial law with $k''$ trials and probabilities $\bm{n}/n = \left(n_s/n\right)_{s \in \{0, 1\}^3}$. Plugging these into equation \ref{eq:hamming_weight_truth_table_single_clause_expectation_f_step_2}, the expectation over variable indices $\bm{l}$, conditioned on choices of negations $\bm\nu$, has been expressed as follows:
\begin{align}
    & \mathbf{E}_{\bm{l}}\left[f\left(\sum_{\left(l, \nu\right) \in \sigma}z^{[1]}_l \oplus \nu, \sum_{\left(l, \nu\right) \in \sigma}z^{[0]}_l \oplus \nu, \sum_{\left(l, \nu\right) \in \sigma}z^{[-1]} \oplus \nu\right)\,\bigg|\,\bm\nu\right]\nonumber\\
    & = \sum_{\substack{\bm{k'} \in \mathcal{P}(k')\\\bm{k''} \in \mathcal{P}(k'')}}\binom{k'}{\bm{k''}}\left\{\prod_{s \in \{0, 1\}^3}\left(\frac{n_s}{n}\right)^{k'_s}\right\}\binom{k''}{\bm{k''}}\left\{\prod_{s \in \{0, 1\}^3}\left(\frac{n_s}{n}\right)^{k''_s}\right\}\nonumber\\
    & \hspace*{80px} \times f\left(\left(\sum_{\substack{z^{[1]},\,z^{[0]},\,z^{[-1]} \in \{0, 1\}\\z^{[t]} = 1}}\hspace*{-20px}k'_{z^{[1]}z^{[0]}z^{[-1]}} + \sum_{\substack{z^{[1]},\,z^{[0]},\,z^{[-1]} \in \{0, 1\}\\z^{[t]} = 0}}\hspace*{-20px}k''_{z^{[1]}z^{[0]}z^{[-1]}}\right)_{t \in \{1, 0, -1\}}\right)\nonumber\\
    & = \sum_{\substack{\bm{k'} \in \mathcal{P}(k')\\\bm{k''} \in \mathcal{P}(k'')}}\binom{k'}{\bm{k''}}\left\{\prod_{s \in \{0, 1\}^3}\left(\frac{n_s}{n}\right)^{k'_s}\right\}\binom{k''}{\bm{k''}}\left\{\prod_{s \in \{0, 1\}^3}\left(\frac{n_s}{n}\right)^{k''_s}\right\}\nonumber\\
    & \hspace*{80px} \times f\left(\left(\sum_{\substack{z^{[1]},\,z^{[0]},\,z^{[-1]} \in \{0, 1\}\\z^{[t]} = 1}}\hspace*{-20px}\left(k'_{z^{[1]}z^{[0]}z^{[-1]}} + \overline{k''}_{z^{[1]}z^{[0]}z^{[-1]}}\right)\right)_{t \in \{1, 0, -1\}}\right)\nonumber\\
    & = \sum_{\substack{\bm{k'} \in \mathcal{P}(k')\\\bm{k''} \in \mathcal{P}(k'')}}\binom{k'}{\bm{k''}}\left\{\prod_{s \in \{0, 1\}^3}\left(\frac{n_s}{n}\right)^{k'_s}\right\}\binom{k''}{\bm{k''}}\left\{\prod_{s \in \{0, 1\}^3}\left(\frac{n_s}{n}\right)^{k''_{\overline{s}}}\right\}\nonumber\\
    & \hspace*{80px} \times f\left(\left(\sum_{\substack{z^{[1]},\,z^{[0]},\,z^{[-1]} \in \{0, 1\}\\z^{[t]} = 1}}\hspace*{-20px}\left(k'_{z^{[1]}z^{[0]}z^{[-1]}} + k''_{z^{[1]}z^{[0]}z^{[-1]}}\right)\right)_{t \in \{1, 0, -1\}}\right),
\end{align}
where in the last-but-one line we introduced the negation $\bm{\overline{k''}}$ of configuration basis numbers $\bm{k''}$ as specified in notation \ref{notation:configuration_basis_numbers_negation} and in the final line we changed summation index $\bm{k''} \longrightarrow \bm{k''}$. To obtain the desired expectation over a single random clause $\sigma$, it remains to average the last quantity over negation choices $\bm\nu$. We observe the quantity only depends on $k'$ and $k''$ (the numbers of $0$ and $1$ negations) which has distribution
\begin{align}
    2^{-k}\binom{k}{k'}\mathbf{1}\left[k' + k'' = k\right]
\end{align}
since negations are independent Bernoulli variables of probability $1/2$. Hence,
\begin{align*}
    & \mathbf{E}_{\sigma}f\left(\sum_{\left(l, \nu\right) \in \sigma}z^{[1]}_l \oplus \nu, \sum_{\left(l, \nu\right) \in \sigma}z^{[0]}_l \oplus \nu, \sum_{\left(l, \nu\right) \in \sigma}z^{[-1]} \oplus \nu\right)\\
    & = \mathbf{E}_{\bm\nu}\left[\mathbf{E}_{\bm{l}}\left[f\left(\sum_{\left(l, \nu\right) \in \sigma}z^{[1]}_l \oplus \nu, \sum_{\left(l, \nu\right) \in \sigma}z^{[0]}_l \oplus \nu, \sum_{\left(l, \nu\right) \in \sigma}z^{[-1]} \oplus \nu\right)\,\bigg|\,\bm\nu\right]\right]\\
    & = 2^{-k}\sum_{\substack{k', k''\\k' + k'' = k}}\binom{k}{k'}\sum_{\substack{\bm{k'} \in \mathcal{P}(k')\\\bm{k''} \in \mathcal{P}(k'')}}\binom{k'}{\bm{k'}}\left\{\prod_{s \in \{0, 1\}^3}\left(\frac{n_s}{n}\right)^{k'_s}\right\}\binom{k''}{\bm{k''}}\left\{\prod_{s \in \{0, 1\}^3}\left(\frac{n_{\overline{s}}}{n}\right)^{k''_s}\right\}\nonumber\\
    & \hspace*{80px} \times f\left(\left(\sum_{\substack{z^{[1]}, z^{[0]}, z^{[-1]} \in \{0, 1\}\\z^{[t]} = 1}}\left(\bm{k'} + \bm{k''}\right)_{z^{[1]}z^{[0]}z^{[-1]}}\right)_{t \in \{1, 0, -1\}}\right)
\end{align*}
Some simplification occurs between multinomial coefficients:
\begin{align*}
    \binom{k}{k'}\binom{k'}{\bm{k''}}\binom{k''}{\bm{k''}} & = \frac{k!}{k'!k''!} \times \frac{k'}{\prod\limits_{s \in \{0, 1\}^3}k'_s!} \times \frac{k''}{\prod\limits_{s \in \{0, 1\}^3}k''_s!}\\
    & = \frac{k!}{\left(\prod\limits_{s \in \{0, 1\}^3}k'_s!\right)\left(\prod\limits_{s \in \{0, 1\}^3}k''_s!\right)}
\end{align*}
We then exploit the fact that the part involving $f$ only depends on $\bm{k'} + \bm{k''}$ by summing over $k'_s, k''_s$ for each fixed value of $k'_s + k''_s$. More precisely, given a prescribed value $k_s$ for $k'_s + k''_s$, we use the following identity resulting from the standard binomial theorem:
\begin{align}
    \sum_{\substack{k'_s, k''_s \geq 0\\k'_s + k''_s = k_s}}\frac{1}{k'_s!k''_s!}\left(\frac{n_s}{n}\right)^{k'_s}\left(\frac{n_{\overline{s}}}{n}\right)^{k''_s} & = \frac{1}{k_s!}\left(\frac{n_s + n_{\overline{s}}}{n}\right)^{k_s}.
\end{align}
We then obtain
\begin{align*}
    & \mathbf{E}_{\sigma}f\left(\sum_{\left(l, \nu\right) \in \sigma}z^{[1]}_l \oplus \nu, \sum_{\left(l, \nu\right) \in \sigma}z^{[0]}_l \oplus \nu, \sum_{\left(l, \nu\right) \in \sigma}z^{[-1]} \oplus \nu\right)\\
    & = 2^{-k}\sum_{\bm{k} \in \mathcal{P}(k)}\binom{k}{\bm{k}}\left\{\prod_{s \in \{0, 1\}^3}\left(\frac{n_s + n_{\overline{s}}}{n}\right)^{k_s}\right\}f\left(\left(\sum_{\substack{z^{[1]}\,z^{[0]},\,z^{[-1]} \in \{0, 1\}\\z^{[t]} = 1}}k_{z^{[1]}z^{[0]}z^{[-1]}}\right)_{t \in \{1, 0, -1\}}\right)
\end{align*}
\end{proof}
\end{lemma}

An explicit formula for the single-clause polynomial is directly deduced from lemma \ref{lemma:averaging_hamming_weights_function_hamming_weight_truth_table}:

\begin{proposition}[Single-clause polynomial for Hamming weight truth table]
\label{prop:single_clause_polynomial_hamming_weight_truth_table_explicit_expression}
The single-clause polynomial (as introduced in definition \ref{def:single_clause_polynomial}) for a Hamming weight truth table $T$ is given by:
\begin{align}
    & P_{\mathrm{single}}\left(\bm{n}\right)\nonumber\\
    & := 2^{-k}\sum_{\bm{k} \in \mathcal{P}(k)}\binom{k}{\bm{k}}\left\{\prod_{s \in \{0, 1\}^3}\left(\frac{n_s + n_{\overline{s}}}{n}\right)^{k_s}\right\}\exp\left(-\frac{i\gamma}{2}\left(T\left(\sum_{z^{[0]},\,z^{[-1]} \in \{0, 1\}}\hspace*{-15px}k_{1z^{[0]}z^{[-1]}}\right) - T\left(\sum_{z^{[1]},\,z^{[0]} \in \{0, 1\}}\hspace*{-15px}k_{z^{[1]}z^{[0]}1}\right)\right)\right)\nonumber\\
    & \hspace*{200px} \times T\left(\sum_{z^{[1]},\,z^{[-1]} \in \{0, 1\}}\hspace*{-15px}k_{z^{[1]}1z^{[-1]}}\right)\label{eq:single_clause_polynomial_hamming_weight_truth_table_explicit_expression}
\end{align}
\begin{proof}
Recalling expression \ref{eq:single_clause_polynomial_hamming_weight_truth_table} of the single-clause polynomial in the case of a Hamming-weight truth table, this results from applying lemma \ref{lemma:averaging_hamming_weights_function_hamming_weight_truth_table} to function:
\begin{align}
    f\left(h^{[1]}, h^{[0]}, h^{[-1]}\right) & := \exp\left(-\frac{i\gamma}{2}\left(T\left(h^{[1]}\right) - T\left(h^{[-1]}\right)\right)\right)T\left(h^{[0]}\right).
\end{align}
\end{proof}
\end{proposition}

\subsection{General truth table}
\label{sec:general_formulae_general_truth_table}

In this paragraph, we compute the single-clause expectation for a general truth table. That is, we assume definition \ref{def:general_truth_table}, whereby the truth value of a $k$-bitstring may depend on the ordering of the $1$ bits unlike in the Hamming weight case treated in the previous section.

To analyze this case, it will be convenient to partition $k$-bitstring triplets jointly according to their truth values and configuration basis numbers. We introduce such a partitioning in the next definition: 

\begin{definition}[Partitioning of bitstring triplets, finer partition]
\label{def:general_truth_table_triplet_fine_partition}
Let be given:
\begin{itemize}
    \item A triplet of truth values $\bm{y} = \left(y^{[1]}, y^{[0]}, y^{[-1]}\right)$.
    \item A partition of $[k]$ into two disjoint subsets $K', K''$: $[k] = K' \sqcup K''$. We denote $k' = |K'|$ and $k'' = |K''|$.
    \item A set of configuration basis numbers $\bm{k'} = \left(k'_s\right)_{s \in \{0, 1\}^3}$ of weight $k'$ and a set of configuration basis numbers $\bm{k''} = \left(k''_s\right)_{s \in \{0, 1\}^3}$ of weight $k''$.
\end{itemize}
From these quantities, we define
\begin{align}
    \mathcal{Z}\left(\bm{y},\,K',\,\bm{k'},\,K'', \bm{k''}\right)
\end{align}
as the set of $k$-bitstrings triplets $\left(\bm{w}^{[1]}, \bm{w}^{[0]}, \bm{w}^{[-1]}\right)$ satisfying:
\begin{itemize}
    \item $T\left(\bm{w}^{[t]}\right) = y^{[l]}$ for $t \in \{1, 0, -1\}$.
    \item The triplet of bitstrings restricted to $K'$: $\left(\bm{w}^{[1]}_{K'},\,\bm{w}^{[0]}_{K'},\,\bm{w}^{[-1]}_{K'}\right)$ satisfies configuration $\bm{k'}$. Similarly, the triplet of bitstrings restricted to $K''$: $\left(\bm{w}^{[1]}_{K''},\,\bm{w}^{[0]}_{K''},\,\bm{w}^{[-1]}_{K''}\right)$ satisfies configuration $\bm{k''}$.
\end{itemize}
\end{definition}

We also introduce a coarser partition of bitstring triplets, only accounting for the joint truth value and the global, weight-$k$, configuration basis numbers:
\begin{definition}[Partitioning of bitstring triplets, coarser partition]
	\label{def:general_truth_table_triplet_coarse_partition}
	Let be given:
	\begin{itemize}
		\item A triplet of truth values $\bm{y} = \left(y^{[1]}, y^{[0]}, y^{[-1]}\right) \in \{0, 1\}^3$.
		\item A set of configuration basis numbers $\bm{k} = \left(k_s\right)_{s \in \{0, 1\}^3}$ of weight $k$.
	\end{itemize}
	From these quantities, we define:
	\begin{align}
		\mathcal{Z}\left(\bm{y},\,\bm{k}\right)
	\end{align}
	as the set of bitstring triplets $\left(\bm{w}^{[1]}, \bm{w}^{[0]}, \bm{w}^{[-1]}\right)$ satisfying:
	\begin{itemize}
		\item $T\left(\bm{w}^{[t]}\right) = y^{[t]}$ for all $t \in \{1, 0, -1\}$.
		\item The triplet of $k$-bit bitstrings $\left(\bm{w}^{[1]}, \bm{w}^{[0]}, \bm{w}^{[-1]}\right)$ has configuration $\bm{k}$.
	\end{itemize}
\end{definition}

The ``coarse" partitioning introduced in definition \ref{def:general_truth_table_triplet_coarse_partition} can be related to the ``fine" ones from definition \ref{def:general_truth_table_triplet_fine_partition} in the following way:

\begin{lemma}[Relating fine and coarse partitioning of bitstrings]
	\label{lemma:general_truth_table_triplet_coarse_fine_partitioning_relation}
	Consider a ``coarse" partitioning of bitstring triplets
	\begin{align}
		\left\{\left(\bm{w}^{[1]}, \bm{w}^{[0]}, \bm{w}^{[-1]}\right)\,:\,\bm{w}^{[t]} \in \{0, 1\}^k\,\,\forall t \in \{1, 0, -1\}\right\}
	\end{align}
	as introduced in definition \ref{def:general_truth_table_triplet_coarse_partition}. Let
	\begin{align}
		[k] = K' \sqcup K''
	\end{align}
	any partition of $[k]$ (fixed in all the statement of this result). Then for all truth assignment
	\begin{align}
		\bm{y} \in \{0, 1\}^3
	\end{align}
	and set of global configuration numbers
	\begin{align}
		\bm{k} \in \mathcal{P}(k),
	\end{align}
	component $\mathcal{Z}\left(\bm{y}, \bm{k}\right)$ of the coarse partitioning decomposes as the following disjoint union of fine components:
	\begin{align}
		\mathcal{Z}\left(\bm{y},\,\bm{k}\right) & = \bigsqcup_{\substack{\bm{k'} \in \mathcal{P}\left(\left|K'\right|\right)\\\bm{k''} \in \mathcal{P}\left(\left|K''\right|\right)\\\bm{k'} + \bm{k''} = \bm{k}}}\mathcal{Z}\left(\bm{y},\,K',\,\bm{k'},\,K'',\,\bm{k''}\right).
	\end{align}
\end{lemma}

Another relation between the ``fine" (introduced in definition \ref{def:general_truth_table_triplet_fine_partition}) and ``coarse" (introduced in definition \ref{def:general_truth_table_triplet_coarse_partition}) partitions of bitstring triplets is the following:

\begin{lemma}[Relating fine and coarse partitioning of bitstrings, variant]
	\label{lemma:general_truth_table_triplet_coarse_fine_partitioning_relation_variant}
	Consider a ``coarse" partitioning of bitstring triplets $\left\{\left(\bm{w}^{[1]}, \bm{w}^{[0]}, \bm{w}^{[-1]}\right)\,:\,\bm{w}^{[t]} \in \{0, 1\}^k\,\,\forall t \in \{1, 0, -1\}\right\}$ as introduced in definition \ref{def:general_truth_table_triplet_coarse_partition}. Let $k', k''$, where $k = k' + k''$, and partitions $\bm{k'} \in \mathcal{P}(k'), \bm{k''} \in \mathcal{P}(\bm{k''})$ of $k', k''$ be fixed. Then the the sum of cardinals of fine partition components over sets $K', K''$ of sizes $k', k''$ is related to the cardinal of the corresponding coarse partition component by a combinatorial factor:
	\begin{align}
		\sum_{\substack{K',\,K''\\ [k] = K' \sqcup K''\\|K'| = k',\,|K''| = k''}}\left|\mathcal{Z}\left(\bm{y},\,K',\,\bm{k'},\,K'',\,\bm{k''}\right)\right| & = \left(\prod_{s \in \{0, 1\}^3}\binom{k'_s + k''_s}{k'_s}\right)\left|\mathcal{Z}\left(\bm{y},\,\bm{k'} + \bm{k''}\right)\right|
	\end{align}
	\begin{proof}
            In the following, we let
            \begin{align}
                \bm{k} & := \bm{k'} + \bm{k''},
            \end{align}
            which forms a set of configuration basis numbers for a $k$-bistrings triplet. We compute the left-hand side of the proposition's equality by injecting the definition of $\mathcal{Z}\left(\bm{y},\,K'\,\bm{k'},\,K'',\,\bm{k''}\right)$ stated in definition \ref{def:general_truth_table_triplet_fine_partition}.
		\begin{align*}
			& \sum_{\substack{K',\,K''\\ [k] = K' \sqcup K''\\|K'| = k',\,|K''| = k''}}\left|\mathcal{Z}\left(\bm{y},\,K',\,\bm{k'},\,K'',\,\bm{k''}\right)\right|\\
			& = \sum_{\substack{K',\,K''\\ [k] = K' \sqcup K''\\|K'| = k',\,|K''| = k''}}\left|\left\{\left(\bm{w}^{[1]}, \bm{w}^{[0]}, \bm{w}^{[-1]}\right) \in \{1, -1\}^{3 \times k}\,:\,\begin{array}{c}\left(\bm{w}_{K'}^{[1]}, \bm{w}_{K'}^{[0]}, \bm{w}_{K'}^{[-1]}\right)\textrm{ satisfies } \bm{k'}\\\left(\bm{w}_{K''}^{[1]}, \bm{w}_{K''}^{[0]}, \bm{w}_{K''}^{[-1]}\right)\textrm{ satisfies } \bm{k''}\\\left(T\left(\bm{w}^{[1]}\right), T\left(\bm{w}^{[0]}\right), T\left(\bm{w}^{[-1]}\right)\right) = \bm{y}\end{array}\right\}\right|\\
			& = \sum_{\substack{K',\,K''\\ [k] = K' \sqcup K''\\|K'| = k',\,|K''| = k''}}\sum_{\left(\bm{w}^{[1]}, \bm{w}^{[0]}, \bm{w}^{[-1]}\right) \in \{1, -1\}^{3 \times k}}\mathbf{1}\left[\left(\bm{w}_{K'}^{[1]}, \bm{w}_{K'}^{[0]}, \bm{w}_{K'}^{[-1]}\right)\textrm{ satisfies }\bm{k'}\right]\mathbf{1}\left[\left(\bm{w}_{K''}^{[1]}, \bm{w}_{K''}^{[0]}, \bm{w}_{K''}^{[-1]}\right)\textrm{ satisfies }\bm{k''}\right]\\[-25px]
			& \hspace*{190px} \times \mathbf{1}\left[\left(T\left(\bm{w}^{[1]}\right), T\left(\bm{w}^{[0]}\right), T\left(\bm{w}^{[-1]}\right)\right) = \bm{y}\right]\\[10pt]
			& = \sum_{\substack{K',\,K''\\ [k] = K' \sqcup K''\\|K'| = k',\,|K''| = k''}}\sum_{\left(\bm{w}^{[1]}, \bm{w}^{[0]}, \bm{w}^{[-1]}\right) \in \{1, -1\}^{3 \times k}}\mathbf{1}\left[\left(\bm{w}_{K'}^{[1]}, \bm{w}_{K'}^{[0]}, \bm{w}_{K'}^{[-1]}\right)\textrm{ satisfies }\bm{k'}\right]\mathbf{1}\left[\left(\bm{w}_{K''}^{[1]}, \bm{w}_{K''}^{[0]}, \bm{w}_{K''}^{[-1]}\right)\textrm{ satisfies }\bm{k''}\right]\\[-25px]
			& \hspace*{190px} \times \mathbf{1}\left[\left(\bm{w}^{[1]}, \bm{w}^{[0]}, \bm{w}^{[-1]}\right)\textrm{ satisfies }\bm{k}\right]\\
                & \hspace*{190px} \times \mathbf{1}\left[\left(T\left(\bm{w}^{[1]}\right), T\left(\bm{w}^{[-]}\right), T\left(\bm{w}^{[-1]}\right)\right) = \bm{y}\right]\\[10pt]
			& = \sum_{\substack{K',\,K''\\ [k] = K' \sqcup K''\\|K'| = k',\,|K''| = k''}}\sum_{\left(\bm{w}^{[1]}, \bm{w}^{[0]}, \bm{w}^{[-1]}\right) \in \{1, -1\}^{3 \times k}}\mathbf{1}\left[\left(\bm{w}_{K'}^{[1]}, \bm{w}_{K'}^{[0]}, \bm{w}_{K'}^{[-1]}\right)\textrm{ satisfies }\bm{k'}\right]\mathbf{1}\left[\left(\bm{w}_{K''}^{[1]}, \bm{w}_{K''}^{[0]}, \bm{w}_{K''}^{[-1]}\right)\textrm{ satisfies }\bm{k''}\right]\\[-25px]
			& \hspace*{190px} \times \mathbf{1}\left[\left(\bm{w}^{[1]}, \bm{w}^{[0]}, \bm{w}^{[-1]}\right)\textrm{ satisfies }\bm{k'} + \bm{k''}\right]\nonumber\\
                & \hspace*{190px} \times \mathbf{1}\left[\left(T\left(\bm{w}^{[1]}\right), T\left(\bm{w}^{[0]}\right), T\left(\bm{w}^{[-1]}\right)\right) = \bm{y}\right]\\[10pt]
			& = \sum_{\left(\bm{w}^{[1]}, \bm{w}^{[0]}, \bm{w}^{[-1]}\right) \in \{1, -1\}^{3 \times k}}\mathbf{1}\left[\left(\bm{w}^{[1]}, \bm{w}^{[0]}, \bm{w}^{[-1]}\right)\textrm{ satisfies } \bm{k}\right]\mathbf{1}\left[\left(T\left(\bm{w}^{[1]}\right), T\left(\bm{w}^{[0]}\right), T\left(\bm{w}^{[-1]}\right)\right) = \bm{y}\right]\\
			& \hspace*{120px} \times \sum_{\substack{K',\,K''\\ [k] = K' \sqcup K''\\|K'| = k',\,|K''| = k''}}\mathbf{1}\left[\left(\bm{w}_{K'}^{[1]}, \bm{w}_{K'}^{[0]}, \bm{w}_{K'}^{[-1]}\right)\textrm{ satisfies }\bm{k'}\right]\mathbf{1}\left[\left(\bm{w}_{K''}^{[1]}, \bm{w}_{K''}^{[0]}, \bm{w}_{K''}^{[-1]}\right)\textrm{ satisfies }\bm{k''}\right]
		\end{align*}
		It remains to evaluate the inner sum over $K', K''$. For that purpose, observe the sum counts the choice of bipartitions of $[k]$ such that bitstring triplet $\left(\bm{w}^{[1]}, \bm{w}^{[0]}, \bm{w}^{[-1]}\right)$ satisfies $\bm{k'}$ when restricted to one index set of the partition and $\bm{k''}$ when restricted to the complementary index set. Thanks to the indicator function before the sum, we may assume the non-restricted triplet $\left(\bm{w}^{[1]}, \bm{w}^{[0]}, \bm{w}^{[-1]}\right)$ satisfies configuration $\bm{k}$. The choice of such a bipartition can then be described as follows: for all $s \in \{0, 1\}^3$, choose $k'_s$ indices $j \in [k]$ such that $\left(w^{[1]}_j,\,w^{[0]}_j,\,w^{[-1]}_j\right)$ is in configuration $s$, among the the $k'_s + k''_s$ such available indices; the (disjoint) union of such indices $j$ over all $s \in \{0, 1\}^3$ then form set $K'$. Next, let $K'' := [k] - K'$, ensuring $\left(\bm{w}^{[1]}_{K''},\,\bm{w}^{[0]}_{K''},\,\bm{w}^{[-1]}_{K''}\right)$ satisfies $\bm{k''}$ given the unrestricted triplet satisfies $\bm{k'} + \bm{k''}$. All in all, this gives
		\begin{align}
			\prod_{s \in \{0, 1\}^3}\binom{k'_s + k''_s}{k'_s}
		\end{align}
		choices of bipartitions of $[k]$. It follows:
		\begin{align*}
			& \sum_{\substack{K',\,K''\\ [k] = K' \sqcup K''\\|K'| = k',\,|K''| = k''}}\left|\mathcal{Z}\left(\bm{y},\,K',\,\bm{k'},\,K'',\,\bm{k''}\right)\right|\\
			& = \sum_{\left(\bm{w}^{[1]},\,\bm{w}^{[0]},\,\bm{w}^{[-1]}\right) \in \{1, -1\}^{3 \times k}}\mathbf{1}\left[\left(\bm{w}^{[1]}, \bm{w}^{[0]}, \bm{w}^{[-1]}\right)\textrm{ satisfies } \bm{k}\right]\mathbf{1}\left[\left(T\left(\bm{w}^{[1]}\right), T\left(\bm{w}^{[0]}\right), T\left(\bm{w}^{[-1]}\right)\right) = \bm{y}\right]\\
			& \hspace*{120px} \times \prod_{s \in \{0, 1\}^3}\binom{k'_s + k''_s}{k'_s}\\
			& = \left(\prod_{s \in \{0, 1\}^3}\binom{k'_s + k''_s}{k'_s}\right)\sum_{\substack{K',\,K''\\ [k] = K' \sqcup K''\\|K'| = k',\,|K''| = k''}}\left|\left\{\left(\bm{w}^{[1]}, \bm{w}^{[0]}, \bm{w}^{[-1]}\right) \in \{1, -1\}^{3 \times k}\,:\,\begin{array}{c}\left(\bm{w}^{[1]}, \bm{w}^{[0]}, \bm{w}^{[-1]}\right)\textrm{ satisfies } \bm{k}\\\left(T\left(\bm{w}^{[1]}\right), T\left(\bm{w}^{[0]}\right), T\left(\bm{w}^{[-1]}\right)\right) = \bm{y}\end{array}\right\}\right|\\
			& =\left(\prod_{s \in \{0, 1\}^3}\binom{k'_s + k''_s}{k'_s}\right)\left|\mathcal{Z}\left(\bm{y},\,\bm{k}\right)\right|.
		\end{align*}
	\end{proof}
\end{lemma}

We now show how to compute the single-clause expectation:
\begin{align}
    \mathbf{E}_{\sigma}f\left(\mathbf{1}\left[\bm{z}^{[1]} \vdash \sigma\right],\,\mathbf{1}\left[\bm{z}^{[0]} \vdash \sigma\right],\,\mathbf{1}\left[\bm{z}^{[-1]} \vdash \sigma\right]\right)
\end{align}
from the cardinals of the ``fine partition" sets of bitstring triplets introduced in definition \ref{def:general_truth_table_triplet_fine_partition}.

\begin{proposition}[Single-clause expectation from partitioning of bitstrings]
\label{prop:single_clause_expectation_general_truth_table}
	Let $\bm{z}^{[1]}, \bm{z}^{[0]}, \bm{z}^{[-1]}$ be $3$ fixed $n$-bit bitstrings. Consider the single-clause expectation
	\begin{align}
		\mathbf{E}_{\sigma}f\left(\mathbf{1}\left[\bm{z}^{[1]} \vdash \sigma\right], \mathbf{1}\left[\bm{z}^{[0]} \vdash \sigma \right], \mathbf{1}\left[\bm{z}^{[-1]} \vdash \sigma\right]\right),
	\end{align}
	where $f$ is an arbitrary function $\{0, 1\}^3 \longrightarrow \mathbf{C}$. Recall the partitioning of triplets of length-$k$ bitstrings introduced in definition \ref{def:general_truth_table_triplet_fine_partition}. Then the above expectation evaluates to:
	\begin{align}
		\sum_{\substack{\bm{y} \in \{0, 1\}^3\\\bm{k} \in \mathcal{P}(k)}}f\left(\bm{y}\right)\frac{1}{2^k}\left(\prod_{s \in \{0, 1\}^3}\left(\frac{n_s + n_{\overline{s}}}{n}\right)^{k_s}\right)\left|\mathcal{Z}\left(\bm{y},\,\bm{k}\right)\right|.
	\end{align}
	\begin{proof}
		First, it will be useful to decompose the single-clause expectation as follows:
		\begin{align}
			\mathbf{E}_{\sigma}\left[\cdot\right] & = \mathbf{E}_{J}\left[\mathbf{E}_{\sigma}\left[\cdot \,|\,J\right]\right],
		\end{align}
		where the outer expectation $\mathbf{E}_J\left[\cdot\right]$ is over the choice of clause variables (that is the ``scope" of the clause) and the inner expectation $\mathbf{E}_{\sigma}\left[\cdot\,|\,J\right]$ is over random clauses conditioned on the scope. In other words, the inner expectation runs over choices of negations. The scope of the clause will be described by an ordered tuple $J$ (order matters since the truth table is now arbitrary and not only dependent on the Hamming weight).

		\begin{align*}
			\mathbf{E}_{\sigma}f\left(\mathbf{1}\left[\bm{z}^{[1]} \vdash \sigma\right],\,\mathbf{1}\left[\bm{z}^{[0]} \vdash \sigma\right],\,\mathbf{1}\left[\bm{z}^{[-1]} \vdash \sigma\right]\right) & = \mathbf{E}_{J}\left[\mathbf{E}_{\sigma}\left[f\left(\mathbf{1}\left[\bm{z}^{[1]} \vdash \sigma\right],\,\mathbf{1}\left[\bm{z}^{[0]} \vdash \sigma\right],\,\mathbf{1}\left[\bm{z}^{[-1]} \vdash \sigma\right]\right)\,\big|\,J\right]\right]\\
			& = \mathbf{E}_{J}\left[\mathbf{E}_{\bm\nu}\left[f\left(T\left(\bm\nu \oplus \bm{z}_J^{[1]}\right), T\left(\bm\nu \oplus \bm{z}_J^{[0]}\right), T\left(\bm\nu \oplus \bm{z}_J^{[-1]}\right)\right)\right]\right].
		\end{align*}
		In the third line, we introduced for all $t \in \{1, 0, -1\}$ the restriction $\bm{z}^{[t]}_J$ of $n$-bit bitstring $\bm{z}^{[t]}$ to index $k$-tuple $J$ (see definition \ref{def:subbitstring_triplet}); if some indices are repeated in this tuple, so are the corresponding bits of the string. We also used that the expectation over random clauses conditioned on scope $J$ is equivalent to averaging over $k$-bit negations $\bm\nu$. Given a restricted bitstring $\bm{z}^{[t]}_J \in \{1, -1\}^k$, $\bm\nu \oplus \bm{z}^{[t]}_J$ denotes $\bm{z}^{[t]}_J$ with negation bits applied (which corresponds to a coordinate-wise XOR when working with $\{0, 1\}$-valued bits). We now partition (negated) restricted bitstrings according to sets $\mathcal{Z}\left(\bm{y}, \bm{k}\right)$ introduced in definition \ref{def:general_truth_table_triplet_coarse_partition}:
		\begin{align*}
			& \mathbf{E}_{J}\left[\mathbf{E}_{\bm\nu}\left[f\left(T\left(\bm\nu \oplus \bm{z}_J^{[1]}\right), T\left(\bm\nu \oplus \bm{z}_J^{[0]}\right), T\left(\bm\nu \oplus \bm{z}_J^{[-1]}\right)\right)\right]\right]\\
			& = \mathbf{E}_J\left[\mathbf{E}_{\bm\nu}\left[\sum_{\substack{\bm{y} \in \{0, 1\}^3\\\bm{k} \in \mathcal{P}\left(k\right)}}\mathbf{1}\left[\left(\bm\nu \oplus \bm{z}^{[1]}_J, \bm\nu \oplus \bm{z}^{[0]}_J, \bm\nu \oplus \bm{z}^{[-1]}_J\right) \in \mathcal{Z}\left(\bm{y},\,\bm{k}\right)\right]f\left(T\left(\bm\nu \oplus \bm{z}_J^{[1]}\right), T\left(\bm\nu \oplus \bm{z}_J^{[0]}\right), T\left(\bm\nu \oplus \bm{z}_J^{[-1]}\right)\right)\right]\right]\\
			& = \mathbf{E}_J\left[\mathbf{E}_{\bm\nu}\left[\sum_{\substack{\bm{y} \in \{0, 1\}^3\\\bm{k} \in \mathcal{P}\left(k\right)}}\mathbf{1}\left[\left(\bm\nu \oplus \bm{z}^{[1]}_J, \bm\nu \oplus \bm{z}^{[0]}_J, \bm\nu \oplus \bm{z}^{[-1]}_J\right) \in \mathcal{Z}\left(\bm{y},\,\bm{k}\right)\right]f\left(\bm{y}\right)\right]\right]\\
			& = \sum_{\substack{\bm{y} \in \{0, 1\}^3\\\bm{k} \in \mathcal{P}\left(k\right)}}\mathbf{E}_J\left[\mathbf{E}_{\bm\nu}\left[\mathbf{1}\left[\left(\bm\nu \oplus \bm{z}^{[1]}_J, \bm\nu \oplus \bm{z}^{[0]}_J, \bm\nu \oplus \bm{z}^{[-1]}_J\right) \in \mathcal{Z}\left(\bm{y},\,\bm{k}\right)\right]f\left(\bm{y}\right)\right]\right]\\
			& = \sum_{\substack{\bm{y} \in \{0, 1\}^3\\\bm{k} \in \mathcal{P}\left(k\right)}}\mathbf{E}_{\bm\nu}\left[\mathbf{E}_J\left[\mathbf{1}\left[\left(\bm\nu \oplus \bm{z}^{[1]}_J, \bm\nu \oplus \bm{z}^{[0]}_J, \bm\nu \oplus \bm{z}^{[-1]}_J\right) \in \mathcal{Z}\left(\bm{y},\,\bm{k}\right)\right]f\left(\bm{y}\right)\right]\right]
		\end{align*}
		It remains to evaluate the expectations for each triplet of truth values $\bm{y} \in \{0, 1\}^3$ and set of weight-$k$ configuration basis numbers $\bm{k} \in \mathcal{P}(k)$. To achieve that, one starts by choosing negations $\bm\nu$ randomly before computing the expectation over tuples $J$ for this fixed $\bm\nu$. A choice of negations $\bm\nu$ can be described by a partition
		\begin{align}
			[k] = K' \sqcup K''
		\end{align}
		of the $k$ bits into two disjoint sets, where $K'$ is the set of bits without negation applied and $K''$ the set of bits with negation applied. In the following, we denote by $k' = \left|K'\right|$ and $k'' = \left|K''\right|$ the sizes of the two sets forming the partition. Given a choice of negations $\bm\nu$ described by partition
        \begin{align}
            [k] = K' \sqcup K''
        \end{align}
        $[k]$, we denote by $J' \sqcup J''$ the corresponding partition of $J$; that is
        \begin{align}
            J = \left(j_0, \ldots, j_{k - 1}\right) \implies J' := \left(j_m\right)_{m \in K'}, \quad J'' := \left(j_m\right)_{m \in K''}.
        \end{align}
        It will be convenient to denote triplets of bitstrings restricted to $J$ over two rows:
		\begin{align}
			\left(\bm{z}^{[1]}_J, \bm{z}^{[0]}_J, \bm{z}^{[1]}_J\right) & = \begin{pmatrix}
				\bm{z}^{[1]}_{J'} & \bm{z}^{[0]}_{J'} & \bm{z}^{[-1]}_{J'}\\
				\bm{z}^{[1]}_{J''} & \bm{z}^{[0]}_{J''} & \bm{z}^{[-1]}_{J''}
			\end{pmatrix},
		\end{align}
		where the top row collects indices in $J'$ and the bottom one indices in $J''$. [However, care should be taken to implicitly reorder $J$ when evaluating the joint truth assignment of the triplet, since the order of bits matters in the case of a general truth table.] Using this notation, the action of the negations can be simply represented as:
		\begin{align}
			\left(\bm\nu \oplus z^{[1]}_j, \bm\nu \oplus \bm{z}^{[0]}_J, \bm\nu \oplus \bm{z}^{[-1]}_J\right) & = \begin{pmatrix}
				\bm{z}^{[1]}_{J'} & \bm{z}^{[0]}_{J'} & \bm{z}^{[-1]}_{J'}\\
				\mathbf{1}_{k''} \oplus \bm{z}^{[1]}_{J''} & \mathbf{1}_{k''} \oplus \bm{z}^{[0]}_{J''} & \mathbf{1}_{k''} \oplus \bm{z}^{[-1]}_{J''}
			\end{pmatrix}.
		\end{align}
		Besides, we denote by $\bm{k'} \in \mathcal{P}\left(k'\right)$ the configuration of the triplet of bitstrings restricted to $J'$:
		\begin{align}
			\left(\bm{z}^{[1]}_{J'}, \bm{z}^{[0]}_{J'}, \bm{z}^{[-1]}_{J'}\right) \qquad \textrm{configuration } \bm{k'},
		\end{align}
		and by $\bm{k''} \in \mathcal{P}\left(k''\right)$ the configuration of the triplet of bitstrings restricted to $J''$:
		\begin{align}
			\left(\bm{z}^{[1]}_{J''}, \bm{z}^{[0]}_{J''}, \bm{z}^{[-1]}_{J''}\right) \qquad \textrm{configuration } \bm{k''}.
		\end{align}
		The requirement $\left(\bm{z}^{[1]}_J, \bm{z}^{[0]}_J, \bm{z}^{[-1]}_J\right) \in \mathcal{Z}\left(\bm{y}, \bm{k}\right)$ implies in particular that the triplet satisfies configuration $\bm{k}$, which enforces $\bm{k} = \bm{k'} + \bm{k''}$.
		Recapitulating, given:
		\begin{itemize}
			\item a choice of negations parametrized by $K', K''$ (let $k' := \left|K'\right|$, $k'' := \left|K''\right|$);
			\item a set of weight-$k'$ configuration numbers $\bm{k'}$ and a set of weight-$k''$ configuration numbers $\bm{k''}$ summing to configuration numbers $\bm{k}$: $\bm{k} = \bm{k'} + \bm{k''}$;
		\end{itemize}
		one wishes to count the $k$-tuples of variable indices $J$ satisfying:
		\begin{itemize}
			\item the indices of $J$ in positions $K'$, forming $k'$-tuple $J'$, satisfy configuration $\bm{k'}$;
			%
			%
			\item the indices of $J$ in positions $K''$, forming $k''$-tuple $J''$, satisfy configuration $\bm{k''}$;
			\item the joint truth values of the full (reorganized) triplet $\begin{pmatrix}
				\bm{z}^{[1]}_{J'} & \bm{z}^{[0]}_{J'} & \bm{z}^{[-1]}_{J'}\\
				\mathbf{1}_{k''} \oplus \bm{z}^{[1]}_{J''} & \mathbf{1}_{k''} \oplus \bm{z}^{[0]}_{J''} & \mathbf{1}_{k''} \oplus \bm{z}^{[-1]}_{J''}
			\end{pmatrix}$ are $\bm{y}$.
		\end{itemize}
		Recalling definition \ref{def:general_truth_table_triplet_fine_partition} for the partitioning of $k$-bitstrings, there are exactly
		\begin{align}
			\left|\mathcal{Z}\left(\bm{y}, K', \bm{k'}, K'', \bm{\overline{k''}}\right)\right|
		\end{align}
		triplets of $k$-bitstrings satisfying these conditions, where we introduced:
		\begin{align}
			\bm{\overline{k''}} & := \left(k''_{\overline{s}}\right)_{s \in \{0, 1\}^3}
		\end{align}
		to denote configuration basis numbers $\bm{k''}$ with assignments swapped between coordinates related by bit flip (see notation \ref{notation:configuration_basis_numbers_negation}). Finally, the probability of making a choice of variable indices $J$ yielding any of the satisfying $k$-bitstrings triplet is
		\begin{align}
			\prod_{s \in \{0, 1\}^3}\left(\frac{n_s}{n}\right)^{k'_s + k''_{{s}}}. \qquad
		\end{align}
		The full expectation (including averaging over negations) now reads:
		\begin{align}
			& \mathbf{E}_{K', K''}\left[\sum_{\substack{\bm{k'} \in \mathcal{P}\left(\left|K'\right|\right)\\\bm{k''} \in \mathcal{P}\left(\left|K''\right|\right)\\\bm{k} = \bm{k'} + \bm{k''}}}\left(\prod_{s \in \{0, 1\}^3}\left(\frac{n_s}{n}\right)^{k'_s + k''_s}\right)\left|\mathcal{Z}\left(\bm{y}, K', \bm{k''}, K'', \bm{\overline{k''}}\right)\right|\right]
		\end{align}
		It remains to make explicit the expectation over negations. Since negations are applied independently and with probability $\frac{1}{2}$ on each variable, each set of negated variables $K''$ has probability $2^{-k}$ of being chosen. All in all, the expectations above can now be transformed:
		\begin{align*}
			\frac{1}{2^k}\sum_{\substack{K', K''\\ [k] = K' \sqcup K''}}\sum_{\substack{\bm{k'} \in \mathcal{P}\left(\left|K'\right|\right)\\\bm{k''} \in \mathcal{P}\left(\left|K''\right|\right)\\\bm{k} = \bm{k'} + \bm{k''}}}\left(\prod_{s \in \{0, 1\}^3}\left(\frac{n_s}{n}\right)^{k'_s + k''_{{s}}}\right)\left|\mathcal{Z}\left(\bm{y}, K', \bm{k'}, K'', \overline{\bm{k''}}\right)\right| \qquad
		\end{align*}
		We now use lemma \ref{lemma:general_truth_table_triplet_coarse_fine_partitioning_relation_variant} over choices of subsets $K', K''$ given the size of these; schematically, this means we reorganize sums as follows:
		\begin{align}
			\sum_{\substack{K',\,K''\\ [k] = K' \sqcup K''}}\sum_{\substack{\bm{k'} \in \mathcal{P}\left(|K'|\right)\\\bm{k''} \in \mathcal{P}\left(|K''|\right)\\\bm{k} = \bm{k'} + \bm{k''}}} & \longrightarrow \sum_{\substack{k',\,k''\\k = k' + k''}}\sum_{\substack{\bm{k'} \in \mathcal{P}(k')\\\bm{k''} \in \mathcal{P}(k'')\\\bm{k} = \bm{k'} + \bm{k''}}}\sum_{\substack{K',\,K''\\ [k] = K' \sqcup K''\\|K'| = k',\,|K''| = k''}}
		\end{align}
		The desired expectations now read:
		\begin{align*}
			\frac{1}{2^k}\sum_{\substack{k',\,k''\\k = k' + k''}}\sum_{\substack{\bm{k'} \in \mathcal{P}(k')\\\bm{k''} \in \mathcal{P}(k'')\\\bm{k} = \bm{k'} + \bm{k''}}}\left(\prod_{s \in \{0, 1\}^3}\left(\frac{n_s}{n}\right)^{k'_s + k''_s}\right)\left(\prod_{s \in \{0, 1\}^3}\binom{k'_s + k''_{\overline{s}}}{k'_s}\right)\left|\mathcal{Z}\left(\bm{y}, \bm{k'} + \overline{\bm{k''}}\right)\right|.
		\end{align*}
		Restoring the summation over $\bm{y} \in \{0, 1\}^3$ and $\bm{k} \in \mathcal{P}(k)$ to get the full single-clause expectation:
		\begin{align*}
			& \mathbf{E}_{\sigma}f\left(\mathbf{1}\left[z^{[1]} \vdash \sigma\right],\mathbf{1}\left[z^{[0]} \vdash \sigma\right], \mathbf{1}\left[z^{[-1]} \vdash \sigma\right]\right)\\
			& = \sum_{\substack{\bm{y} \in \{0, 1\}^3\\\bm{k} \in \mathcal{P}(k)}}f(\bm{y})\frac{1}{2^k}\sum_{\substack{k',\,k''\\k = k' + k''}}\sum_{\substack{\bm{k'} \in \mathcal{P}(k')\\\bm{k''} \in \mathcal{P}(k'')\\\bm{k} = \bm{k'} + \bm{k''}}}\left(\prod_{s \in \{0, 1\}^3}\left(\frac{n_s}{n}\right)^{k'_s + k''_s}\right)\left(\prod_{s \in \{0, 1\}^3}\binom{k'_s + k''_{\overline{s}}}{k'_s}\right)\left|\mathcal{Z}\left(\bm{y}, \bm{k'} + \overline{\bm{k''}}\right)\right|\\
			& = \sum_{\bm{y} \in \{0, 1\}^3}\sum_{\substack{\bm{k'},\,\bm{k''}\\\sum\limits_{s \in \{0, 1\}^3}k'_s + \sum\limits_{s \in \{0, 1\}^3}k''_s = k}}f(\bm{y})\frac{1}{2^k}\left(\prod_{s \in \{0, 1\}^3}\left(\frac{n_s}{n}\right)^{k'_s + k''_s}\right)\left(\prod_{s \in \{0, 1\}^3}\binom{k'_s + k''_{\overline{s}}}{k'_s}\right)\left|\mathcal{Z}\left(\bm{y}, \bm{k'} + \overline{\bm{k''}}\right)\right|\\
			& = \sum_{\substack{\bm{y} \in \{0, 1\}^3\\\bm{k'''} \in \mathcal{P}(k)}}f(\bm{y})\frac{1}{2^k}\left(\prod_{s \in \{0, 1\}^3}\left(\frac{n_s + n_{\overline{s}}}{n}\right)^{k'''_s}\right)\left|\mathcal{Z}\left(\bm{y},\,\bm{k'''}\right)\right|,
		\end{align*}
		where in the final line we introduced
		\begin{align}
			k'''_{s} & := k'_s + k''_{\overline{s}},
		\end{align}
		and converted the sum over $\bm{k'}, \bm{k''}$ to one over this new variable thanks to the binomial theorem.
	\end{proof}
\end{proposition}

As a simple corollary of proposition \ref{prop:single_clause_expectation_general_truth_table}, we get the single-clause polynomial for a general truth table:

\begin{proposition}[Single-clause polynomial for general truth table]
The single-clause polynomial for a general truth table is well-defined and given by:
\begin{align}
    P_{\mathrm{single}}\left(\bm{n}\right) & = 2^{-k}\sum_{\substack{\bm{y} \in \{0, 1\}^3\\\bm{k} \in \mathcal{P}(k)}}\left|\mathcal{Z}(\bm{y}, \bm{k})\right|\left(\prod_{s \in \{0, 1\}^3}\left(\frac{n_s + n_{\overline{s}}}{n}\right)^{k_s}\right)\exp\left(-\frac{i\gamma}{2}\left(y^{[1]} - y^{[-1]}\right)\right)y^{[0]}
\end{align}
In particular, the single-clause polynomial only depends explicitly on configuration basis numbers
\begin{align}
    \bm{n} & = \left(n_s\right)_{s \in \{0, 1\}^3}.
\end{align}
through reduced configuration basis numbers:
\begin{align}
    \bm{n'} & = \left(n'_{s^{[0]}s^{[-1]}}\right)_{s^{[0]},\,s^{[-1]} \in \{0, 1\}} = \left(n_{0\,s^{[0]}\,s^{[-1]}} + n_{1\,\overline{s^{[0]}}\,\overline{s^{[-1]}}}\right)_{s^{[0]},\,s^{[-1]} \in \{0, 1\}}.
\end{align}
\end{proposition}

\section{Numerical results}

In this section, we apply the formulae in Section \ref{sec:generalformulae} to compute expected success probabilities of QAOA for general truth tables.

\subsection{QAOA Runtimes}

Given a truth table, we can compute the success probability (now denoted $p(n)$) from Proposition~\ref{prop:success_probability_general_formula} by using the results from Section~\ref{sec:general_formulae_general_truth_table} to calculate $P_{\mathrm{single}}\left(\bm{n}\right)$. We can then model the expected running time of QAOA $T_q(n)$ as  $1/p(n)$, though note that formally this is a {\em lower} bound on the expected running time of QAOA by Jensen's inequality. In order to compute this probability, for each instance we first need to specify the clauses-to-variables ratio $r$, and the QAOA angles $\beta$ and $\gamma$.

\begin{figure}[t]
\centering
\includegraphics[width=12cm]{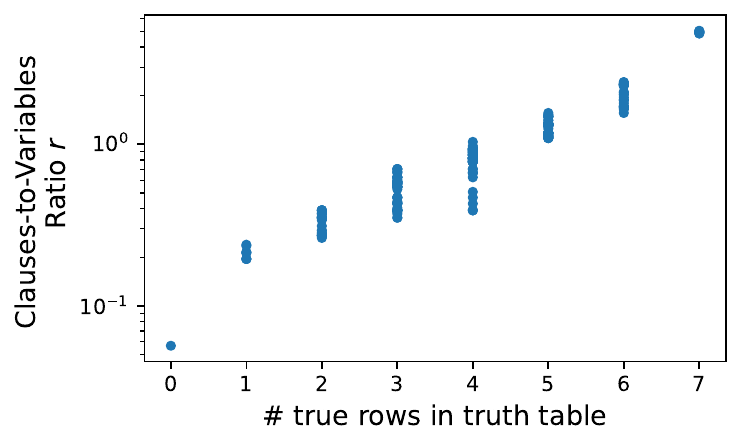}
\caption{The clauses-to-variables ratio for all $2^{2^k}$ truth tables with $k=3$, grouped by the number of true rows in the truth table. Note that statistical noise has resulted in slightly different ratios for equivalent truth tables.}
\label{fig:clauses_to_variables_k3}
\end{figure}

\begin{figure}[t]
\centering
    \begin{subfigure}{0.45\linewidth}
        \includegraphics[width=\linewidth]{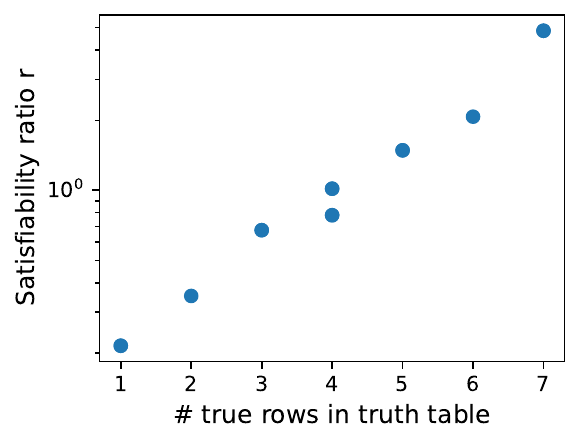}
        \caption{$k=3$}
    \end{subfigure}
    \begin{subfigure}{0.45\linewidth}
        \includegraphics[width=\linewidth]{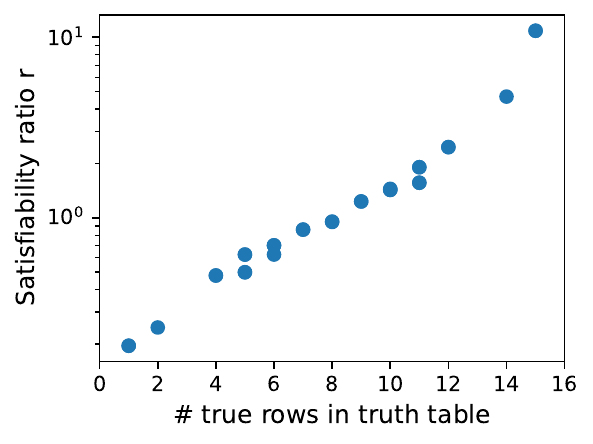}
        \caption{$k=4$}
    \end{subfigure}
    \begin{subfigure}{0.45\linewidth}
        \includegraphics[width=\linewidth]{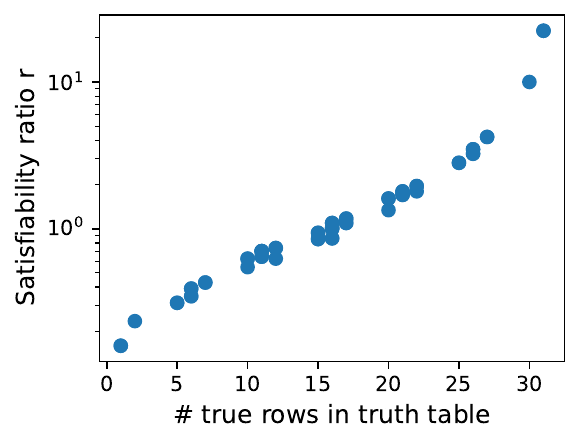}
        \caption{$k=5$}
    \end{subfigure}
    \begin{subfigure}{0.45\linewidth}
        \includegraphics[width=\linewidth]{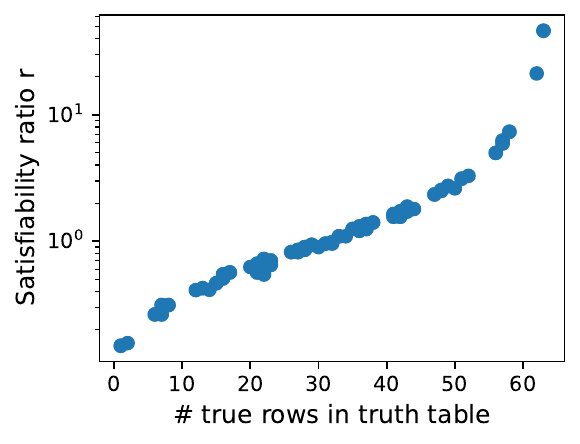}
        \caption{$k=6$}
    \end{subfigure}
\caption{The clauses to variables ratio for Hamming weight truth tables with $k \in \{ 3, 4, 5, 6\}$. The same rough exponential trend from Figure \ref{fig:clauses_to_variables_k3} remains as we increase $k$, with problems with many true rows requiring even more clauses to have a one half probability of the random formula being satisfiable.}
\label{fig:clauses_to_variables_hamming}
\end{figure}

We choose the clauses-to-variables ratio $r$ so that the probability an instance is satisfiable is equal to $\frac{1}{2}$. We can approximately estimate this probability by fixing a small $n$, in this case 12, generating a large number of random instances, in this case 200, and computing the number which are satisfiable by brute force. We can then perform binary search on $r$ until we get this probability to be close to one half. The clauses-to-variables ratios for $k=3$ are shown in Figure~\ref{fig:clauses_to_variables_k3}, where a roughly exponential relationship between the number of true rows in the truth table is demonstrated. This is to be expected as the more true rows there are, the more likely a random assignment is to satisfy any clause, and so more clauses are required to make the overall instance unsatisfiable. Truth tables that correspond to $1$-SAT or $2$-SAT problems also have a larger ratio. For larger values of $k$ we plot the Hamming weight truth tables in Figure \ref{fig:clauses_to_variables_hamming}, as there are a very large number of truth tables. The same trend appears, where the more true rows there are in the truth table, the higher the clauses-to-variables ratio is.

Once we have this ratio, we need to estimate optimal values of the QAOA angle parameters $\beta$ and $\gamma$. To do this, we uniformly sample 50 values for both $\beta$ and $\gamma$ from $[0, 2 \pi)$ and choose the pair which maximise the probability. There is no guarantee that these parameters will be optimal; however, this is sufficient for us to compare the difficulty of varying CSPs.

\begin{figure}[t]
\centering
\includegraphics[width=5.5cm]{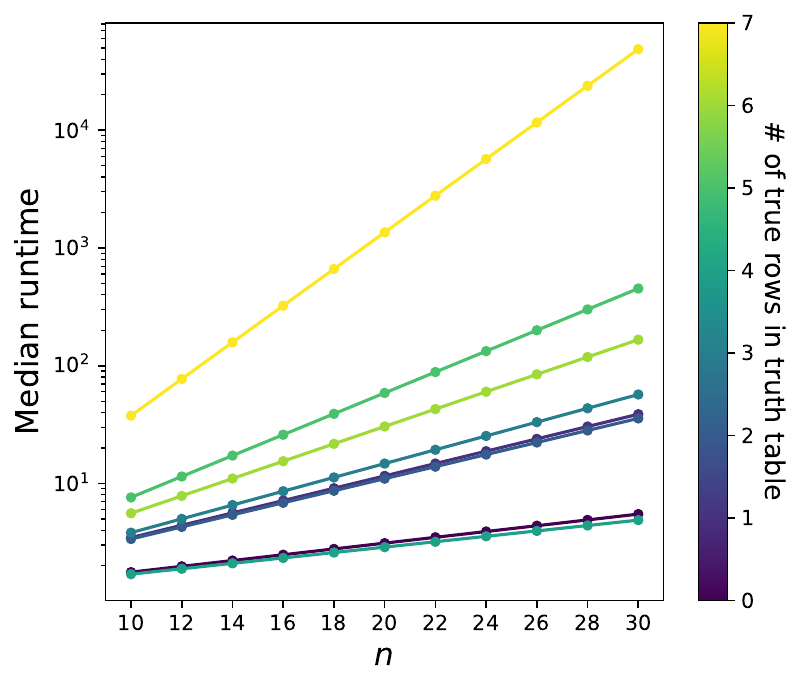}
\hfill
\includegraphics[width=5.5cm]{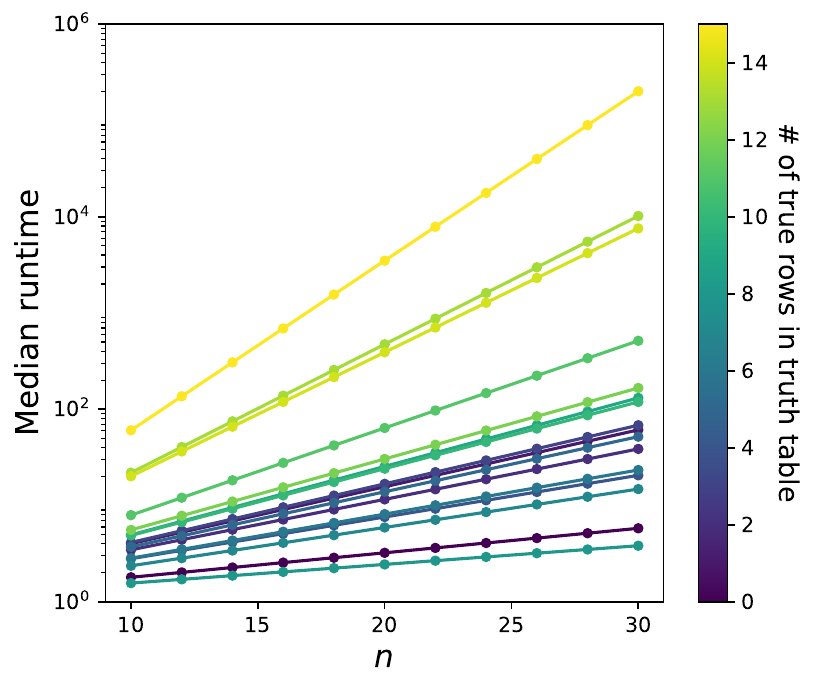}
\hfill
\includegraphics[width=5.5cm]{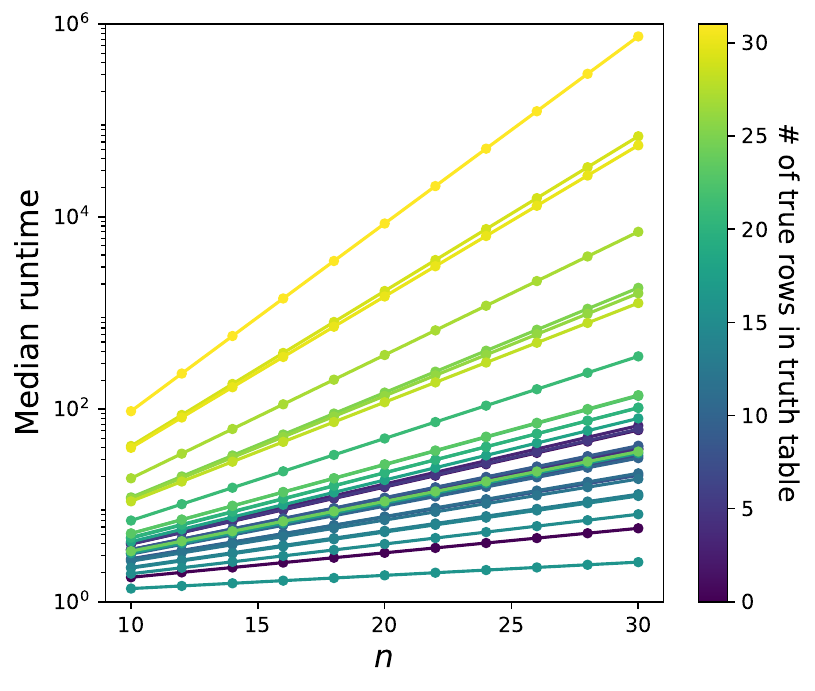}
\caption{The median runtime for QAOA lower bounded by the inverse of the success probability for a subset of truth tables for $k\in \{3,4,5\}$ described in the text, at the satisfiability threshold. 
}
\label{fig:qaoa_k3_runtimes}
\end{figure}

Using these values, we can plot the inverse success probability $T_q(n)$ as a function of $n$ for a given truth table. We have done this for all truth tables for $k=3$, and have plotted this function in Figure \ref{fig:qaoa_k3_runtimes} for a choice of one truth table with a given number of true rows to avoid an overly busy plot. Explicitly, we have chosen the truth table corresponding to the first $i$ rows being false and the other $2^k - i$ rows being true. We have also done the same for $k=4$, $k=5$ for this family of truth tables, and included these in the same plot. The graphs demonstrate a strong exponential relationship between the number of variables and the runtime of QAOA.

\begin{figure}[h!]
\centering
    \begin{subfigure}{0.45\linewidth}
        \includegraphics[width=\linewidth]{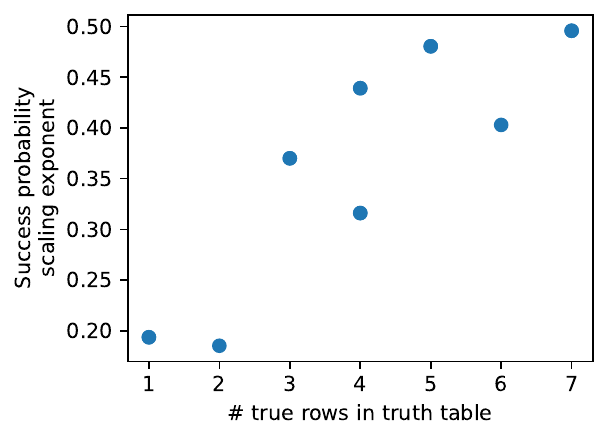}
        \caption{$k=3$}
    \end{subfigure}
    \begin{subfigure}{0.45\linewidth}
        \includegraphics[width=\linewidth]{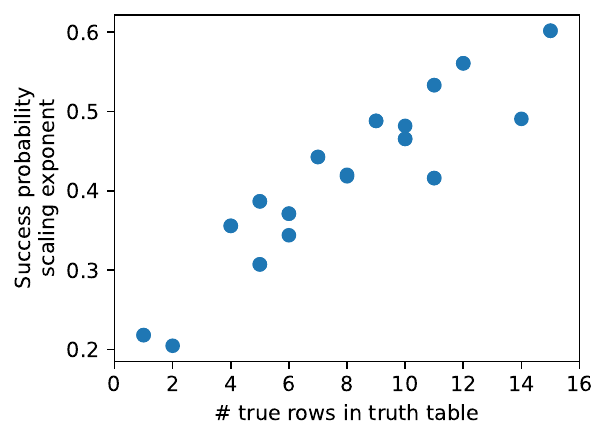}
        \caption{$k=4$}
    \end{subfigure}
    \begin{subfigure}{0.45\linewidth}
        \includegraphics[width=\linewidth]{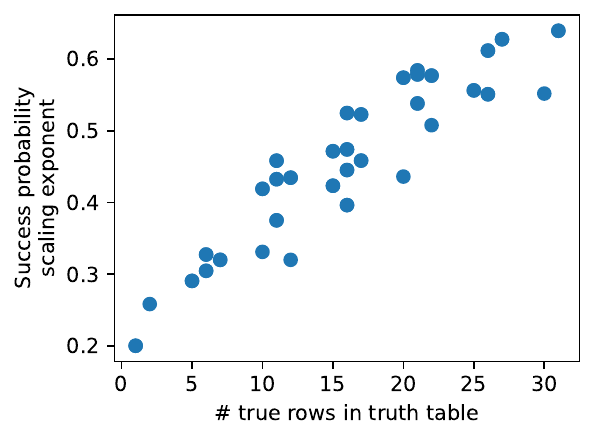}
        \caption{$k=5$}
    \end{subfigure}
    \begin{subfigure}{0.45\linewidth}
        \includegraphics[width=\linewidth]{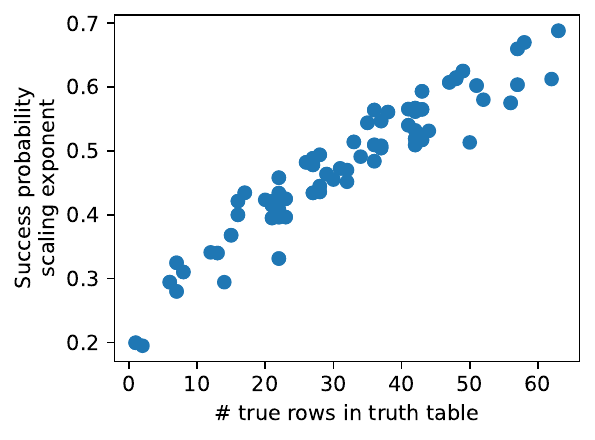}
        \caption{$k=6$}
    \end{subfigure}
\caption{The runtime scaling exponent for QAOA for all Hamming weight truth tables for $k \in \{3, 4, 5, 6\}$. The success probability tends to increase as the number of true values increases.}
\label{fig:qaoa_exp_hamming}
\end{figure}

Finally, we can find the gradient to determine the scaling exponent for QAOA. Restricting to Hamming weight truth tables in Figure \ref{fig:qaoa_exp_hamming}, we can see a trend of the more true values there are in the truth table, the worse the scaling is, with $3$-SAT being the hardest problem.

\subsection{Classical Solver Runtimes}

In order to estimate the classical resource requirements for solving these constraint satisfaction problems, we use the classical SAT solver MapleSAT~\cite{DBLP:conf/sat/LiangGPC16}. Given an instance:
$$
\phi = \sigma_{1}\wedge ... \wedge \sigma_p
$$
we can use the semantics of the truth table to convert this into a $k$-CNF formula with standard semantics in the naive way, where if there are $f$ false rows in the truth table, we convert each of these $p$ clauses into $f$ $k$-clauses. We can then pass this resultant formula to MapleSAT and estimate the running time $T_c(n)$ by adding the total number of decisions and propagations that were made in the execution of the algorithm. We do this for 500 instances of this constraint satisfaction problem and take the median runtime.

\begin{figure}[h!]
\centering
\includegraphics[width=10cm]{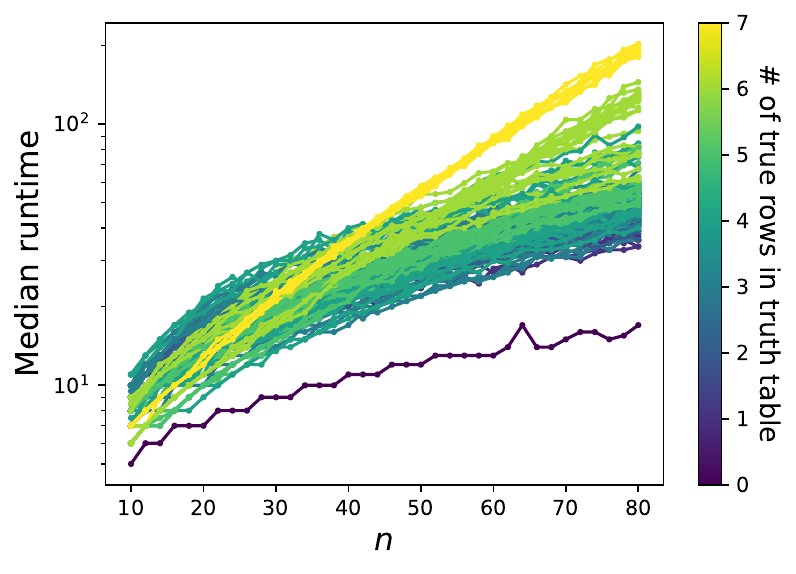}
\caption{The median runtime of MapleSAT for all $k=3$ truth tables. Similarly to QAOA, MapleSAT finds problems with more true rows in the truth table, and hence more clauses per variable, more difficult. }
\label{fig:maplesat_k3_runtimes}
\end{figure}

For $k=3$, we plot the runtimes for all of the truth tables in Figure \ref{fig:maplesat_k3_runtimes}. In order to compare to the quantum runtime, modelled by the inverse of the success probability, we use the same clauses-to-variables ratio for each truth table. The truth tables are coloured by their Hamming weight, defined by the Hamming distance of the truth table from the origin when considered as a vector in $\mathbb{B}^{2^k}$. For the values of $n$ computed here, it seems the only Hamming weight truth tables that demonstrates clearly exponential scaling are those with only one false value, which are problems similar to $k$-SAT. More computation would be required to validate that the other truth tables also show an exponential runtime, as expected from NP-completeness of most of the problems considered. 

The darkest line in Figure \ref{fig:maplesat_k3_runtimes} represents the truth table with all false rows and so all instances will be unsatisfiable, which means that this runtime is the time it takes for MapleSAT to realise the instance is unsatisfiable. More generally, one possible reason why truth tables with more false values take less time on average is because the algorithm is able to determine if the instance is unsatisfiable faster and doesn't have to search as large a space. In addition to the number of true/false values in the truth table, the structure of the problem also plays a key part in the MapleSAT runtime. For example a problem may superficially look like 3-SAT but may only depend on one or two variables making it easier to solve than 3-SAT, this is the cause of the spread of lines when fixing the number of true values.

\begin{figure}[h!]
\centering
\includegraphics[width=8cm]{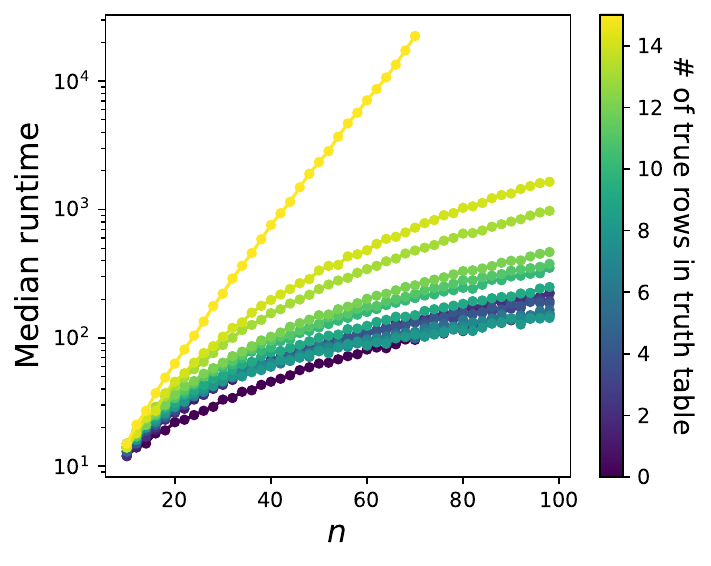}
\includegraphics[width=8cm]{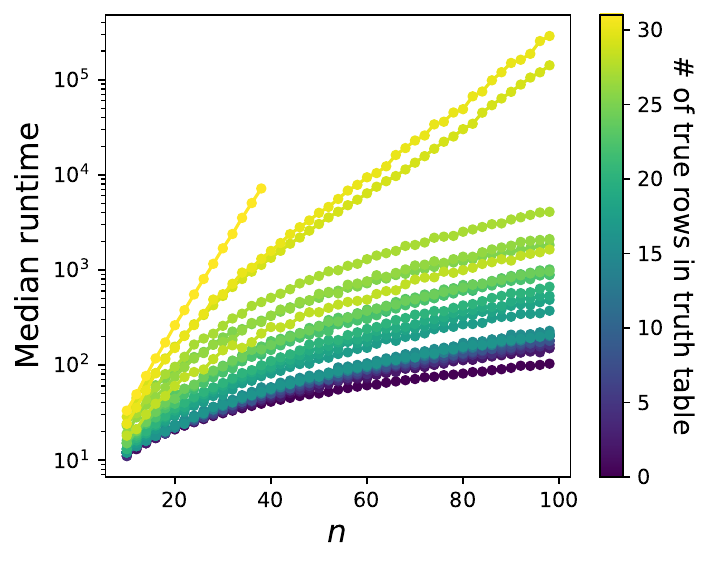}
\caption{The median runtime for MapleSAT for truth tables for $k=4$ (left) and $k=5$ (right) of the form $i$ false rows followed by $2^k-i$ true rows. } 
\label{fig:maplesat_k45_runtimes}
\end{figure}

This pattern also seems to hold for $k=4$ and $k=5$ in Figure~\ref{fig:maplesat_k45_runtimes}, where problems with only a few false rows in the truth table seem to take exponential time. In this case, we only plot one truth table for each Hamming weight as the number of truth tables is now very large. As $k$ increases, we expect more truth tables to begin to show exponential behaviour.

\subsection{Classical vs. Quantum Comparison}

\begin{figure}[t]
\centering
\includegraphics[width=10cm]{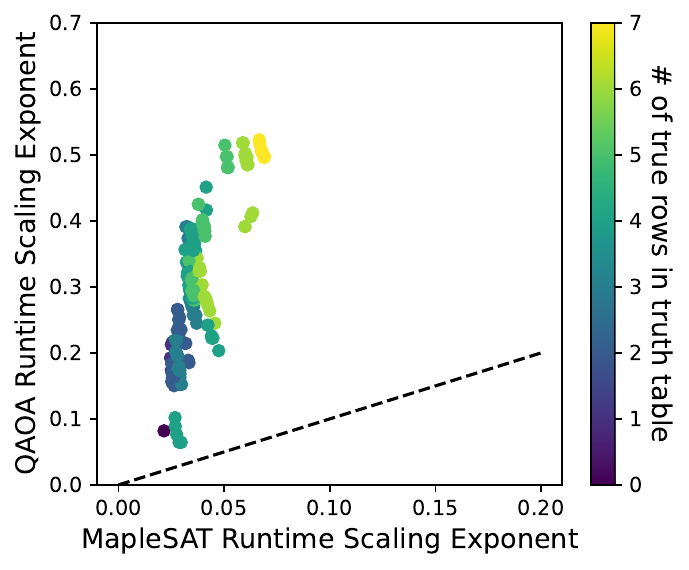}
\caption{Comparison of the QAOA and MapleSAT runtime scaling exponents for all $k=3$ truth tables, where the black dotted line corresponds to equal scaling. }
\label{fig:classical_vs_quantum_k3}
\end{figure}

\begin{figure}[t]
\centering
\includegraphics[width=8cm]{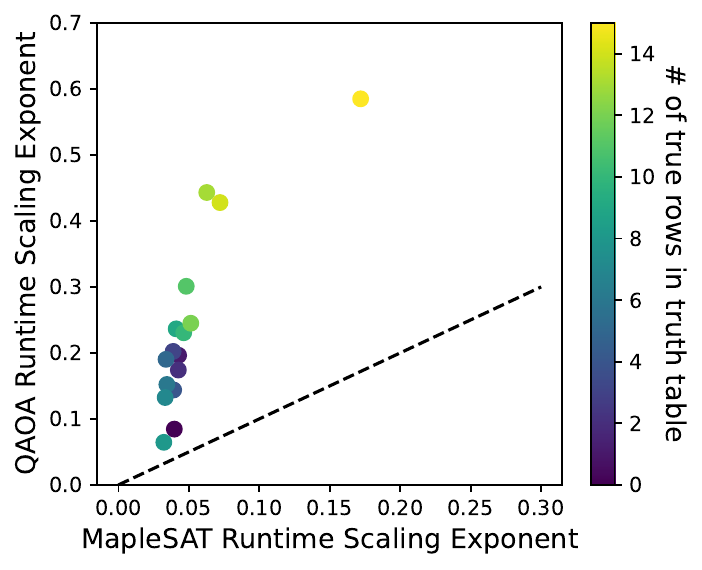}
\includegraphics[width=8cm]{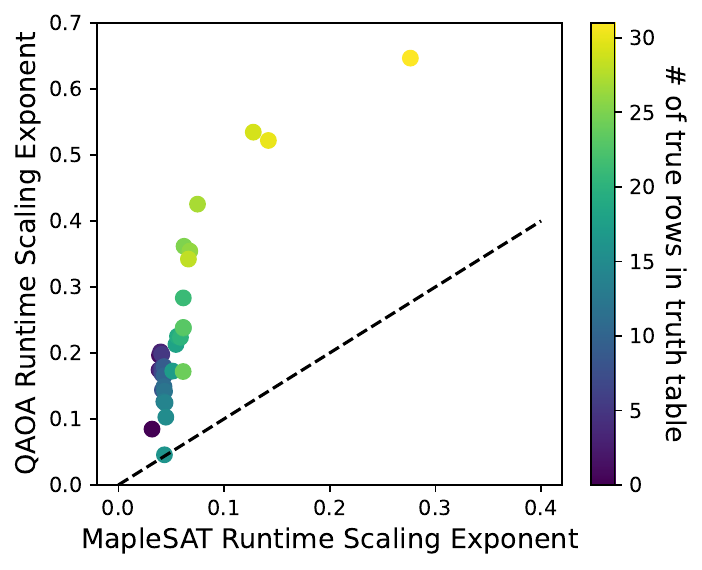}
\caption{Comparison of the QAOA and MapleSAT runtime scaling exponents for $k=4$ (left) and $k=5$ (right) truth tables. }
\label{fig:classical_vs_quantum_k45}
\end{figure}

For both the classical and quantum algorithms, we assume the running time is exponential of the form $T(n) = c \cdot 2^{\alpha n}$, and we can then compare the different values of $\alpha$ in each case, which we call the runtime scaling exponent. A smaller value is desirable as the running time then increases more slowly with $n$.
The scaling exponents for $k=3$ are shown in Figure \ref{fig:classical_vs_quantum_k3}. The black line shows where $\alpha_{classical} = \alpha_{quantum}$, and the colours are set by the Hamming weight of the truth table. In this case, most of the problems are very easy for MapleSAT to solve and do not require exponential time resulting a very small coefficient.
Similarly the scaling exponents for $k=4$ and $k=5$ are shown in Figure \ref{fig:classical_vs_quantum_k45}, where only one truth table for each Hamming weight is shown. In all of these cases QAOA has a larger scaling exponent than MapleSAT, however for larger $k$ and more QAOA layers $p$ it is possible that QAOA may outperform classical solvers as was demonstrated for $k$-SAT in~\cite{2208.06909}.

\begin{figure}[t]
\centering
\includegraphics[width=8cm]{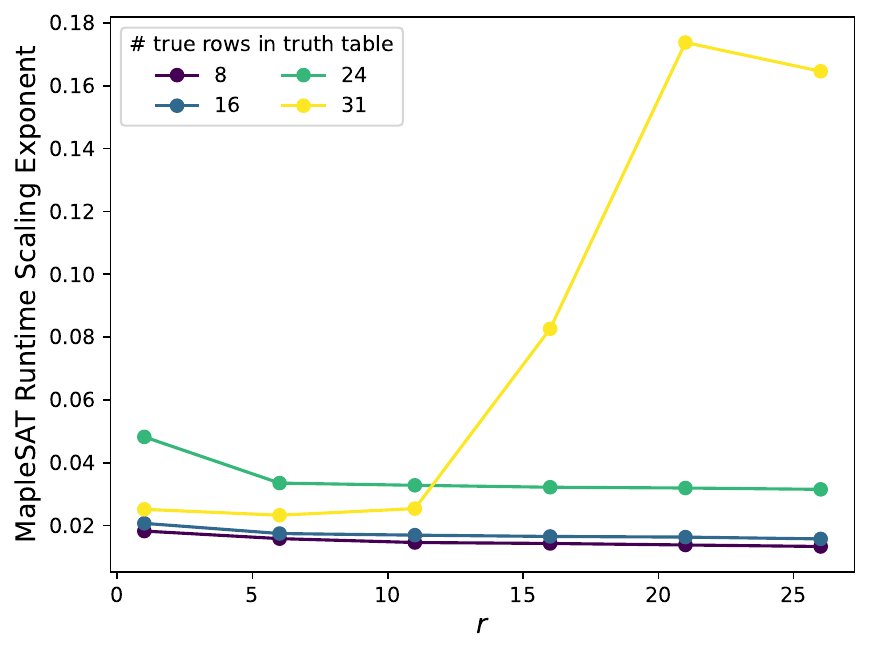}
\includegraphics[width=8cm]{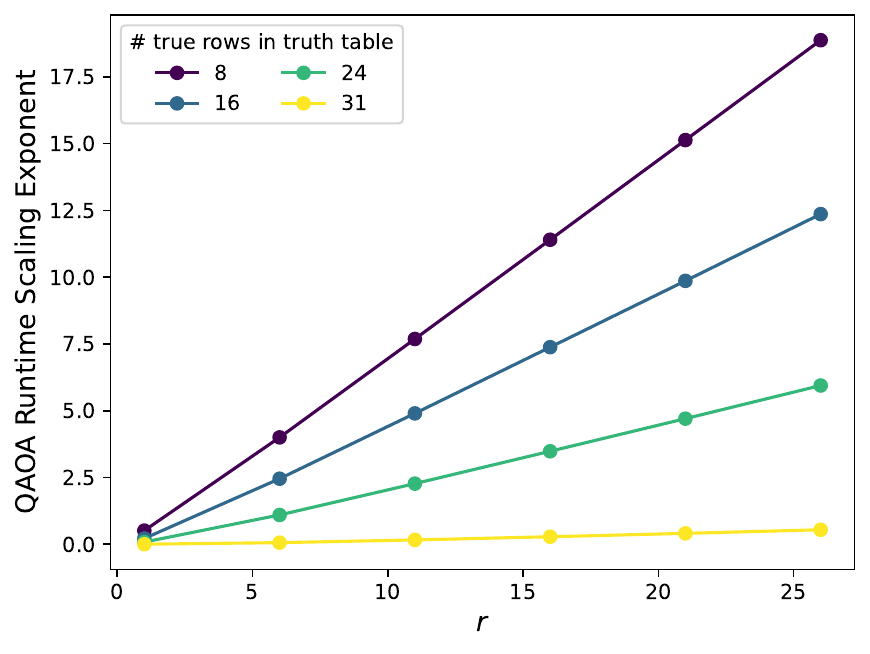}
\caption{The classical and quantum scaling exponents of a sample of truth tables for $k=5$ and varying values of the clause-to-variable ratio $r$. The quantum scaling exponents are much larger (note different scales).}
\label{fig:exponents_vary_r}
\end{figure}

Finally, for a selection of $k=5$ truth tables, we calculate the scaling exponents for various clause-to-variable ratios $r$. Figure~\ref{fig:exponents_vary_r} shows that, as expected, the QAOA exponent depends linearly on the value of $r$, likely due to the $e^{rn}$ factor in the success probability equation in Proposition~\ref{prop:success_probability_general_formula}. More unexpectedly, when using MapleSAT, for the small instances that we were able to consider, for most of the problem types, there did not seem to be a significant dependence on the value of $r$. This is another component to consider when comparing scaling exponents and determining whether QAOA could outperform classical solvers.

\section{Conclusions}
In this work, we have developed a theoretical formula that allows one to predict the $p=1$ QAOA success probability for constraint satisfaction problems. This formula has been applied to a variety of problems specified by truth tables of the constraints, for $k$-ary constraints with $k \le 5$, where the clauses are chosen at random. Based on our results, $k$-SAT appears to be the problem with the highest potential to deliver a quantum-classical separation. None of the problems we considered demonstrated an improved running time scaling of QAOA when compared with MapleSAT, which is as expected given that we were only able to analyse $p=1$, which even in the case of $k$-SAT does not achieve a speedup~\cite{2208.06909}.

A key question which cannot be answered conclusively with our techniques is whether the relative difficulty of the families of CSPs considered would be maintained for higher $p$. If this is the case, then random $k$-SAT may be the most promising constraint satisfaction problem of the form that we consider to demonstrate a quantum speedup.

While the complexity of our algorithm for determining QAOA success probabilities for $p=1$ is formally polynomial in $n$ and $k$, in practice its cubic dependence on $n$ and 7th power dependence on $k$ limits us to the approximate range $n\le 30$, $k \le 5$. It would be interesting to be able to extend our results to larger $n$ and $k$ via a more efficient algorithm.

Several improvements to these complexities could be considered. First, all other things equal, one may consider extracting the scaling exponents of success probability analytically as $n \longrightarrow \infty$, rather than from empirical fits over finite $n$ calculations. This would likely require to develop a \textit{generalized multinomial theorem}, similar to the approach of recent works on QAOA \cite{Basso2022,2208.06909}. Besides the infinite-size limit, another interesting extension of this work would be to decrease the complexity of our formulae in $k$ for certain choices of constraint satisfaction problems. As an example, for the $1$-in-$k$-SAT and NAE-SAT special cases treated in the appendix, as well as for $k$-SAT \cite{2208.06909}, the complexity is independent of $k$ rather than $\mathcal{O}\left(k^7\right)$, showing potential for major improvements. Given this better understanding of the infinite-size limit and the complexity in $k$, one may then consider extending our results to the $p > 1$ QAOA. While this was proven relatively tractable in the special case of $k$-SAT \cite{2208.06909}, with a complexity $4^p$ independent of $k$, for more general problems we expect a complexity scaling exponentially in $pk$ based on related work~\cite{Basso2022}.

\subsection*{Acknowledgements}

This project was supported by the European Research Council (ERC) under the European Union’s Horizon 2020 research and innovation programme (grant agreement No.\ 817581) and was supported by InnovateUK grant 10031626, ``Near-term quantum computing for solving hard industrial optimisation problems''.

\appendix

\section{Two special examples: $1$-in-$k$-SAT and NAE-SAT}

In section \ref{sec:generalformulae}, we introduced general formulae to compute the instance-averaged success probability of QAOA for an arbitrary random constraint satisfaction problem as specified in definition \ref{def:csp_random_instance}. A special case, where the truth table is ``of Hamming weight type'', was studied in section \ref{sec:general_formulae_hamming_weight_truth_table}. In this section, we specialize the discussion even further by considering two concrete examples of problems with a Hamming weight truth tables: $1$-in-$k$-SAT and NAE-SAT. These examples provide a more concrete illustration of the combinatorial calculations from sections \ref{sec:general_formulae_hamming_weight_truth_table}, \ref{sec:general_formulae_general_truth_table}. Besides, the formulae obtained in these cases for the single-clause polynomial can be evaluated with complexity independent of $k$, unlike the complexity $\mathcal{O}\left(k^7\right)$ complexity generic formulae from propositions \ref{prop:single_clause_polynomial_hamming_weight_truth_table_explicit_expression}, \ref{prop:single_clause_expectation_general_truth_table}. In this appendix we will change the distribution on random clauses slightly to consider clauses whose literals are chosen without replacement.

\begin{definition}[Random problem instance, without variables repetition in clauses]
	\label{def:csp_random_instance_without_repetition}
	Let be given a truth table $T$ on $k$ bits as introduced in definition \ref{def:general_truth_table} and integers $n \geq 1$ (number of variables) and $m \geq 0$ (number of clauses). A random instance of a constraint satisfaction problem with $n$ variables and $m$ clauses is constructed by sampling $m$ clauses $\sigma_j$, $0 \leq j < m$, identically independently. Each clause
	\begin{align}
		\sigma_j = \left(\left(l_{j,\,0},\,\nu_{j,\,0}\right),\,\left(l_{j,\,1},\,\nu_{j,\,1}\right),\,\ldots,\,\left(l_{j,\,k - 1},\,\nu_{j,\,k - 1}\right)\right)
	\end{align}
	is constructed by choosing variable indices
        \begin{align}
            l_{j, 0},\,l_{j, 1},\,\ldots,\,l_{j, k - 1}
        \end{align}
        uniformly among all combinations of $k$ elements from $\{0, 1, \ldots, n - 1\}$ (without repetition), and each negation $\nu_{j,\,0}$ independently with equal probabilities for $0$ (no negation applied) or $1$ (negation applied). We usually denote a single clause randomly sampled from this distribution by $\sigma$, and denote by
	\begin{align}
		\mathbf{E}_{\sigma}f\left(\sigma\right) \qquad \sigma: \mathrm{random\,clause}
	\end{align}
	the expectation of a function $f(\sigma)$ of this random clause.
\end{definition}

\subsection{$1$-in-$k$ SAT}
\label{oneinksat}

In this section, we evaluate the instance-averaged success probability of $p = 1$ QAOA on $1$-in-$k$-SAT according to the formula stated in proposition \ref{prop:success_probability_general_formula}. This amounts to computing the single-clause polynomial, i.e.
\begin{align}
    P_{\mathrm{single}}\left(n'_{000}, n'_{001}, n'_{010}, n'_{011}\right) & = \mathbf{E}_{\sigma}\left[\exp\left(\frac{i\gamma}{2}\left(\mathbf{1}\left[\bm{z}^{[1]} \vdash \sigma\right] - \mathbf{1}\left[\bm{z}^{[-1]} \vdash \sigma\right]\right)\right)\mathbf{1}\left[\bm{z}^{[0]} \vdash \sigma\right]\right],
\end{align}
where $\mathbf{E}_{\sigma}$ denotes a single random clause and $\left(\bm{z}^{[1]}, \bm{z}^{[0]}, \bm{z}^{[-1]}\right)$ is any bitstring triplet satisfying reduced configuration basis numbers $\left(n'_{000}, n'_{001}, n'_{010}, n'_{011}\right)$. For the example treated in this section, it will be most convenient to expand the single-clause polynomial as a Boolean polynomial in the indicator functions:
\begin{align*}
    & P_{\mathrm{single}}\left(n'_{000}, n'_{001}, n'_{010}, n'_{011}\right)\\
    & = \mathbf{E}_{\sigma}\left[\left(1 + \left(e^{i\gamma/2} - 1\right)\mathbf{1}\left[\bm{z}^{[1]} \vdash \sigma\right]\right)\left(1 + \left(e^{-i\gamma/2} - 1\right)\mathbf{1}\left[\bm{z}^{[-1]} \vdash \sigma\right]\right)\mathbf{1}\left[\bm{z}^{[0]} \vdash \sigma\right]\right]\\
    & = \mathbf{E}_{\sigma}\left[\mathbf{1}\left[\bm{z}^{[0]} \vdash \sigma\right]\right] + \left(e^{i\gamma/2} - 1\right)\mathbf{E}_{\sigma}\left[\mathbf{1}\left[\bm{z}^{[1]} \vdash \sigma\right]\mathbf{1}\left[\bm{z}^{[0]} \vdash \sigma\right]\right] + \left(e^{-i\gamma/2} - 1\right)\mathbf{E}_{\sigma}\left[\mathbf{1}\left[\bm{z}^{[0]} \vdash \sigma\right]\mathbf{1}\left[\bm{z}^{[-1]} \vdash \sigma\right]\right]\nonumber\\
    & \hspace*{10px} + 4\sin^2\frac{\gamma}{4}\mathbf{E}_{\sigma}\left[\mathbf{1}\left[\bm{z}^{[1]} 
    \vdash \sigma\right]\mathbf{1}\left[\bm{z}^{[0]} 
    \vdash \sigma\right]\mathbf{1}\left[\bm{z}^{[-1]} 
    \vdash \sigma\right]\right]\\
    & = \mathbf{E}_{\sigma}\mathbf{1}\left[\bm{z}^{[0]} \vdash \sigma\right] + \left(e^{i\gamma/2} - 1\right)\mathbf{1}\left[\bm{z}^{[1]}, \bm{z}^{[0]} \vdash \sigma\right] + \left(e^{-i\gamma/2} - 1\right)\mathbf{E}_{\sigma}\mathbf{1}\left[\bm{z}^{[0]}, \bm{z}^{[-1]} \vdash \sigma\right] + 4\sin^2\frac{\gamma}{4}\mathbf{E}_{\sigma}\mathbf{1}\left[\bm{z}^{[1]}, \bm{z}^{[0]}, \bm{z}^{[-1]} \vdash \sigma\right]
\end{align*}

 In order to evaluate the above, we need to compute the quantities
\begin{align}
\label{eq:one_in_ksat_to_compute_bitstring_clause_expectations}
    \mathbf{E}_{\sigma}\mathbf{1}\left[\bm{z}^{[0]}\vdash \sigma\right], && \mathbf{E}_{\sigma}\mathbf{1}\left[\bm{z}^{[0]}, \bm{z}^{[1]}\vdash \sigma \right], && \mathbf{E}_{\sigma}\mathbf{1}\left[\bm{z}^{[0]}, \bm{z}^{[1]}, \bm{z}^{[-1]}\vdash \sigma \right].
\end{align}
First we will do this analytically for the $1$-in-$k$ SAT problem. This problem corresponds to the function $T:\{0,1\}^k \to \{0,1\}$ where $T(x) = 1 \Leftrightarrow |x|=1$, and $|x|$ denotes the Hamming weight. That is, each clause is satisfied if and only if exactly one literal evaluates to true. Here, we assume that each clause has exactly $k$ literals, which is often known as the Exact 1-in-$k$ SAT problem in the literature. Recall that we consider random instances where the literals in each clause are uniformly random.

To compute the average success probability of QAOA, we need to evaluate equation \ref{eq:qaoa_success_probability}, giving the instance-averaged success probability. Despite our general definition \ref{def:csp_random_instance} allowing for repetition of variables in the construction of random clause, we will for this specific problem consider a choice without repetition. To evaluate equation \ref{eq:qaoa_success_probability}, we need to compute the following quantities:

\begin{align}
\label{eq:one_in_ksat_to_compute_bitstring_clause_expectations}
    \mathbf{E}_{\sigma}\mathbf{1}\left[\bm{z}^{[0]}\vdash \sigma\right], && \mathbf{E}_{\sigma}\mathbf{1}\left[\bm{z}^{[0]}, \bm{z}^{[1]}\vdash \sigma \right], && \mathbf{E}_{\sigma}\mathbf{1}\left[\bm{z}^{[0]}, \bm{z}^{[1]}, \bm{z}^{[-1]}\vdash \sigma \right],
\end{align}
and we compute these quantities in terms of $n_s$ where in our case ($p=1$), $s\in \{0,1\}^3 \equiv (s_2,s_1,s_0)$. Thus, the first quantity will depend on $n_0,n_1$, the second quantity on $n_{00},n_{01},n_{10},n_{11}$ (where $n_{ij} = \sum_{l=0,1} n_{ijl}$) and so on, fulfilling that $\sum_{i,j,l=0,1} n_{ijl} = n$.\\\mbox{}

\textbf{1 bitstring case}\\\mbox{}\\
In this case, we have to consider the different possibilities for the true variable knowing that we have $k$ literals. Hence, the probability is the number of possible positions for the true variable divided by the total number of possible configurations for $k$ literals:

\begin{equation}
\mathbf{E}_{\sigma}\mathbf{1}\left[\bm{z}^{[1]} \vdash \sigma\right]= \frac{k}{2^k}
\end{equation}

\textbf{2 bitstring case}\\\mbox{}\\
In this case (as in the following with 3 bitstrings), a distinction must be made between the case in which the position of the true literal coincides in both bitstrings and the case in which it differs. First, let's consider the case in which the true literal in the clause $\sigma$ is at the same position in $z^{[1]}$ and $z^{[0]}$.
This implies that the values of the $k$ literals in $\sigma$ coincide at all positions in $z^{[1]}$ and $z^{[0]}$. Then:

\begin{equation}
\mathbf{E}_{\sigma}\left[\mathbf{1}\left[\bm{z}^{[1]},\bm{z}^{[0]} \vdash \sigma\right]\,:\,\textrm{identical positions}\right] = \frac{k}{2^k} \frac{\binom{n_{00} + n_{11}}{k}}{\binom{n}{k}},
\end{equation}
where we are multiplying the probability that the $k$ variables involved in the clause coincide with the probability that these variables satisfy the clause. 

In the second case, the position of the true literal differs in the two bitstrings. This implies that, when the clause is satisfied, two positions differ and the rest coincide. Hence:

\begin{equation}
    \mathbf{E}_{\sigma}\left[\mathbf{1}\left[\bm{z}^{[1]},\bm{z}^{[0]} \vdash \sigma\right]\,;\,\textrm{different positions}\right]=  \frac{2\binom{n_{00} + n_{11}}{k-2}\binom{n_{01}+n_{10}}{2}}{\binom{n}{k}}\frac{1}{2^k},
\end{equation}
where we are multiplying the probability of choosing two bits that don't coincide and $(k-2)$ bits that coincide times the probability that this clause is satisfied. 

All in all, we have:

\begin{equation}
    \mathbf{E}_{\sigma}\mathbf{1}\left[\bm{z}^{[1]},\bm{z}^{[0]} \vdash \sigma\right] = \frac{1}{2^k}\frac{1}{\binom{n}{k}}\left[k\binom{n_{00}+n_{11}}{k}+2\binom{n_{00}+n_{11}}{k-2}\binom{n_{01}+n_{10}}{2}\right].
\end{equation}

\textbf{3 bitstring case}\\\mbox{}\\

In this case, we have 5 disjoint cases: the one in which the positions of the true literals coincide in the 3 bitstrings, the one in which they don't coincide between any pair of bitstrings and 3 cases in which 2 bitstrings have the true literal in the same position and the third one differs. These are: $\{1\} \sqcup \{0\} \sqcup \{-1\}$, $\{1, 0\} \sqcup \{-1\}$, $\{1, -1\} \sqcup \{0\}$, $\{0, -1\} \sqcup \{-1\}$, $\{1, 0, -1\}$. Let's study the different cases:

\begin{itemize}
    \item $\{1, 0, -1\}$. This case corresponds to having the same position for the true literal in the three bitstrings. This implies that all the literals coincide in the three bitstrings. Then, analogously to the 2 bitstrings case, the probability would be:
    \begin{equation}
        \mathbf{E}_{\sigma}\mathbf{1}\left[\bm{z}^{[1]} \vdash \sigma, \bm{z}^{[0]} \vdash \sigma, \bm{z}^{[-1]} \vdash \sigma\,:\,\{1, 0, -1\}\right] = \frac{\binom{n_{000} + n_{111}}{k}}{\binom{n}{k}}\frac{k}{2^k}
    \end{equation}
    \item $\{1, 0\} \sqcup \{-1\}$. This means that $z^{[1]}$ and $z^{[0]}$ satisfy the clause by means of a true literal in the same position while $z^{[-1]}$ satisfies it at a different position. Hence, the configuration is 001/110. The number of possibilities with this configuration is $\binom{n_{001}+n_{110}}{2}\binom{n_{000}+n_{111}}{k-2}$ and the number of possible assignments of negations to satisfy the clause is 2. Then:
    \begin{equation}
        \mathbf{E}_{\sigma}\mathbf{1}\left[\bm{z}^{[1]} \vdash \sigma, \bm{z}^{[0]} \vdash \sigma, \bm{z}^{[-1]} \vdash \sigma\,:\,\{1, 0\} \sqcup \{-1\}\right] = \frac{2\binom{n_{001}+n_{110}}{2}\binom{n_{000} + n_{111}}{k-2}}{\binom{n}{k}}\frac{1}{2^k}
    \end{equation}
    \item $\{1, -1\} \sqcup \{0\}$. Following an analogous procedure:
    \begin{equation}
        \mathbf{E}_{\sigma}\left[\mathbf{1}\left[\bm{z}^{[1]} \vdash \sigma, \bm{z}^{[0]} \vdash \sigma, \bm{z}^{[-1]} \vdash \sigma\right]\,:\,\{1, -1\} \sqcup \{0\}\right] = \frac{2\binom{n_{010} + n_{101}}{2}\binom{n_{000} + n_{111}}{k - 2}}{\binom{n}{k}}\frac{1}{2^k}
    \end{equation}
    \item $\{0, -1\} \sqcup \{-1\}$. Again, in a similar way:
    \begin{equation}
        \mathbf{E}_{\sigma}\left[\mathbf{1}\left[\bm{z}^{[1]} \vdash \sigma, \bm{z}^{[0]} \vdash \sigma, \bm{z}^{[-1]} \vdash \sigma\right]\,:\,\{1, -1\} \sqcup \{0\}\right] = \frac{2\binom{n_{011} + n_{100}}{2}\binom{n_{000} + n_{111}}{k - 2}}{\binom{n}{k}}\frac{1}{2^k}
    \end{equation}
    \item $\{1\} \sqcup \{0\} \sqcup \{-1\}$. In this case, the three bitstrings satisfy the clause in different positions. Therefore, the configuration would be 001/110,010/101,011/100. The number of possibilities for the positions of true literals are, therefore, $(n_{001}+n_{110})(n_{010}+n_{101})(n_{011}+n_{100})$. In the end, we have:
    \begin{align}
        \mathbf{E}_{\sigma}\left[\mathbf{1}\left[\bm{z}^{[1]} \vdash \sigma,\bm{z}^{[0]} \vdash \sigma, \bm{z}^{[-1]} \vdash \sigma\right]\,:\,\{1, -1\} \sqcup \{0\}\right] &=\frac{(n_{001}+n_{110})(n_{010}+n_{101})(n_{011}+n_{100})\binom{n_{000} + n_{111}}{k - 3}}{\binom{n}{k}}\frac{1}{2^k}
    \end{align}
\end{itemize}
Taking everything into account, we have:

\begin{align*}
    &
    \mathbf{E}_{\sigma}\mathbf{1}\left[\bm{z}^{[1]} \vdash \sigma, \bm{z}^{[0]} \vdash \sigma, \bm{z}^{[-1]} \vdash \sigma\right]\\
    &=\frac{1}{2^k\binom{n}{k}}k\binom{n_{000}+n_{1111}}{k}+\frac{2}{2^k\binom{n}{k}}\binom{n_{000}+n_{111}}{k-2}\left[\binom{n_{001}+n_{110}}{2}+\binom{n_{010}+n_{101}}{2}+\binom{n_{011}+n_{100}}{2}\right]\\
    &+\frac{1}{2^k\binom{n}{k}}(n_{001}+n_{110})(n_{010}+n_{101})(n_{011}+n_{100})\binom{n_{000}+n_{111}}{k-3}
\end{align*}

\begin{figure}[t]
\centering
\includegraphics[width=11cm]{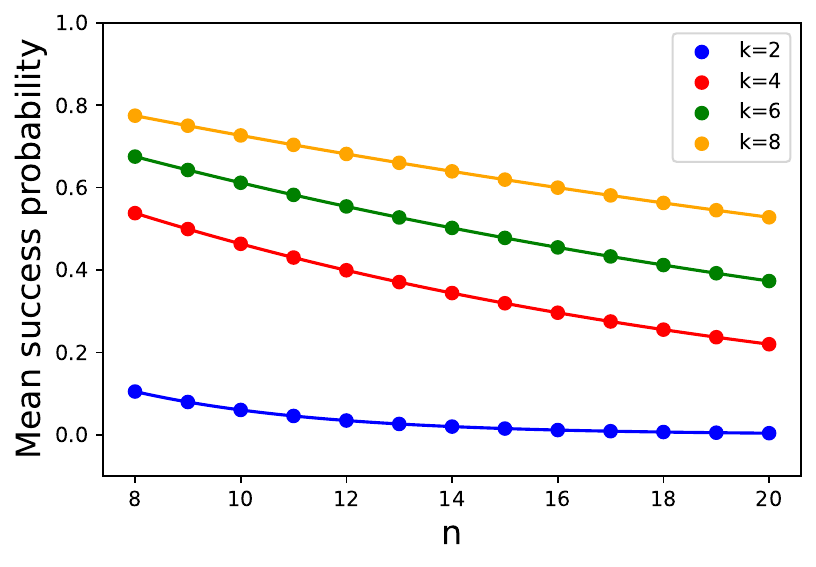}
\caption{Representation of the  prediction for the success probability for $p=1$ QAOA. The dots correspond to the data computed for the prediction of the success probability and the lines correspond to the exponential fittings of these data given in Table \ref{table1}.}
\label{fig:oneink_ks}
\end{figure}

Figure \ref{fig:oneink_ks} shows the prediction for the success probability with QAOA $p=1$ for different values of $k$. The QAOA parameters are chosen to achieve the optimal success probability. For each of these $k$, we let the clauses-to-variables ratio equal the satisfiability threshold \cite{ratio1ksat}:
\begin{align}
    r & = r\left(k\right) = \frac{2}{k(k - 1)}.
\end{align}

\begin{table}
\centering
\begin{tabular}{|l|l|l|l|l|l|}
\hline
$k$ & $r$            & $\beta$ & $\gamma$ & $a$          & $b$           \\
\hline
2 & 1            & 8.54159265               & 1.9                       & -0.04480352458 & -0.4009893105  \\
4 & 0.16666      & 5.74159265               & 3.2                       & -0.03328918685 & -0.1075926183  \\
6 & 0.06666      & 4.54159265               & 4.7                       & 0.00388386931 & -0.07125997391  \\
8 & 0.0357142857 & 4.84159265               & 5.6                       & 0.0004854429 & -0.04609253403 \\
\hline
\end{tabular}
\caption{Parameters for the computations done for the prediction of the $p=1$ QAOA success probability. r is the satisfiability threshold, $\beta$ and $\gamma$ are the QAOA angles optimized for the corresponding $k$ and $a$ and $b$ are the parameters of the exponential fitting for the previous parameters (where $p = 2^{a+bn}$).}
\label{table1}
\end{table}

If we compare with the results for k-SAT \cite{2208.06909}, where the values for the exponential fit for $k=8$ are $a = -0.2191994921686201$ and $b = -0.658520375254191$, we can see that success probability for the k-SAT problem decays much faster than for the $1$-in-k SAT problem ($b = -0.658520375254191$ respect $b= -0.04609253403$). In fact, we can see that the decay for $1$-in-k SAT is almost linear in the regime considered.

\begin{figure}[t]
\centering
\includegraphics[width=11cm]{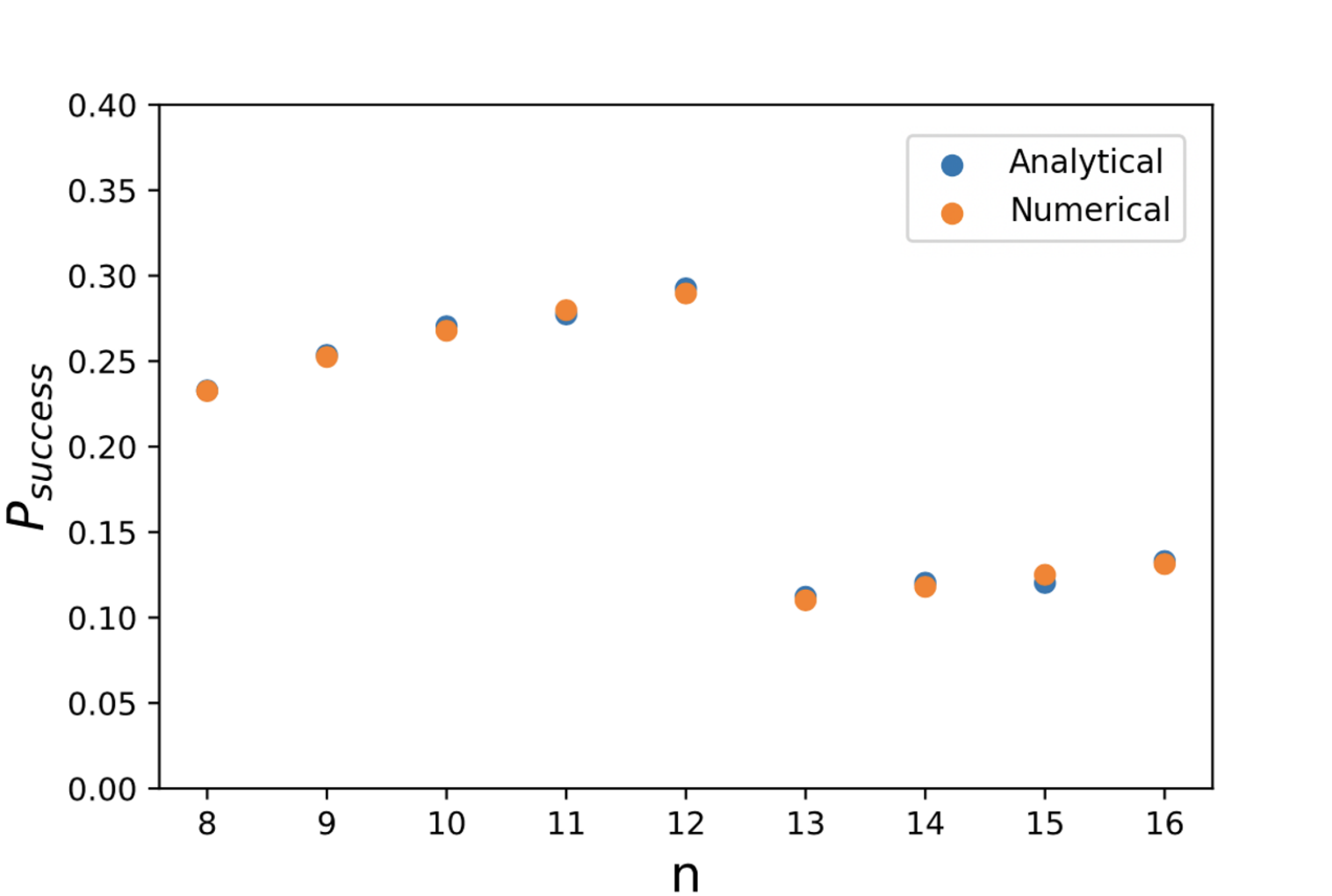}
\caption{Representation of the analytical and numerical prediction for the success probability for $p=1$ QAOA for ``fixed'' $m$.}
\label{fig:compfixedm}
\end{figure}

Until now, we have compared the analytical and numerical prediction when we choose $m$ from a Poisson distribution of mean $nr$, i.e. $m\sim \mathrm{Poisson}(nr)$. However, we can also study the comparison when we have a ``fixed'' $m$; in this case, the instance-averaged success probability is evaluated through equation \ref{eq:success_probability_general_formula_no_poisson} rather than equation \ref{eq:success_probability_general_formula}. If we choose $m$ and $n$, $r$ would be determined. Then, in order to be at the satisfiability threshold, we choose $m$ as $\lceil nr \rceil$. The results found for $k=4$ and 1000 instances are shown in Figure \ref{fig:compfixedm}.

\subsection{NAE-SAT}
\label{naesat}
We now study the instance averaged success probability of QAOA on a different constraint satisfaction problem: NAE-SAT (not-all-equal SAT). This problem corresponds to the function $T:\{0,1\}^k \to \{0,1\}$ where $T(x) = 1 \Leftrightarrow 0<|x|<k$, and $|x|$ denotes the Hamming weight. That is, each clause fails to be satisfied if all literals are false, or all literals are true.

Similar to the study of $1$-in-$k$-SAT (section \ref{oneinksat}), we will slightly amend definition \ref{def:csp_random_instance} of a random constraint satisfaction problem to consider clauses constructed from choices of variables without repetition. More precisely, in this case we will provide formulae for both the repetition and no-repetition case to illustrate the flexibility of our methods.\\\mbox{}

\textbf{1 bitstring case}\\\mbox{}\\

Given a fixed bitstring, there are 2 ways of negating the $k$ bits in the clause's scope so these bits become $0^k$ or $1^k$, i.e. violate the clause. Therefore:
\begin{align}
\label{eq:nae_sat_single_clause_expectation_1_bitstring}
    \mathbf{E}_{\sigma}\mathbf{1}\left[\bm{z}^{[1]} \not\vdash \sigma\right] & = \frac{2}{2^k}.
\end{align}
Notice that this single-clause expected value only depends on $k$ and, then, 

\begin{align}
    \mathbf{E}_{\sigma}\left\{\mathbf{1}\left[\bm{z}^{[1]} \not\vdash \sigma\right]\right\} = \mathbf{E}_{\sigma}\left\{\mathbf{1}\left[\bm{z}^{[0]} \not\vdash \sigma\right]\right\} = \mathbf{E}_{\sigma}\left\{\mathbf{1}\left[\bm{z}^{[-1]} \not\vdash \sigma\right]\right\}
\end{align}

\textbf{2 bitstring case}\\\mbox{}\\

We first consider the conditional probability of $\bm{z}^{[1]}, \bm{z}^{[0]}$ simultaenously violating the clause conditioned on the clause's scope $\mathcal{V}_{\sigma}$. If $\mathcal{V}_{\sigma}$ includes both indices in configuration $00/11$ and indices in configurations $01/10$, the conditional probability is $0$ since no choice of negations can make the bitstrings (restricted to $\mathcal{V}_{\sigma}$) simultaneously violate the clause. On the other hand, if all indices in $\mathcal{V}_{\sigma}$ are in configuration $00/11$ or all are in configuration $01/10$, there still are $2$ choices of negations making the clause violated, giving a conditional probability:
\begin{align}
\label{eq:nae_sat_single_clause_expectation_2_bitstrings}
    \mathbf{E}_{\sigma}\left[\mathbf{1}\left[\bm{z}^{[1]}, \bm{z}^{[0]} \not\vdash \sigma\right]\,\bigg|\,\mathcal{V}_{\sigma}\right] & = \frac{2}{2^k}.
\end{align}
Multiplying this by the probability of choosing the scope in the way described: $\frac{\binom{n_{00} + n_{11}}{k} + \binom{n_{01} + n_{10}}{k}}{\binom{n}{k}}$ without repetition, $\left(\frac{n_{00} + n_{11}}{n}\right)^k + \left(\frac{n_{01} + n_{10}}{n}\right)^k$ with repetition yields the $2$-bitstrings expectation:
\begin{align}
\label{eq:nae_sat_single_clause_expectation_3_bitstrings}
    \mathbf{E}_{\sigma}\mathbf{1}\left[\bm{z}^{[1]}, \bm{z}^{[0]} \not\vdash \sigma\right] & = \left\{\begin{array}{cc}
        \frac{\binom{n_{00} + n_{11}}{k} + \binom{n_{01} + n_{10}}{k}}{\binom{n}{k}}\frac{2}{2^k} & \textrm{without repetition}\\
        \left(\left(\frac{n_{00} + n_{11}}{n}\right)^k + \left(\frac{n_{01} + n_{10}}{n}\right)^k\right)\frac{2}{2^{k}} & \textrm{with repetition}
    \end{array}\right..
\end{align}

\textbf{3 bitstring case}\\\mbox{}\\

For the 3-bitstrings case, a similar reasoning gives:
\begin{align}
    & \mathbf{E}_{\sigma}\mathbf{1}\left[\bm{z}^{[1]}, \bm{z}^{[0]}, \bm{z}^{[-1]} \not\vdash \sigma\right]\nonumber\\
    & = \left\{\begin{array}{cc}
        \frac{\binom{n_{000} + n_{111}}{k} + \binom{n_{001} + n_{110}}{k} + \binom{n_{010} + n_{101}}{k} + \binom{n_{011} + n_{100}}{k}}{\binom{n}{k}}\frac{2}{2^k} & \textrm{without repetition}\\
        \left(\left(\frac{n_{000} + n_{111}}{n}\right)^k + \left(\frac{n_{001} + n_{110}}{n}\right)^k + \left(\frac{n_{010} + n_{101}}{n}\right)^k + \left(\frac{n_{011} + n_{100}}{n}\right)^k\right)\frac{2}{2^{k}} & \textrm{with repetition}
    \end{array}\right..
\end{align}

\begin{figure}[t]
\centering
\includegraphics[width=12cm]{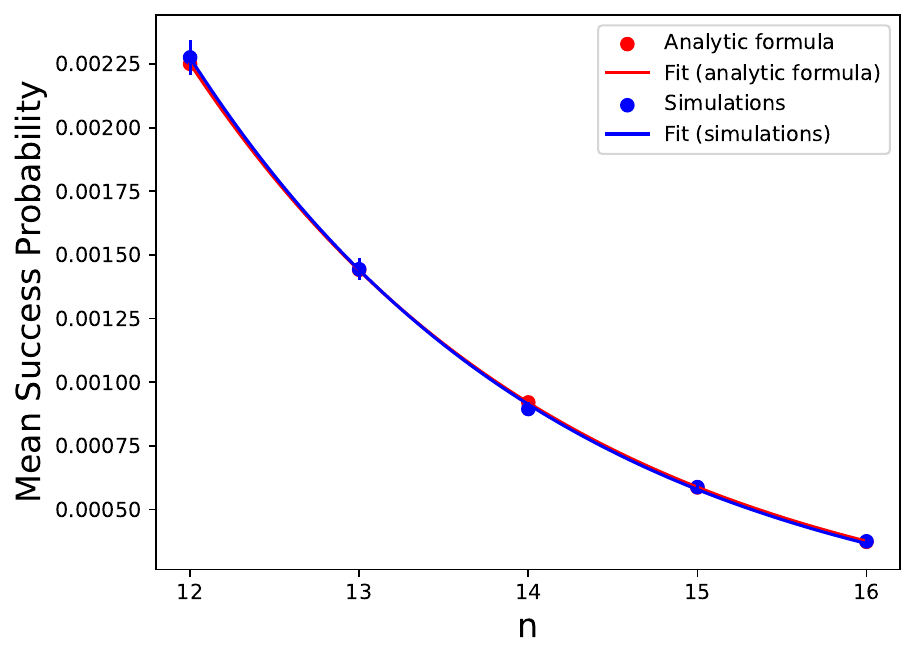}
\caption{Representation of the numerical and theoretical prediction for the success probability of NAE-SAT for $k = 8$.}
\label{fig:naesat}
\end{figure}

The numerical and theoretical prediction for the success probability averaging instances over Poisson distribution are shown in Figure \ref{fig:naesat}. The numerical evaluation of analytic formulae, and statevector simulations of QAOA, use the satisfiability threshold for the ratio obtained from \cite{ratioNAE} and the optimal values for the QAOA angles. For statevector simulations, 2500 random instances were generated for each instance size $12 \leq n \leq 16$.

\begin{figure}[t]
\centering
\includegraphics[width=12cm]{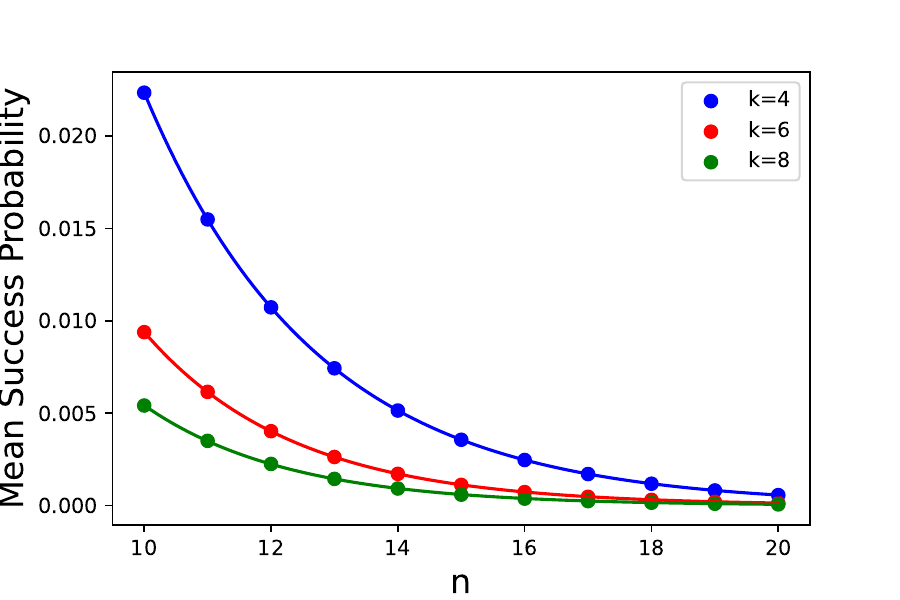}
\caption{Representation of the  prediction for the success probability for $p=1$ QAOA for different values of $k$. The dots represent the data computed for the prediction of the success probability and the lines the exponential fittings of this data (given in table \ref{table2}).}
\label{fig:naesat_ks}
\end{figure}

After testing the validity of our formula, Figure \ref{fig:naesat_ks} shows the prediction for the success probability for QAOA $p=1$ for different values of $k$.

\begin{table}
\[
\centering
\begin{array}{|l|l|l|l|l|l|}
\hline
k  & r                  & \beta & \gamma & a           & b            \\
\hline
4  & 4.972710556317915  & 5.5                  & 1.1                   & -0.1770989003   & -0.5304733689   \\
6  & 21.583456938459364 & 5.6                  & 0.9                   & -0.5625334157  & -0.6162079454    \\
8  & 88.12349051732973  & 5.7                  & 0.8                   & -1.044659548  & -0.6458301102  \\
\hline
\end{array}
\]
\caption{Parameters for the computations done for the prediction of the $p=1$ QAOA success probability. $r$ is the satisfiability threshold, $\beta$ and $\gamma$ are the QAOA angles optimized for the corresponding $k$ and $a$ and $b$ are the parameters of the exponential fitting for the previous parameters (where $p = 2^{a+bn}$).}
\label{table2}
\end{table}
If we compare with the results of $k$-SAT \cite{2208.06909}, where the values for the exponential fit for $k=8$ are $a = -0.2191994921686201$ and $b = -0.658520375254191$, we can see that success probability for the k-SAT problem has a similar decay to the one in NAE-SAT problem ($b = -0.658520375254191$ resp.\ $b=-0.6458301102$), although the values for the probability are approximately half.

\bibliographystyle{mybibstyle}
\bibliography{bibliography}

\end{document}